\documentclass[]{scrartcl}

\usepackage{amsmath}
\usepackage{amssymb}
\usepackage{algorithm}
\usepackage{algorithmicx}
\usepackage{algpseudocode}
\usepackage{multirow}
\usepackage{colortbl}
\usepackage{booktabs}
\usepackage{url}
\usepackage{epstopdf}
\usepackage{xcolor}
\usepackage{changes}
\usepackage{verbatim}
\usepackage{graphicx}
\usepackage{subfig}

\title{Bayesian Image Super-Resolution with Deep Modeling of Image Statistics}
\author{Shangqi~Gao and~Xiahai Zhuang$^*$\\
	School of Data Science, Fudan University, Shanghai, China}

\begin{document}

\maketitle

\begin{abstract}
Modeling statistics of image priors is useful for image super-resolution, but little attention has been paid from the massive works of deep learning-based methods. In this work, we propose a Bayesian image restoration framework, where natural image statistics are modeled with the combination of smoothness and sparsity priors. Concretely, firstly we consider an ideal image as the sum of a smoothness component and a sparsity residual, and model real image degradation including blurring, downscaling, and noise corruption. Then, we develop a variational Bayesian approach to infer their posteriors. Finally, we implement the variational approach for single image super-resolution (SISR) using deep neural networks, and propose an unsupervised training strategy. The experiments on three image restoration tasks, \textit{i.e.,} ideal SISR, realistic SISR, and real-world SISR, demonstrate that our method has superior model generalizability against varying noise levels and degradation kernels and is effective in unsupervised SISR. The code and resulting models are released via \url{https://zmiclab.github.io/projects.html}.
\end{abstract}

\section{Introduction}\label{sec:introduction}
Single image super-resolution (SISR), aiming to recover high-resolution (HR) images from low-resolution (LR) observations, is a typical task of image restoration. Image restoration (IR) has many significant applications, such as low-level image processing \cite{Celebi/2014}, medical imaging \cite{Jan/2006}, and remote sensing \cite{Jensen/1995}. Thanks to the advance of deep learning, studying IR becomes more popular in computer vision. Particular efforts have been made to explore the end-to-end IR frameworks for many applications \cite{Dong/2016,Zhang_beyond/2017,Lim/2017,Tong/2017,Lai/2017}. Although the approaches deliver promising performance on synthetic data, directly transferring them to real-world images often undergoes a great decrease in performance, meaning the resulting models could suffer from poor generalization ability. In reality, the ground truth of images is unavailable, and thus unsupervised learning is more challenging. 

Current methods could be categorized into two groups, i.e., the model-based and the learning-based schemes \cite{Dong/2019}. Model-based IR represents image degradation as analytical or statistical models \cite{Tikhonov/1977,Rudin/1992}, and it aims to restore a degraded image without using any further data. This problem is known as being ill-posed. Therefore, many image priors were proposed to model the domain knowledge of natural images, such as Gaussian priors \cite{Hunt/1977}, Markov random field (MRF) \cite{Qian/1993}, sparsity priors \cite{Portilla/2015}, and low-rank priors \cite{Candes_tight/2011}. Many of them could not perfectly model image priors due to the complex structure of real-world images. Therefore, modeling image structure is still active and challenging.  

Learning-based IR aims to learn the mappings from degraded spaces to the original space \cite{Dong/2016,Zhang_beyond/2017}. Deep neural networks (DNNs) are widely used to learn the mappings due to their powerful ability in modeling complex functions. One of such networks is the deep convolutional neural networks (CNNs), which were widely adopted in image denoising \cite{Zhang_beyond/2017,Zhang_ffdnet/2018}, deblurring \cite{Nah/2017} as well as super-resolution \cite{Dong/2016,Lim/2017}, and achieved promising performance. For example, the residual networks (ResNet) were firstly proposed for the task of classification \cite{He/2016}, which were then successfully applied in SISR and achieved superior performance against previous works \cite{Lim/2017}. 

The great majority of SISR models trained on ideal data \cite{Dong/2016,Lim/2017,Kim/2016}, e.g., synthesized by bicubic interpolation, cannot generalize well when LR images include noise. To rectify the weakness, one can model image priors explicitly, and then restore them via Bayesian inference. Bigdeli et al. \cite{Bigdeli/2017} proposed to estimate image priors using pre-trained denoising autoencoders and restored images via maximum a \textit{posteriori} (MAP). Their restoration problem was solved iteratively, which could be computationally expensive. The multivariate Gaussian prior was adopted to model clean images in the recent four denoising works, including self-supervised Bayesian image denoising \cite{Laine/2019}, variational denoising network \cite{Yue/2019}, blind universal Bayesian image denoising \cite{Helou/2020}, and patch-based non-local Bayesian networks \cite{Izadi/2020}. However, the methods cannot deal with the problem of SISR, since they did not involve blurring and downscaling in their modeling. 

Many of SISR models were developed for supervised SISR \cite{Zhangyulun/2018,Ledig/2017,Lugmayr/2020}, and thus cannot be used in real-world scenarios where the ground truth is unavailable. To tackle the difficulty, Shocher et al. \cite{Shocher/2018} used the information of a single image itself for internal learning, but the method requires long inference time due to thousands of gradient updates. To improve its efficiency, Soh et al. \cite{Soh/2020} used meta-learning to find suitable initial parameters. Besides, Ulyanov et al. \cite{Ulyanov/2018} showed that the deep image prior extracted by randomly initialized DNNs could be used as a handcrafted prior for SISR, but its inference is time-consuming due to thousands of iterations. Recently, several models based on generative adversarial network (GAN) were developed to super-resolve real-world images using unpaired LR and HR images \cite{Bulat/2018,Maeda/2020,Lugmayr/2019}, but the training of these DNN models \textit{per se} can be challenging.

\begin{figure}[!t]
	\centering
	\includegraphics[width=0.9\linewidth]{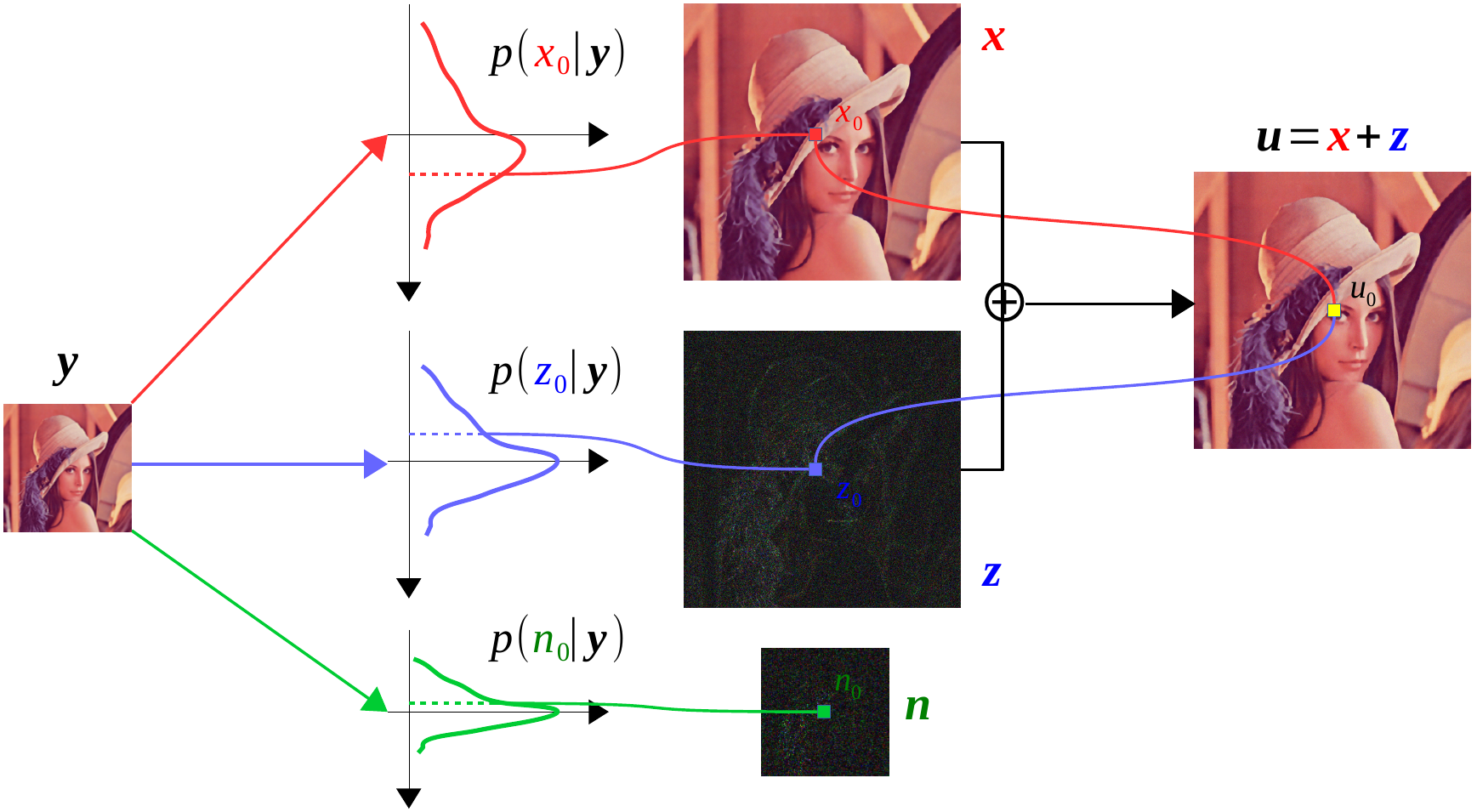}
	\caption{Diagram of super-resolving a low-resolution image. We first infer the pixel-wise distributions of a smoothness component $ \mathbf x $, a sparsity residual $ \mathbf z $, and a noise $ \mathbf n $ from an observation $ \mathbf y $, where we only visualize the distribution of one pixel for each component. Then, we reconstruct a high-resolution image $ \mathbf u $ by randomly sampling $ \mathbf x $ and $ \mathbf z $ from their distributions. One can refer to the text in introduction section for details.}
	\label{fig:diagram}
\end{figure}

In this work, we aim to build a Bayesian image restoration framework by explicit modeling of image priors. Most of learning-based methods do not model image priors, and often use the mean squared error (MSE) or mean absolute error (MAE) for discriminative learning, which could result in models with poor generalization ability once observations contain noise. In this work, we propose to formulate two priors. One is the smoothness prior, and the other is the sparsity prior. The former is aimed to model locally similar components of images, while the latter is introduced to fit non-smooth details of images. Since any image could be decomposed into the sum of a piece-wisely smooth component and a residual error which is more likely to be sparse \cite{Osher/2005}, we can build DNNs to restore the smoothness component and the sparsity residual for SISR.

We propose a Bayesian image super-resolution network, referred to as BayeSR, by implementing the IR framework via DNNs. Concretely, we first model each image as the sum of a smoothness component and a sparsity residual, and its observation can be degraded by blurring, downscaling, and noise. Then, we build DNNs to infer variational posterior distributions, i.e., pixel-wise Gaussian distributions, of the smoothness component, the sparsity residual, and the noise, as shown in Fig. \ref{fig:diagram}. Finally, we sample the smoothness component and the sparsity residual from their distributions, respectively, and the sum of them is considered as a restoration. Note that we could repeat the last step to generate many restorations for a given degraded image, and therefore \textit{BayeSR is a stochastic restoration method, instead of a deterministic one}.

The contributions of this work are summarized as follows:
\begin{itemize}
	\item Firstly, we propose a Bayesian image restoration framework by explicit modeling of image priors. This framework infers variational posterior distributions given observations, and can restore stochastic images by randomly sampling from the resulting distributions.
	\item  Secondly, we build the BayeSR embedded with downsampling, upsampling, and inferring modules for SISR. The dowmsampling module is aimed to learn image degradation; the upsampling module is developed to upscale image space; and the inferring module is built to infer the variational parameters of posteriors.
	\item  Finally, we develop an unsupervised learning strategy of training BayeSR when only LR images are available, and extend it for pseudo-supervised and supervised learning if unpaired and paired HR images are provided, respectively. Moreover, we show the generalization ability and unsupervised performance of BayeSR via three tasks, i.e., ideal, realistic, and real-world SISR.
\end{itemize}

The rest of our paper is organized as follows. In Section 2, we introduce the related works about model-based and learning-based IR. Section 3 presents the framework of BayeSR, including the network architecture and the training strategies. Section 4 provides the implementation details of BayeSR and the experimental results on three SISR tasks. We finally conclude this work in Section 5.

\section{Related works}
\subsection{Model-based image restoration}
Conventional IR is based on mathematical and statistical models which are designed to model the domain knowledge of images \cite{Rudin/1992,Qian/1993}. Both of them aim to explicitly model domain knowledge, and therefore are often referred to as model-based IR \cite{Dong/2019}. A typical image degradation model could be expressed as 
$ \mathbf y = \mathbf A\mathbf u + \mathbf n $,
where, $ \mathbf y $, $ \mathbf A $, $ \mathbf u $, and $ \mathbf n $ respectively denote the degraded image, degradation operator, natural image, and the addictive noise \cite{Jalobeanu/2004}. IR could be categorized into specific tasks based on the forms of $ \mathbf A $. For example, $ \mathbf A $ is an identity matrix for image denoising \cite{Osher/2005}, a blurring operator for image deblurring \cite{Jalobeanu/2004}, and a downsampling operator for SISR \cite{Yang/2008}. 

From a mathematical perspective, IR aims to solve an inverse problem,
e.g., \[\min_{\mathbf u} \left\| \mathbf A\mathbf u - \mathbf y \right\|_2^2/2 + \lambda \mathcal R(\mathbf u), \]
where, $ \mathcal R(\mathbf u) $ denotes a regularization term, and $ \lambda $ is a hyperparameter \cite{Tikhonov/1977}. Many efforts have been made to explore appropriate regularization terms. Tikhonov et al. \cite{Tikhonov/1977} proposed the classical regularization for solving ill-posed inverse problems, and showed its application in IR. Rudin et al. \cite{Rudin/1992} introduced the total variation (TV) regularization to keep images piece-wisely smooth, and it was widely applied in image denoising \cite{Chambolle/1997,Osher/2005,Chan/2006}. Figueiredo et al. \cite{Figueiredo/2003} and Chan et al. \cite{Chan/2003} explored the sparsity of images based on wavelet transform, and demonstrated the effectiveness of sparsity regularization in reconstructing HR images. Koltchinskii et al. \cite{Koltchinskii/2011} and Candes et al. \cite{Candes/2011} showed the low-rank property of images, and developed efficient algorithms of recovering low-rank matrix. These methods solve inverse problems iteratively, which can be computationally expensive for large-scale images. Besides, manually selecting regularization parameters can be a practical issue. 

From a statistical perspective, IR aims to infer the distribution of an image, $ \mathbf u $, given an observation, $ \mathbf y $, by maximizing the posterior probability $ p(\mathbf u|\mathbf y) \propto p(\mathbf y|\mathbf u) p(\mathbf u) $,
where, $ p(\mathbf u) $ represents the prior knowledge of images \cite{Qian/1993}. Many works have been done to model image priors. Hunt et al. \cite{Hunt/1977} used Gaussian prior to keep images smooth. Qian et al. \cite{Qian/1993} introduced Markov random field (MRF) to preserve the edges of textural images. Molina et al. \cite{Molina/1994} proposed a hierarchical Bayesian approach to model the structural form of the noise and local characteristics of images. After that, they introduced the compound Gaussian MRF \cite{Molina/2003} to model the multichannel image prior. Jalobeanu et al. \cite{Jalobeanu/2004} proposed the inhomogeneous Gaussian MRF to model the spatially variant characteristics of real satellite images. Pan et al. \cite{Pan/2006} developed the Huber-MRF to preserve the edges of images and improved the computational efficiency. Guerrero et al. \cite{Guerrero/2008} proposed the space-variant Gaussian scale mixtures to provide an effective local statistical description of images. Babacan et al. \cite{Babacan/2009} adopted TV prior to describe statistical characteristics of images, and used a hierarchical Bayesian model to estimate the hyperparameter of the prior. Ayasso et al. \cite{Ayasso/2012} adopted the Markovian prior and Student's-t prior to model the smooth part and point sources of astrophysical images, respectively. Many of these models are iteratively solved, which can be computationally expensive for large-scale images. However, they have the advantages of sampling a stochastic restoration instead of a deterministic one and quantifying the uncertainty of restorations. 

\subsection{Learning-based image restoration}
Modern IR aims to learn mappings from degraded image spaces to the original image space via dictionaries \cite{Timofte/2014} or neural networks \cite{Dong/2016,Lim/2017,Zhang_beyond/2017}. Different from the conventional IR, the methods use data for learning, and therefore are referred to as learning-based IR \cite{Dong/2019}.

Many works have been done to learn deterministic mappings, i.e., the outputs of IR models are deterministic \cite{Dong/2016,Zhang_beyond/2017}. In image denoising, Burger et al. \cite{Burger/2012} adopted a multi-layer perceptron and achieved comparable performance with the conventional methods. Zhang et al. \cite{Zhang_beyond/2017,Zhang_ffdnet/2018} trained CNN-based residual networks, and their method delivered a promising performance in removing Gaussian noise. Lehtinen et al. \cite{Lehtinen/2018} only used noisy image pairs to train networks without requiring clean targets. Krull et al. \cite{Krull/2019} developed a blind-spot masking scheme to train networks using a single noisy image. Ulyanov et al. \cite{Ulyanov/2018} proposed to directly extract image prior by randomly initializing CNNs, and then used the deep image prior for unsupervised denoising. Batson et al. \cite{Batson/2019} proposed a self-supervised method for blind denoising by exploiting noise independence between pixels. Chen et al. \cite{Chen/2018} proposed to first estimate the distribution of noise from noisy images by GAN, and then to generate noise samples to construct paired training data. In SISR, Yang et al. \cite{Yang/2008,Yang/2010} proposed to learn the sparse representation of images patches via dictionaries, and the resulting model showed good performance in reconstructing details. Dong et al. \cite{Dong/2016} developed three-layer CNNs to super-resolve images, and the resulting models delivered much better performance than the conventional methods. Following with this, deep neural networks, such as residual networks \cite{Lim/2017,Sajjadi/2017,Kim1/2016,Yu/2018}, recursive networks \cite{Kim/2016}, dense networks \cite{Zhang/2018,Tong/2017}, and pyramid networks \cite{Lai/2017}, were studied to improve the Peak Signal-to-Noise Ratio (PSNR) value of SR images. Besides, Ledig et al. \cite{Ledig/2017} developed a super-resolution GAN (SRGAN) and included perceptual loss \cite{Johnson/2016} for training, which could improve the visual quality of SR images. Wang et al. \cite{Wang/2018} further enhanced the performance of SRGAN by improving its network architecture. 
Recently, Chen et al. \cite{Chen/2020} developed an image processing transformer by introducing self-attention, which delivers superior performance in image denoising and SISR.

To be the best of our knowledge, limited works have been reported to learn stochastic mappings, i.e., the outputs of IR models could be random samples \cite{Lugmayr/2020}. Bigdeli et al. \cite{Bigdeli/2017} built a Bayesian deep learning framework using a deep mean-shift prior, but the approach of restoring images is iterative and can be computationally expensive. Laine et al. \cite{Laine/2019} proposed a self-supervised Bayesian denoising framework using the multivariate Gaussian prior. Yue et al. \cite{Yue/2019} developed a variation denoising network using the conjugate Gaussian prior. Helou et al. \cite{Helou/2020} built a blind image denoiser using the Gaussian prior and a fusion network architecture. Izadi et al. \cite{Izadi/2020} developed non-local Bayesian networks using the multivariate Gaussian prior and the non-local mean filtering. These denoising methods cannot deal with the problem of SISR, since the blurring and downscaling are not involved in their modeling. Recently, Lugmayr et al. \cite{Lugmayr/2020} explored the SR space using normalizing flow to reconstruct diverse SR images given an observation. However, the method does not explicitly model image priors, and therefore the description of statistical characteristics is unclear. Different from deterministic learning, stochastic learning could produce diverse restorations from an observation by random sampling, which may follow the property of ill-posed inverse problems that the number of solutions could be infinite.

\section{Methodology}\label{sec:method}
\begin{table}[!t]
	\centering
	\caption{Summarization of notions and notations. Here, VDs denote variational distributions.}
	\label{tab:notation}
		\begin{tabular}{|c|c|}
			\hline
			Notion&  Notation\\
			\hline
			Scalar&  lowercase letter, e.g., $ a $\\
			Vector&  boldface lowercase letter, e.g., $ \mathbf a $\\
			Matrix&  boldface capital letter, e.g., $ \mathbf A $\\
			\hline
			Observation/Reference&  $ \mathbf y\in \mathbb R^{d_y} $/$ \mathbf u^*\in \mathbb R^{d_u} $\\
			Restoration&  $ \mathbf u \in \mathbb R^{d_u} $\\
			Smoothness component& $ \mathbf x \in \mathbb R^{d_u} $\\
			Sparsity residual& $ \mathbf z \in \mathbb R^{d_u} $\\
			Gaussian noise& $ \mathbf n \in \mathbb R^{d_y} $\\
			\hline
			Spatial correlation w.r.t. $ \mathbf x $ & $ \boldsymbol \upsilon \in \mathbb R^{d_u} $\\
			Sparsity precision w.r.t. $ \mathbf z $ &  $ \boldsymbol \omega \in \mathbb R^{d_u} $\\
			Mean/Strength w.r.t. $ \mathbf n $& $ \mathbf m \in \mathbb R^{d_y} $/$ \boldsymbol \rho \in \mathbb R^{d_y} $\\
			\hline
			Mean/Deviation of VDs& $ \breve{\boldsymbol \mu}_{\cdot} $/$ \breve{\boldsymbol \sigma}_{\cdot} $\\
			Normal/Gamma distribution& $ \mathcal N(\cdot, \cdot) $/$ \mathcal G(\cdot, \cdot) $\\
			Hyperparameters& $ s, \mathbf k, \boldsymbol \mu_0, \sigma_0, \boldsymbol \phi_{\cdot}, \boldsymbol \gamma_{\cdot}, \lambda, \tau $\\
			\hline
	\end{tabular}
\end{table}

This work is aimed to build a Bayesian image restoration framework, and implement it by DNNs for SISR. Image restoration is particularly challenging when only a few degraded and noisy observations are available. To tackle the difficulty, we first impose on smoothness and sparsity priors to describe statistical characteristics of the original images, and then estimate the smoothness component and the sparsity residual by MAP. Although the iterative variational Bayesian approaches could be used to infer the posteriors \cite{Chantas/2008,Babacan/2009,Ayasso/2012}, they are computationally expensive, due to many steps of iteration for SR images with large size. Motivated by the advance of deep learning which has great potential for real-time SISR, in this work we develop a Bayesian image super-resolution method via deep modeling of image priors. 

\begin{figure*}[!t]
	\centering	
	\subfloat[Probabilistic graphical model]{\includegraphics[width=0.4\linewidth]{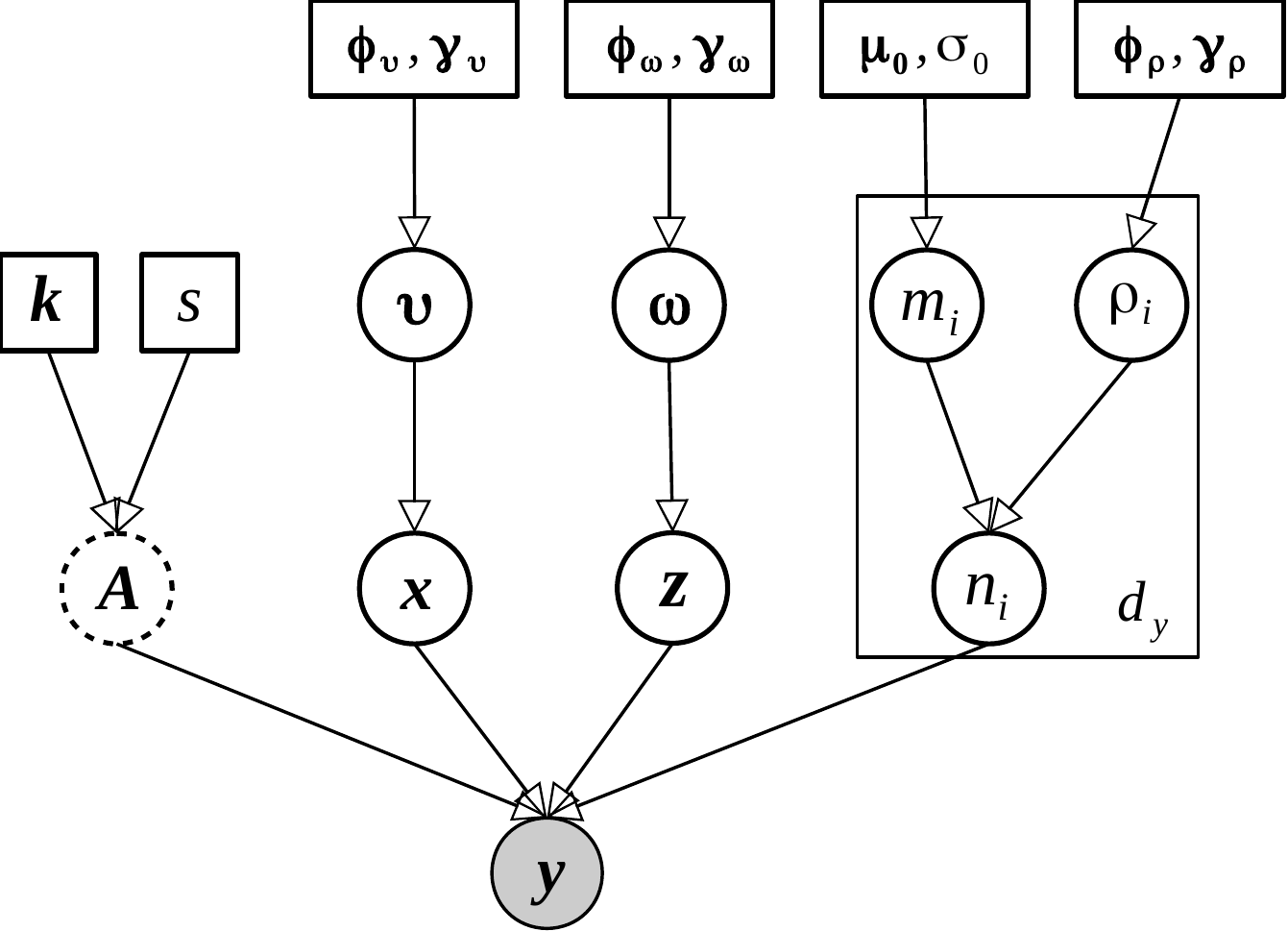}}
	\hfil
	\subfloat[Bayesian image super-resolution]{\includegraphics[width=0.5\linewidth]{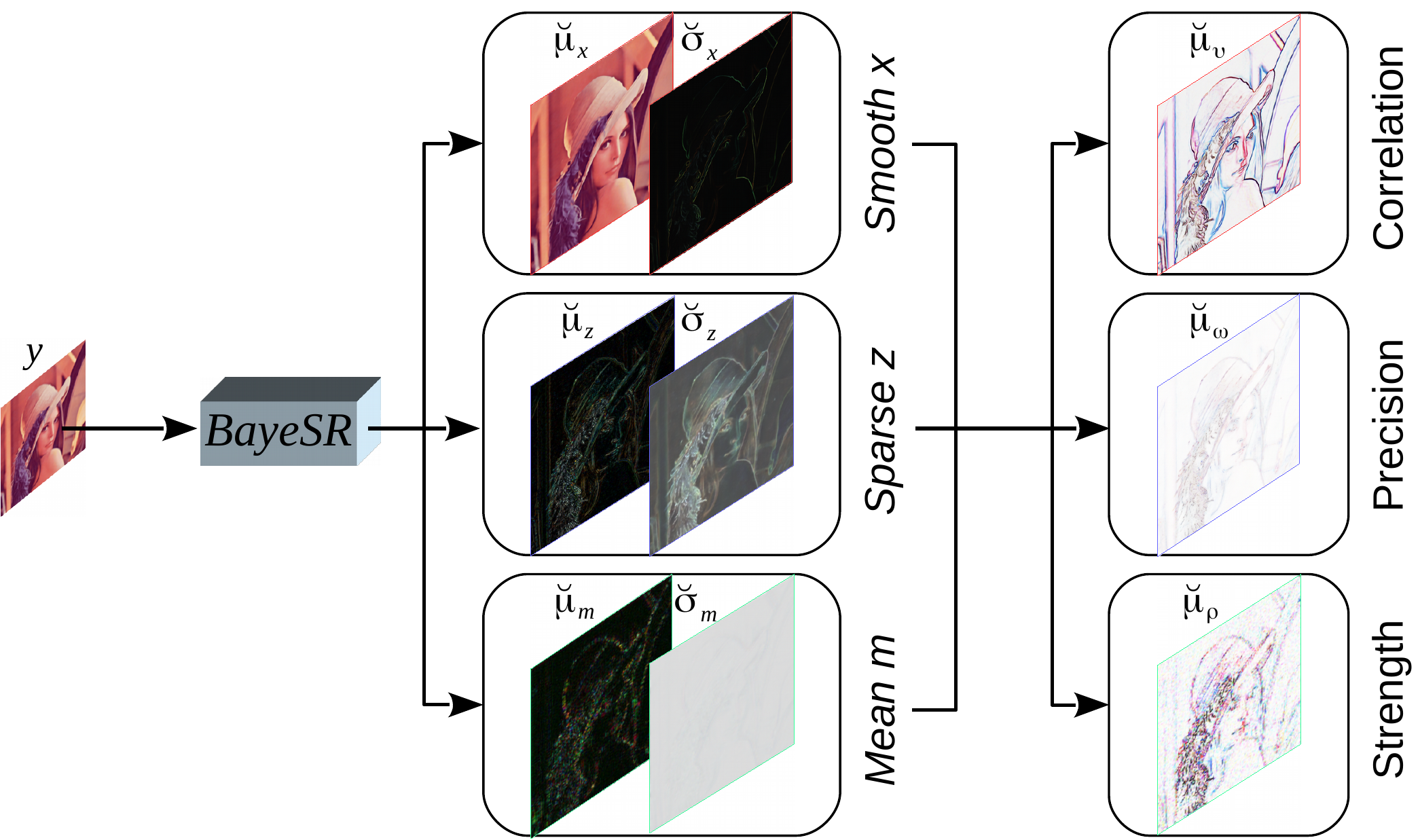}}
	\caption{Probabilistic graphical model and Bayesian image super-resolution. (a) shows the graphical model of image degradation. (b) shows the pipeline of Bayesian image super-resolution (BayeSR). Here, light gray circles denote observed variables, white circles denote unobserved variables, dashed circles denote deterministic functions, and rectangles denote hyperparameters. One can refer to the text in Section \ref{sec:method} for details. } 
	\label{fig:pipeline}
\end{figure*}

For convenience, \textit{raw tensor data of images are vectorized in this paper}, unless stated otherwise. Fig. \ref{fig:pipeline} (a) shows the probabilistic graphical model, which is also known as Bayesian belief network, of modeling an observation $ \mathbf y $. Concretely, $ \mathbf y $ can be modeled as the composition of a smoothness component $ \mathbf x $, a sparsity residual $ \mathbf z $, a Gaussian noise $ \mathbf n $, and a deterministic downsampling operator $ \mathbf A $, where the sum of $ \mathbf x $ and $ \mathbf z $ is considered as the restoration of $ \mathbf y $, denoted as $ \mathbf u $. Besides, $ \mathbf x $ depends on a variable of spatial correlation $ \boldsymbol \upsilon $, $ \mathbf z $ depends on sparsity precision $ \boldsymbol \omega $, and $ \mathbf n $ depends on mean $ \mathbf m $ and noise strength $ \boldsymbol \rho $. Fig. \ref{fig:pipeline} (b) shows the pipeline of Bayesian image super-resolution. To be specific, we first develop DNNs to infer the variational posterior distributions of $ \mathbf x $, $ \mathbf z $, and $ \mathbf m $. For example, $ \breve{\boldsymbol \mu}_{x} $ and $ \breve{\boldsymbol \sigma}_{x} $ denote the mean and standard deviation of the variational Gaussian distribution of $ \mathbf x $. We further explicitly compute the variational parameters of $ \boldsymbol \upsilon $, $ \boldsymbol \omega $, and $ \boldsymbol \rho $. For instance, $ \breve{\boldsymbol \mu}_{\upsilon} $ denotes the mean of the variational Gamma distribution of $ \boldsymbol \upsilon $. Finally, we sample a smoothness component and a sparsity residual following their variational posterior distributions, and the sum of them is considered as a restoration of $ \mathbf y $.

Table \ref{tab:notation} summarizes the notions and notations used in this paper. Besides, $ \left\| \cdot \right\|_1  $ denotes the $ \ell_1 $ norm of vectors; $ \left\| \cdot \right\|_2  $ denotes the $ \ell_2 $ norm of vectors; $ \left\| \cdot \right\|_{\mathbf M}  $, where $ \mathbf M $ is a symmetric positive definite matrix, denotes the  $ \mathbf M $-norm of vectors, i.e., $ \left\| \mathbf x \right\|_{\mathbf M} = \sqrt{\mathbf x^\top \mathbf M \mathbf x}  $; and $ \left\langle \cdot, \cdot \right\rangle  $ denotes the inner-product of vectors. The rest of this section is organized as follows. We specify the graphical model of modeling image degradation in Section \ref{sec:BBN}, and develop the approach of inferring variational distributions in Section \ref{sec:VB}. After that, we interpret the variational loss in Section \ref{sec:interpretVP}. Section \ref{sec:SmSpNet} illustrates the details of building neural networks. Section \ref{sec:train} describes the training and test strategies.

\subsection{Statistical modeling of image degradation}\label{sec:BBN}
\subsubsection{Formulation of degradation}
Modeling smoothness and sparsity is crucial to IR. In real-world, noise is inevitably introduced by imaging systems. Therefore, denoising could be a basic task. Estimating the piece-wisely smooth components based on the TV prior has shown to be effective in denoising \cite{Osher/2005}, but image details can be missed. Recent works showed that the sparsity prior has the potential of capturing more details \cite{Chantas/2008}. Motivated by this, we propose to infer the smoothness component and sparsity residual of images for restoration. 

Suppose corrupted observations are sampled from some variable $ \mathbf y \in \mathbb R^{d_y} $, where $ d_y $ denotes the dimension of $ \mathbf y $, and clean images are sampled from a variable, $ \mathbf u^* $, where $ \mathbf u^*\in \mathbb R^{d_u} $, then the degradation process of images could be models as,
\begin{equation}\label{eq:model}
	\mathbf y = \mathbf A(\mathbf x + \mathbf z) + \mathbf n,
\end{equation}
where, $ \mathbf x \in \mathbb R^{d_u} $ denotes a variable of smoothness prior; $ \mathbf z\in \mathbb R^{d_u} $ represents another variable of sparsity prior; $ \mathbf n\in \mathbb R^{d_y} $ is a Gaussian noise; and $ \mathbf A\in \mathbb R^{d_y\times d_u} $ denotes a deterministic downsampling matrix related to a convolutional kernel $ \mathbf k\in \mathbb R^{d_k} $ and a downscaling factor $ s $. For example, $ \mathbf A\mathbf x $ equals to the vectorization of $ (\mathbf X \ast \mathbf K)\downarrow_s $ for SISR, where, $ \mathbf X $ and $ \mathbf K $ are the matrix forms of $ \mathbf x $ and $ \mathbf k $, respectively, and $ \downarrow_s (s>1) $ denotes a downscaling operator. Next, we will select detailed statistical models for these variables.

\subsubsection{Modeling of priors in detail}
The observation likelihood of $ \mathbf y $ can be expressed as
\begin{equation}\small\label{eq:ypdf}
	p(\mathbf y|\mathbf A, \mathbf x, \mathbf z, \mathbf m, \boldsymbol \rho) = \mathcal N ( \mathbf y|\mathbf A(\mathbf x+\mathbf z)+\mathbf m, \mbox{diag}(\boldsymbol \rho)^{-1}).  
\end{equation}
Here, we model $ \mathbf n $ as a spatially-variant Gaussian noise with a mean $ \mathbf m \in \mathbb R^{d_y} $ and a variance $ \mbox{diag}(\boldsymbol \rho)^{-1}\in \mathbb R^{d_y\times d_y} $, namely,
\begin{equation}\label{eq:npdf}
	p(\mathbf n| \mathbf m, \boldsymbol \rho) = \mathcal N( \mathbf n|\mathbf m, \mbox{diag}(\boldsymbol \rho)^{-1}).
\end{equation}
Moreover, we assign Gaussian prior to $ \mathbf m $ and Gamma prior to $ \boldsymbol \rho $, i.e.,
\begin{align}
	p(\mathbf m| \boldsymbol \mu_0, \sigma_0) &= \mathcal N(\mathbf m|\boldsymbol \mu_0, \sigma_0^{-1}\mathbf I),\label{eq:mpdf}\\
	p(\boldsymbol \rho| \boldsymbol \phi_{\rho}, \boldsymbol \gamma_{\rho}) &= \textstyle\prod_{i=1}^{d_y} \mathcal G(\rho_i|\phi_{\rho i}, \gamma_{\rho i}), \label{eq:rhopdf}
\end{align}
where, $ \mathbf I $ denotes an identity matrix; $ \boldsymbol \mu_0 $, $ \sigma_0 $, $ \boldsymbol \phi_{\rho} $, and $ \boldsymbol \gamma_{\rho} $ are user-defined hyperparameters, and $ \mathcal G(\cdot, \cdot) $ denotes Gamma distribution.

To account for the piecewise smoothness of $ \mathbf x $, we adopt the TV or Markovian prior which could be expressed as follows,
\begin{equation}\label{eq:xpdf}
	p(\mathbf x| \boldsymbol \upsilon) = \mathcal N(\mathbf x|\mathbf 0, [ \mathbf D_h^T \mbox{diag}(\boldsymbol \upsilon)\mathbf D_h + \mathbf D_v^T \mbox{diag}(\boldsymbol \upsilon)\mathbf D_v ]^{-1} ), 
\end{equation}
where, $ \mathbf D_h $ and $ \mathbf D_v $ denote the finite-difference matrix in the horizontal and vertical directions, respectively, and $ \boldsymbol \upsilon $ is a variable describing the spatial correlation of $ \mathbf x $, which follows the Gamma prior,
\begin{align}
	p(\boldsymbol \upsilon|\boldsymbol \phi_{\upsilon}, \boldsymbol \gamma_{\upsilon}) &= \textstyle\prod_{i=1}^{d_u} \mathcal G(\upsilon_i|\phi_{\upsilon i}, \gamma_{\upsilon i}),\label{eq:upsilonpdf} 
\end{align}
where, $ \boldsymbol \phi_{\upsilon} $ and $ \boldsymbol \gamma_{\upsilon} $ are hyperparameters.

To account for the sparsity of $ \mathbf z $, we adopt the Student's t prior which could be obtained by marginalizing a three-parameter Normal-Gamma distribution as follows,
\begin{equation}\label{eq:zpdf}
	\begin{aligned}
		p(\mathbf z| \boldsymbol \phi_{\omega}, \boldsymbol \gamma_{\omega}) &= \int_{\mathbb R^{d_u}} p(\mathbf z|\boldsymbol \omega) p(\boldsymbol \omega| \boldsymbol \phi_{\omega}, \boldsymbol \gamma_{\omega})d\boldsymbol \omega\\
		&= \textstyle\prod_{i=1}^{d_u}\int_{\mathbb R} \mathcal N(z_i|0, \omega_i^{-1}) \mathcal G(\omega_i|\phi_{\omega i}, \gamma_{\omega i})d\omega_i, 
	\end{aligned}
\end{equation}
where, $ \boldsymbol \phi_{\omega} $ and $ \boldsymbol \gamma_{\omega} $ are hyperparameters; and $ \boldsymbol \omega \in \mathbb R^{d_u} $ is a variable of conducting the sparsity precision of $ \mathbf z $, which follows the Gamma prior,
\begin{equation}\label{eq:omegapdf}
	p(\boldsymbol \omega | \boldsymbol \phi_{\omega}, \boldsymbol \gamma_{\omega} ) = \textstyle\prod_{i=1}^{d_u} \mathcal G(\omega_i|\phi_{\omega i}, \gamma_{\omega i}).
\end{equation}
Specially, $ p(z_i, \omega_i) = \mathcal N(z_i|0, \omega_i^{-1}) \mathcal G(\omega_i|\phi_{\omega i}, \gamma_{\omega i}) $ is known as Normal-Gamma distribution. One can refer to Appendix A of Appendices for the details of the sparsity prior.

Fig. \ref{fig:pipeline} (a) illustrates the architecture of a probabilistic graphical model, which represents the observation $ \mathbf y $ as the composition of a deterministic linear operator $ \mathbf A $ and three variables $ \mathbf n $, $ \mathbf x $, and $ \mathbf z $, where the downsampling operator $ \mathbf A $ is determined by the blur kernel $ \mathbf k $ and the downscaling factor $ s $. Moreover, the noise $ \mathbf n $ depends on the mean $ \mathbf m $ and the variance $\mbox{diag}(\boldsymbol \rho)^{-1} $, where, $ \mathbf m $ is related to the Gaussian hyperparameters $ \boldsymbol \mu_0 $ and $ \sigma_0 $, and $ \boldsymbol \rho $ is related to the Gamma hyperparameters $ \boldsymbol \phi_{\rho} $ and $ \boldsymbol \gamma_{\rho} $; the smoothness component $ \mathbf x $ depends on the spatial correlation $ \boldsymbol \upsilon $, where, $ \boldsymbol \upsilon $ is related to the Gamma hyperparameters $ \boldsymbol \phi_{\upsilon} $ and $ \boldsymbol \gamma_{\upsilon} $; and the sparsity residual $ \mathbf z $ depends on the sparsity precision $ \boldsymbol \omega $, where, $ \boldsymbol \omega $ is related to the Gamma hyperparameters $ \boldsymbol \phi_{\omega} $ and $ \boldsymbol \gamma_{\omega} $. Next, we will estimate the distributions of these variables given $ \mathbf y $ via variational Bayesian inference.

\subsection{Variational inference of posterior distributions}\label{sec:VB}

\begin{figure*}
	\centering
	\includegraphics[width=1\linewidth]{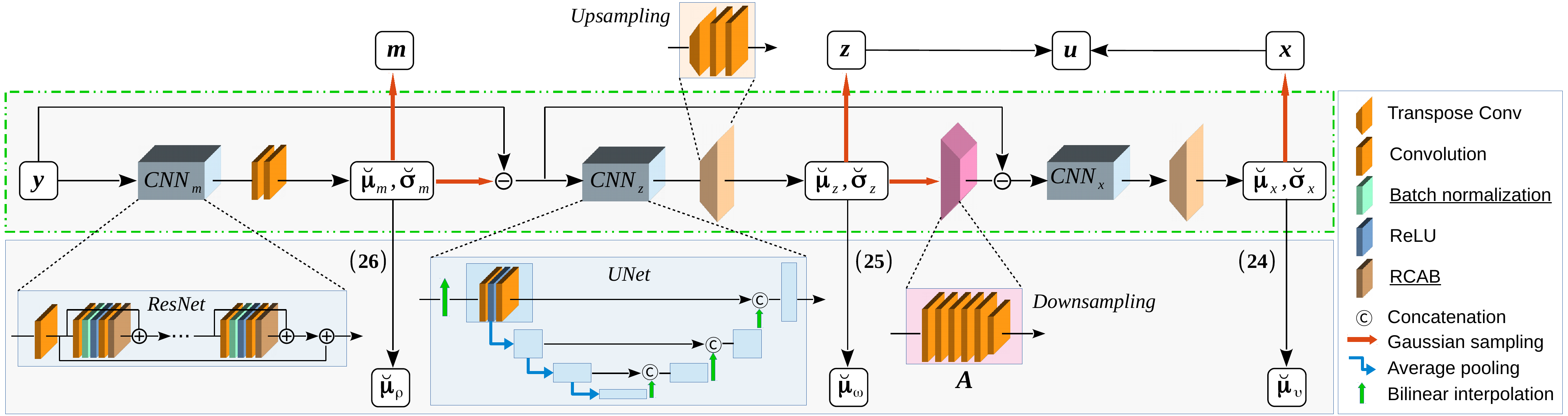}
	\caption{Architecture of BayeSR. Given an observation $ \mathbf y $ (not vectorized), we first build three modules to successively infer the variational parameters w.r.t. the noise mean $ \mathbf m $, the sparsity residual $ \mathbf z $, and the smoothness component $ \mathbf x $. Then, we explicitly compute the variational parameters w.r.t. the spatial correlation $ \boldsymbol \upsilon $, the sparsity precision $ \boldsymbol \omega $, and the noise strength $ \boldsymbol \rho $ by the formulas in (\ref{eq:muupsilon}), (\ref{eq:muomega}), and (\ref{eq:murho}), respectively. Finally, A stochastic sample $ \mathbf u = \mathbf x + \mathbf z $ is considered as a reconstruction of $ \mathbf y $. Here, \textit{ResNet} and \textit{UNet} are two examples of achieving the CNN modules. The underlined font denotes this module is optional in our experiments.}
	\label{fig:architecture}
\end{figure*}

Our aim is to infer the distributions of latent variables given an observation $ \mathbf y $, i.e., to estimate the posterior distribution of each variable in $ \boldsymbol \psi = \left\lbrace \mathbf m, \boldsymbol \rho, \mathbf x, \boldsymbol \upsilon, \mathbf z, \boldsymbol \omega \right\rbrace  $. One could compute the posteriors via the Bayesian rule, i.e., $ p(\boldsymbol \psi|\mathbf y) \propto p(\mathbf y|\boldsymbol \psi)p(\boldsymbol \psi) $, and the marginalization, which is however intractable since some of the variables are conditionally dependent. To tackle the difficulty, we propose to use the variational Bayesian (VB) approach. The VB method approximates $ p(\boldsymbol \psi|\mathbf y) $ via a variational posterior distribution $ q(\boldsymbol \psi) $. Generally, the variables in $ \boldsymbol \psi $ are often enforced to be independent, namely,
\begin{equation}\label{eq:variationalpdf}
	q(\boldsymbol \psi)  = q(\mathbf m) q(\boldsymbol \rho)\textstyle\prod_{i=1}^{d_u}q(x_i) q(\boldsymbol \upsilon) \textstyle\prod_{i=1}^{d_u}q(z_i) q(\boldsymbol \omega).
\end{equation}
One method of obtaining the variational approximations is to minimize the Kullback-Leibler (KL) divergence between $ q(\boldsymbol \psi) $ and $ p(\boldsymbol \psi|\mathbf y) $, as follows, 
\begin{equation}\label{eq:maxKL}
	\breve{q}(\boldsymbol \psi) \in \mathop{\arg\min}_{q(\boldsymbol \psi)} \mbox{KL}(q(\boldsymbol \psi)||p(\boldsymbol \psi|\mathbf y)).
\end{equation}
Since we assigned the conjugate priors \cite{Diaconis/1979} to all variables, the variational posterior approximations of the marginal distributions of $ \mathbf m $, $ \boldsymbol \rho $, $ \mathbf x $, $ \boldsymbol \upsilon $, $ \mathbf z $, and $ \boldsymbol \omega $ could be successively expressed as follows,
\begin{align}\footnotesize
	\breve{q}(\mathbf m) &= \mathcal N(\mathbf m|\breve{\boldsymbol\mu}_m, \mbox{diag}(\breve{\boldsymbol\sigma}_m^2))\label{eq:mvarpdf}\\
	\breve{q}(\boldsymbol \rho) &= \textstyle\prod_{i=1}^{d_y}\mathcal G(\rho_i | \breve{\beta}_{\rho i}, \breve{\alpha}_{\rho i})\label{eq:rhovarpdf}\\
	\breve{q}(\mathbf x) &= \mathcal N(\mathbf x|\breve{\boldsymbol \mu}_x, \mbox{diag}(\breve{\boldsymbol \sigma}_x^2))\label{eq:xvarpdf}\\
	\breve{q}(\boldsymbol \upsilon) &= \textstyle\prod_{i=1}^{d_u}\mathcal G(\upsilon_i | \breve{\beta}_{\upsilon i}, \breve{\alpha}_{\upsilon i})\label{eq:upsilonvarpdf}\\
	\breve{q}(\mathbf z) &= \mathcal N(\mathbf z|\breve{\boldsymbol \mu}_z, \mbox{diag}(\breve{\boldsymbol \sigma}_z^2))\label{eq:zvarpdf}\\
	\breve{q}(\boldsymbol \omega) &= \textstyle\prod_{i=1}^{d_u}\mathcal G(\omega_i | \breve{\beta}_{\omega i}, \breve{\alpha}_{\omega i})\label{eq:omegavarpdf}
\end{align}
where, $ \breve{\boldsymbol \mu}_{\cdot} $, $ \breve{\boldsymbol \sigma}_{\cdot} $, $ \breve{\boldsymbol \alpha}_{\cdot} $ and $ \breve{\boldsymbol \beta}_{\cdot} $ respectively denote the parameters of the variational distributions to be further computed, and the variational posteriors $ \breve{q}(\cdot) $ in (\ref{eq:mvarpdf})-(\ref{eq:omegavarpdf}) are corresponding to the priors $ p(\cdot) $ in (\ref{eq:mpdf}), (\ref{eq:rhopdf}), (\ref{eq:xpdf}), (\ref{eq:upsilonpdf}), (\ref{eq:zpdf}), and (\ref{eq:omegapdf}).

In practice, we do not directly compute the KL divergence, but convert it to an easily derived formula, 
\begin{align}\footnotesize
	\mbox{\footnotesize KL}(\breve{q}(\boldsymbol \psi)||p(\boldsymbol \psi|\mathbf y)) &= \mathbb E\left[ \log \breve{q}(\boldsymbol \psi)\right]  - \mathbb E\left[ \log p(\boldsymbol \psi|\mathbf y)\right] \label{eq:KLtoELBO} \\
	&= \mathbb E\left[ \log \breve{q}(\boldsymbol \psi)\right]  - \mathbb E\left[ \log p(\boldsymbol \psi,\mathbf y)\right]  + \log p(\mathbf y),\nonumber
\end{align}
where, all expectations are taken with respect to $ \breve{q}(\boldsymbol \psi) $, and the evidence $ p(\mathbf y) $ only depends on the priors. This formula shows that minimizing KL divergence is equivalent to
\begin{align}\footnotesize
	& \min_{\breve{q}(\boldsymbol \psi)} \mathbb E\left[ \log \breve{q}(\boldsymbol \psi)\right] - \mathbb E\left[ \log p(\boldsymbol \psi,\mathbf y)\right]\nonumber\\
	=& \min_{\breve{q}(\boldsymbol \psi)} \mbox{\footnotesize KL}(\breve{q}(\boldsymbol \psi)||p(\boldsymbol \psi)) - \mathbb E\left[ \log p(\mathbf y|\boldsymbol \psi)\right]. \label{eq:ELBO}
\end{align}
The second term of (\ref{eq:ELBO}) could be expressed as
\begin{equation}\label{eq:paralikelihood}
	-\mathbb E\left[ \log p(\mathbf y|\boldsymbol \psi)\right]= -\mathbb E_{\breve{q}(\boldsymbol \rho)}\left[ \mathbb E_{\breve{q}(\boldsymbol \psi \setminus \rho)} \left[ \log p(\mathbf y|\boldsymbol \psi)\right] \right]. 
\end{equation}
Since directly computing $ \mathbb E_{\breve{q}(\boldsymbol \psi \setminus \rho)} \left[ \log p(\mathbf y|\boldsymbol \psi)\right] $ is difficult, we adopt the widely used reparameterization technique \cite{Kingma/2019}. Concretely, let $ \boldsymbol \epsilon $ denote white Gaussian noise sampled from $ \mathcal N(\mathbf 0, \mathbf I) $, then we have $ \mathbf x = \breve{\boldsymbol \sigma}_x \odot \boldsymbol \epsilon + \breve{\boldsymbol \mu}_x $, $ \mathbf z = \breve{\boldsymbol \sigma}_z \odot \boldsymbol \epsilon + \breve{\boldsymbol \mu}_z $, and $ \mathbf m = \breve{\boldsymbol \sigma}_m \odot \boldsymbol \epsilon + \breve{\boldsymbol \mu}_m $, where $ \odot $ denotes the element-wise multiplication. Moreover, we consider $ \log p(\mathbf y|\boldsymbol \psi) $ as an approximation of $ \mathbb E_{\breve{q}(\boldsymbol \psi \setminus \rho)} \left[ \log p(\mathbf y|\boldsymbol \psi)\right] $, and therefore the formula (\ref{eq:paralikelihood}) could be converted to
\begin{equation}\label{eq:relaxedlikelihood}
	-\mathbb E_{\breve{q}(\boldsymbol \rho)}\left[ \mathbb E_{\breve{q}(\boldsymbol \psi \setminus \rho)} \left[ \log p(\mathbf y|\boldsymbol \psi)\right] \right] \approx -\mathbb E_{\breve{q}(\boldsymbol \rho)}\left[ \log p(\mathbf y|\boldsymbol \psi) \right] . 
\end{equation}
Finally, we infer variational posteriors by optimizing the following problem,
\begin{equation}\label{eq:relaxation}
	\min_{\breve{q}(\boldsymbol \psi)} \mbox{\footnotesize KL}(\breve{q}(\boldsymbol \psi)||p(\boldsymbol \psi)) - \mathbb E_{\breve{q}(\boldsymbol \rho)}\left[ \log p(\mathbf y|\boldsymbol \psi) \right].
\end{equation}

\begin{algorithm}[!t]
	\caption{Training and test of BayeSR}
	\label{pseudocode}
	\hspace*{0.02in}{\bfseries Input:}
	Training and test datasets\\
	\hspace*{0.02in}{\bfseries Output:}
	A stochastic restoration $ \mathbf u $
	\begin{algorithmic}[1]
		\If {\textit{Preliminary stage}}
		\State Estimate the downsampling $ \mathbf A $ from training data.
		\State Generate a pool of noise patches from training data.
		\EndIf
		\If {\textit{Unsupervised training stage}}
		\State Freeze the parameters of the downsampling module.
		\State Stop the back propagation through $ \breve{\boldsymbol\mu}_{\upsilon} $, $ \breve{\boldsymbol\mu}_{\omega} $, and $ \breve{\boldsymbol\mu}_{\rho} $.
		\While{\textit{not up to total training steps}}
		\State Sample patches $ \mathbf y_i $ and $ \mathbf u_i^{lr} $ from training data.
		\State Sample patches $ \mathbf n_i $ from noise pool. 
		\State Generate pseudo degradations $ \mathbf y_i^{lr} $ via $ \mathbf A\mathbf u_i^{lr} + \mathbf n_i $.
		\State Update the parameters of BayeSR by (\ref{eq:totalunsuploss}). 
		\State Update the parameters of $ D_y $ and $ D_u $ by (\ref{eq:advloss}).
		\EndWhile
		\EndIf
		\If {\textit{Test stage}}
		\State Sample an LR image $ \mathbf y $ from test data.
		\State Infer the variational distribution of $ \breve{q}(\mathbf x) $ and $ \breve{q}(\mathbf z) $.
		\State Sample $ \mathbf x $ and $ \mathbf z $ from their variational distributions.
		\State \Return $ \mathbf u = \mathbf x + \mathbf z $
		\EndIf
	\end{algorithmic}
\end{algorithm}

\subsection{Interpretation of the objective function}\label{sec:interpretVP}

In this section, we decompose the objective function in (\ref{eq:relaxation}) into computational details according to the modeling variables for intuitive interpretation. The variational posteriors of $ \boldsymbol \upsilon $, $ \boldsymbol \omega $, and $ \boldsymbol \rho $ can be explicitly formulated using that of $ \mathbf x $, $ \mathbf z $, and $ \mathbf m $. Concretely, the first term of (\ref{eq:relaxation}) could be expressed as,
\begin{subequations}
	\begin{align}\footnotesize
		\mbox{\footnotesize KL}(\breve{q}(\boldsymbol \psi)||p(\boldsymbol \psi)) &= \mbox{\footnotesize KL}(\breve{q}(\mathbf x)\breve{q}(\boldsymbol \upsilon)||p(\mathbf x|\boldsymbol \upsilon)p(\boldsymbol \upsilon))\label{eq:KLsa}\\
		& + \mbox{\footnotesize KL}(\breve{q}(\mathbf z)\breve{q}(\boldsymbol \omega)||p(\mathbf z|\boldsymbol \omega)p(\boldsymbol \omega))\label{eq:KLsb}\\
		&+ \mbox{\footnotesize KL}(\breve{q}(\mathbf m)||p(\mathbf m)) + \mbox{\footnotesize KL}(\breve{q}(\boldsymbol \rho)||p(\boldsymbol \rho)).\label{eq:KLsc} 
	\end{align}
\end{subequations}

Minimizing (\ref{eq:KLsa}), related to $\mathbf x$ and $\boldsymbol \upsilon$, could induce the formula of computing $ \breve{\boldsymbol \mu}_{\upsilon} $, namely,
\begin{equation}\footnotesize\label{eq:muupsilon}
	\breve{\boldsymbol \mu}_{\upsilon} = \frac{\breve{\boldsymbol \alpha}_{\upsilon}}{\breve{\boldsymbol \beta}_{\upsilon}} = \frac{2 \boldsymbol \gamma_{\upsilon} + 1}{(\mathbf D_h\breve{\boldsymbol \mu}_x)^2 + (\mathbf D_v\breve{\boldsymbol \mu}_x)^2 + 4\breve{\boldsymbol\sigma}_x^2  + 2\boldsymbol\phi_{\upsilon}},
\end{equation}
where, the operations in the above formula are element-wise. 

Minimizing (\ref{eq:KLsb}), related to $\mathbf z$ and $\boldsymbol \omega$, results in the formula of computing $ \breve{\boldsymbol \mu}_{\omega} $,
\begin{equation}\footnotesize\label{eq:muomega}
	\breve{\boldsymbol \mu}_{\omega} = \frac{\breve{\boldsymbol \alpha}_{\omega}}{\breve{\boldsymbol \beta}_{\omega}} = \frac{2 \boldsymbol \gamma_{\omega} + 1}{\breve{\boldsymbol \mu}_z^2 + \breve{\boldsymbol\sigma}_z^2  + 2\boldsymbol\phi_{\omega}}.
\end{equation} 

Finally, minimizing (\ref{eq:KLsc}) and the second term of (\ref{eq:relaxation}), related to $\mathbf m$ and $\boldsymbol \rho$, leads to the formula of computing $ \breve{\boldsymbol \mu}_{\rho} $, as follows,
\begin{equation}\footnotesize\label{eq:murho}
	\breve{\boldsymbol \mu}_{\rho} = \frac{\breve{\boldsymbol \alpha}_{\rho}}{\breve{\boldsymbol \beta}_{\rho}} = \frac{2 \boldsymbol \gamma_{\rho} + 1}{(\mathbf y - \mathbf A (\mathbf x + \mathbf z) - \mathbf m)^2 + 2\boldsymbol\phi_{\rho}}.
\end{equation}

The variational posteriors of $ \mathbf x $, $ \mathbf z $, and $ \mathbf m $ can be inferred from $ \mathbf y $, \textit{given} $ \breve{\boldsymbol \mu}_{\upsilon} $, $ \breve{\boldsymbol \mu}_{\omega} $, $ \breve{\boldsymbol \mu}_{\rho} $, $ \boldsymbol \mu_0 = \mathbf 0 $, and $ \sigma_0 $. Concretely, the formulas in (\ref{eq:relaxation}) induces a variational loss function with respect to $ \left\lbrace \breve{\boldsymbol \mu}_x, \breve{\boldsymbol \sigma}_x \right\rbrace $, $ \left\lbrace \breve{\boldsymbol \mu}_z, \breve{\boldsymbol \sigma}_z \right\rbrace $ and $ \left\lbrace \breve{\boldsymbol \mu}_m, \breve{\boldsymbol \sigma}_m \right\rbrace $ as follows,
\begin{equation}\label{eq:VBloss}
	\mathcal L_{var}(\mathbf y) = \mathcal L_y + \mathcal L_{\breve{\mu}_x} + \mathcal L_{\breve{\sigma}_x} + \mathcal L_{\breve{\mu}_z} + \mathcal L_{\breve{\sigma}_z} + \mathcal L_{\breve{\mu}_m} + \mathcal L_{\breve{\sigma}_m}.
\end{equation}
The computational details and interpretation of each term are summarized in Table \ref{tab:varloss}. Note that the variational loss in (\ref{eq:VBloss}) is a derivation from (\ref{eq:relaxation}). Therefore, all these terms are adaptively balanced by MAP. 
This is different from conventional regularization methods, which use multiple terms and thus require to manually set the balancing weights for different terms.
For details of the derivation of formulas (\ref{eq:muupsilon})-(\ref{eq:VBloss}), please refer to Appendix C of Appendices. 

\begin{table*}[!htbp]
	\centering
	\caption{Computational details and interpretation of the variational terms in  (\ref{eq:VBloss}). Here, $ \mathbf 1 $ denotes a vector with all elements to be ones.}
	\label{tab:varloss}
	\resizebox{1\linewidth}{!}{
		\begin{tabular}{|l|l|c|l|}
			\hline
			Notation&  Formula&  Adaptive weight&  Interpretation\\
			\hline
			$ \mathcal L_y $ & $ \frac{1}{2} \|\mathbf y - \mathbf A (\mathbf x + \mathbf z) - \mathbf m \|^2_{\mathbf M_{\rho}} $&  $ \mathbf M_{\rho} = \mbox{diag}(\breve{\boldsymbol \mu}_{\rho}) $&  Ensures the consistency between restorations and observations\\
			\hline
			$ \mathcal L_{\breve{\mu}_x} $ & $ \frac{1}{2} [\| \mathbf D_h\breve{\boldsymbol\mu}_x \|^2_{\mathbf M_{\upsilon}} + \| \mathbf D_v\breve{\boldsymbol\mu}_x \|^2_{\mathbf M_{\upsilon}}] $&  $ \mathbf M_{\upsilon} = \mbox{diag}(\breve{\boldsymbol \mu}_{\upsilon}) $ &  Encourages $ \breve{\boldsymbol\mu}_x $ to be piece-wisely smooth\\
			$ \mathcal L_{\breve{\sigma}_x} $ &  $\frac{1}{2} [ \langle 4\breve{\boldsymbol \mu}_{\upsilon},\breve{\boldsymbol\sigma}_x^2 \rangle  - \langle \mathbf 1, \log(\breve{\boldsymbol \sigma}_x^2)\rangle  ] $& --&  Prevents $ \breve{q}(\mathbf x) $ from degrading to a one-point distribution\\
			\hline
			$ \mathcal L_{\breve{\mu}_z} $ & $ \frac{1}{2} \left\| \breve{\boldsymbol\mu}_z \right\|^2_{\mathbf M_{\omega}} $& $ \mathbf M_{\omega} = \mbox{diag}(\breve{\boldsymbol \mu}_{\omega}) $ &  Encourages $ \breve{\boldsymbol\mu}_z $ to be sparse\\
			$ \mathcal L_{\breve{\sigma}_z} $ & $ \frac{1}{2} [ \langle \breve{\boldsymbol \mu}_{\omega},\breve{\boldsymbol\sigma}_z^2\rangle  - \langle \mathbf 1, \log(\breve{\boldsymbol \sigma}_z^2)\rangle] $ & --&  Prevents $ \breve{q}(\mathbf z) $ from degrading to a one-point distribution\\
			\hline
			$ \mathcal L_{\breve{\mu}_m} $ & $ \frac{\sigma_0}{2} \left\|\breve{\boldsymbol\mu}_m \right\|_2^2 $  &  --&  Constraints the energy of $ \breve{\boldsymbol\mu}_m $.\\
			$ \mathcal L_{\breve{\sigma}_m} $ & $ \frac{1}{2} [ \langle \sigma_0 \mathbf 1, \breve{\boldsymbol\sigma}_m^2\rangle  - \langle \mathbf 1, \log(\breve{\boldsymbol \sigma}_m^2)\rangle ] $ &-- &  Prevents $ \breve{q}(\mathbf m) $ from degrading to a one-point distribution\\
			\hline
	\end{tabular}}
\end{table*}

\subsection{Deep learning of variational parameters}\label{sec:SmSpNet} 
We develop deep neural networks to implement the Bayesian image restoration framework described in Section \ref{sec:VB} for SISR. In practice, the posterior parameters cannot be explicitly formulated since solving the resulting nonlinear equations is intractable, as shown in Appendix B of Appendices. The previous work showed that iterative VB algorithms could be applied to tackle the difficulty \cite{Babacan/2009}, but they are computationally expensive due to the need for many iterations on high-dimensional parameters. Thanks to the promising performance of DNNs in learning non-linear mappings and the efficient platforms of deploying DNNs in parallel, we build deep neural networks to achieve Bayesian image super-resolution.

Fig. \ref{fig:architecture} illustrates the architecture of BayeSR, which mainly consists of three types of modules, i.e., CNN, upsampling, and downsampling. The CNN module could be designed using the backbone of ResNet \cite{Zhangyulun/2018} or UNet \cite{Ronneberger/2015}. For ResNet, the upsampling module comprises two convolutional layers for $ s=1 $, and one (two) transpose convolutional layer(s) followed by two convolutional layers for $ s=2,3 $ ($ s=4 $). For UNet, since we have adopted bilinear interpolation to upscale its inputs, the transpose convolutional layers of the upsampling module will be removed. The downsampling module, which is removed for $ s=1 $, consists of six convolutional layers, and the strides of the last layer are equal to $ s $. In our experiments, CNN$ _m $ will be fixed as ResNet, while CNN$ _z $ and CNN$ _x $ can be either ResNet or UNet.

The three modules, i.e., CNN$ _m $, CNN$ _z $, and CNN$ _x $, are successively developed to estimate the distribution parameters of $ \breve{q}(\mathbf m) $, $ \breve{q}(\mathbf z) $, and $ \breve{q}(\mathbf x) $. Given an observation $ \mathbf y $, we first use CNN$ _m $ followed by two convolutional layers to estimate $ \breve{\boldsymbol\mu}_m $ and $ \breve{\boldsymbol\sigma}_m $ from $ \mathbf y $. Then, we compute the residual $ \mathbf y - \mathbf m $, and use CNN$ _z $ followed by an upsampling module to infer $ \breve{\boldsymbol \mu}_z $ and $ \breve{\boldsymbol\sigma}_z $ from the residual. Finally, we downsample $ \mathbf z $ by a downsampling module, which is developed to implement the downsampling operator $ \mathbf A $, and compute another residual $ \mathbf y - \mathbf m - \mathbf A \mathbf z $. Similarly, we use CNN$ _x $ followed by another upsampling module to estimate $ \breve{\boldsymbol \mu}_x $ and $ \breve{\boldsymbol\sigma}_x $ from the residual. Once these parameters have been estimated from the observation, we could explicitly compute $ \breve{\boldsymbol \mu}_{\upsilon} $, $ \breve{\boldsymbol \mu}_{\omega} $, and $ \breve{\boldsymbol \mu}_{\rho} $ via the formulas (\ref{eq:muupsilon})-(\ref{eq:murho}), respectively. \textit{Note that the distribution parameters $ \breve{\boldsymbol \mu}_{\cdot} $ and $ \breve{\boldsymbol \sigma}_{\cdot} $ are feature maps parameterized by the network parameters, $ \boldsymbol \theta_G $, of BayeSR.}

\subsection{Training and test strategies}\label{sec:train}
\subsubsection{Preliminary stage}\label{sec:prestage}
We pre-train the dowmsampling operator $ \mathbf A $ before training BayeSR. If the ground truth $ \mathbf u_i^* $ of an observation $ \mathbf y_i $ is available, we will train the dowmsampling module via minimizing MSE, $ \frac{1}{N}\textstyle\sum_{i=1}^N\left\| \mathbf A\mathbf u_i^* - \mathbf y_i \right\|_2^2  $, where $ N $ denotes the number of training samples. Otherwise, we will adopt KernelGAN \cite{Bell/2019} to train the module. Concretely, we discriminate the patch distributions between observations $ \mathbf y_i $ and their degradations $ \mathbf A \mathbf y_i $ via a discriminator, to make the downsampling module learn image degradation from $ \mathbf y_i $. Once the downsampling module is pre-trained, its parameters will be fixed in the following training.

We extract noise patches from observations before training BayeSR. Similar to the noise block extraction in \cite{Chen/2018}, if the mean and variance of any sub-patch, $ \mathbf y_i^p $, of an observation, $ \mathbf y_i $, satisfy $ \left| mean(\mathbf y_i) - mean(\mathbf y_i^p)\right| \le 0.05\cdot mean(\mathbf y_i) $ and $ \left| var(\mathbf y_i) - var(\mathbf y_i^p)\right| \le 0.1\cdot var(\mathbf y_i) $, we will add $ \mathbf n_i = \mathbf y_i - mean(\mathbf y_i) $ into the pool of noise patches, notated as $ \mathcal S_n = \lbrace \mathbf n_i \rbrace $.  

\subsubsection{Unsupervised training}\label{sec:unsuptrain}
BayeSR could be training by combining generative learning (GL), discriminative learning (DL), and generative adversarial learning (GAL). Concretely, training BayeSR by GL induces a variational loss notated as $ \mathcal L_{var}(\boldsymbol{\theta}_G) $; training BayeSR by DL induces a self-supervised loss notated as $ \mathcal L_{self}(\boldsymbol{\theta}_G) $, and training BayeSR by GAL induces a generative loss notated as $ \mathcal L_{gen}(\boldsymbol{\theta}_G) $. Therefore, our unsupervised strategy of training BayeSR is 
\begin{equation}\label{eq:totalunsuploss}
	\min_{\boldsymbol \theta_G} \mathcal L_{var}(\boldsymbol{\theta}_G) + \tau \mathcal L_{self}(\boldsymbol{\theta}_G) + \lambda \mathcal L_{gen}(\boldsymbol{\theta}_G),
\end{equation}
where, $ \tau $ and $ \lambda $ are hyperparameters. The details of this strategy are showed as follows.

BayeSR could be trained via GL when only LR images are available. Suppose $ \mathbf y_i $ and $ \mathbf u_i^{lr} $ are two randomly cropped patches from LR images, we could generate a pseudo degradation from $ \mathbf u_i^{lr} $, i.e., $ \mathbf y_i^{lr} = \mathbf A\mathbf u_i^{lr} + \mathbf n_i $, where $ \mathbf n_i $ denotes a sample from $ \mathcal S_n $. After that, we consider the concatenation of $ \mathbf y_i $ and $ \mathbf y_i^{lr} $ as an input, and infer the distribution parameters as shown in Fig. \ref{fig:architecture}. Finally, we compute the variational loss as shown in (\ref{eq:VBloss}) for $ \mathbf y_i $ and $ \mathbf y_i^{lr} $, and the resulting loss of training BayeSR is
\begin{equation}\label{eq:varloss}
	\mathcal L_{var}(\boldsymbol{\theta}_G) = \frac{1}{N} \textstyle\sum_{i=1}^N \mathcal [\mathcal L_{var}(\mathbf y_i) + \mathcal L_{var}(\mathbf y_i^{lr})].
\end{equation}

BayeSR could be trained via DL when the observation likelihood $ p(\mathbf u_i^{lr}| \mathbf y_i^{lr}, \boldsymbol{\theta}_G) $ is given. If $ p(\mathbf u_i^{lr}| \mathbf y_i^{lr}, \boldsymbol{\theta}_G) = \mathcal N(\mathbf x_i^{lr} + \mathbf z_i^{lr}, \frac{1}{2\tau} \mathbf I) $, maximum log-likelihood will induce the squared $ \ell_2 $ norm, $ \|  \mathbf u_i^{lr} - \mathbf x_i^{lr} - \mathbf z_i^{lr} \|_2^2  $, where $ \mathbf x_i^{lr} $ and $ \mathbf z_i^{lr} $ are smoothness component and sparsity residual parameterized by $ \boldsymbol{\theta}_G $. If $ p(\mathbf u_i^{lr}| \mathbf y_i^{lr}, \boldsymbol{\theta}_G) = \prod_{j=1}^{d_u} \mathcal La(x_{ij}^{lr} + z_{ij}^{lr}, \frac{1}{\tau}) $, where $ \mathcal La $ denotes the Laplace distribution, maximum log-likelihood will induce the $ \ell_1 $ norm, $ \|  \mathbf u_i^{lr} - \mathbf x_i^{lr} - \mathbf z_i^{lr} \|_1  $. Overall, the self-supervised loss of training BayeSR can be expressed as
\begin{equation}\label{eq:selfloss}
	\mathcal L_{self}(\boldsymbol{\theta}_G)= \frac{1}{N} \textstyle\sum_{i=1}^N \|  \mathbf u_i^{lr} - \mathbf x_i^{lr} - \mathbf z_i^{lr} \|_p^p,
\end{equation} 
where, $ p=2 $ if the downscaling factor $ s $ equals to 1, and $ p=1 $ otherwise. Note that the ``self'' means we use \textit{the LR image dataset itself} for discriminative learning, instead of using an LR image itself for internal learning \cite{Shocher/2018}.   

BayeSR could be trained via GAL. Concretely, we use a discriminator, referred to as $ D_u $, to discriminate the patch distributions between the restoration $ \mathbf x_i^{lr} + \mathbf z_i^{lr} $ and the reference $ \mathbf u_i^{lr} $. Moreover, we use another discriminator, referred to as $ D_y $, to discriminate the patch distributions between $ \mathbf A(\mathbf x_i^{lr} + \mathbf z_i^{lr}) $ and $ \mathbf A\mathbf u_i^{lr} $. Therefore, a generative loss of training BayeSR could be expressed as,
\begin{equation}\small\label{eq:genloss}
	\begin{aligned}
		\mathcal L_{gen}(\boldsymbol{\theta}_G) &= \frac{1}{N}\textstyle\sum_{i=1}^N\log [ 1 - D_u(\mathbf x_i^{lr} + \mathbf z_i^{lr})]\\
		&+ \frac{1}{N}\textstyle\sum_{i=1}^N\log [ 1 - D_y(\mathbf A(\mathbf x_i^{lr} + \mathbf z_i^{lr}))].
	\end{aligned}
\end{equation}
Besides, the discriminator $ D_u $ and $ D_y $ are trained by
\begin{equation}\footnotesize\label{eq:advloss}
	\begin{aligned}
		&\max_{\boldsymbol\theta_{D_u}} \frac{1}{N}\textstyle\sum_{i=1}^N \left[   \log D_u (\mathbf u_i^{lr}) + \log [ 1 - D_u(\mathbf x_i^{lr} + \mathbf z_i^{lr})] \right]\\
		&\max_{\boldsymbol\theta_{D_y}} \frac{1}{N}\textstyle\sum_{i=1}^N \left[  \log D_y (\mathbf A\mathbf u_i^{lr}) + \log [ 1 - D_y(\mathbf A(\mathbf x_i^{lr} + \mathbf z_i^{lr}))] \right]
	\end{aligned}
\end{equation}
where, $ \boldsymbol \theta_{D_u} $ and $ \boldsymbol \theta_{D_y} $ denote the parameters of $ D_u $ and $ D_y $, respectively.

\subsubsection{Pseudo-supervised and supervised training}\label{sec:psuptrain}
BayeSR could be trained via \textit{pseudo-supervised} learning, if unpaired LR and HR images are available. Suppose $ \mathbf y_i $ and $ \mathbf u_i^{hr} $ are two randomly cropped patches from LR and HR images, respectively, then we generate a pseudo degradation $ \mathbf y_i^{hr} $ from $ \mathbf u_i^{hr} $ using the same strategy as the unsupervised case, and replace $ \mathbf y_i^{lr} $ with $ \mathbf y_i^{hr} $ to compute the losses for GL, DL, and GAL. The only difference is the loss for DL becomes a pseudo-supervised one notated as $ \mathcal L_{pseudo}(\boldsymbol \theta_G) $, instead of the self-supervised loss. Therefore, the pseudo-supervised strategy of training BayeSR is
\begin{equation}\label{eq:totalpsuploss}
	\min_{\boldsymbol \theta_G} \mathcal L_{var}(\boldsymbol{\theta}_G) + \tau \mathcal L_{pseudo}(\boldsymbol{\theta}_G) + \lambda \mathcal L_{gen}(\boldsymbol{\theta}_G).
\end{equation} 

BayeSR could be trained via \textit{supervised} learning, if paired LR and HR images are available. Suppose $ \mathbf y_i $ is randomly cropped patches from LR images, and $ \mathbf u_i^* $ is its ground truth, then we replace $ \mathbf y_i^{lr} $ and $ \mathbf u_i^{lr} $ of the unsupervised case with $ \mathbf y_i $ and $ \mathbf u_i^* $, and compute the losses for GL and DL. Being different from the unsupervised case, $ \mathcal L_{var}(\boldsymbol \theta_G) $ is computed only for $ \mathbf y_i $, and the loss for DL becomes a supervised one notated as $ \mathcal L_{sup}(\boldsymbol \theta_G) $. Therefore, the supervised strategy of training BayeSR is
\begin{equation}\label{eq:totalsuploss}
	\min_{\boldsymbol \theta_G} \mathcal L_{var}(\boldsymbol{\theta}_G) + \tau \mathcal L_{sup}(\boldsymbol{\theta}_G).
\end{equation}  

\subsubsection{Test stage}\label{sec:teststage}
In the test stage, we could obtain many HR restorations from one LR observation by the proposed BayeSR. Concretely, an LR image $ \mathbf y $ (not vectorized) is fed into BayeSR, and the distribution parameters, $ \lbrace \breve{\boldsymbol \mu}_x, \breve{\boldsymbol \sigma}_x \rbrace $, w.r.t. the smoothness component $ \mathbf x $ and the distribution parameters, $ \lbrace \breve{\boldsymbol \mu}_z, \breve{\boldsymbol \sigma}_z \rbrace $, w.r.t. the sparsity residual $ \mathbf z $ are inferred from $ \mathbf y $. For evaluating the performance of restoration, we directly consider $ \breve{\boldsymbol \mu}_x + \breve{\boldsymbol \mu}_z $ as the deterministic restoration of $ \mathbf y $. For quantifying the diversity of restorations, we repeatedly sample $ \mathbf x $ and $ \mathbf z $ from their variational distributions for 10 times to generate a set of stochastic restorations $ \lbrace \mathbf u_i = \mathbf x_i + \mathbf z_i \rbrace_{i=1}^{10}  $.

\section{Experiments}
In this section, we first performed preliminary studies to obtain proper architectures and settings for the proposed BayeSR, and to interpret the functionality of BayeSR. After that, we validated the generalization ability of BayeSR, and evaluated the unsupervised performance using three tasks, i.e., ideal SISR, realistic SISR, and real-world SISR.
\subsection{Implementation details}\label{sec:implementation}
Three datasets were used to train BayeSR, i.e., DIV2K, Flickr2K, and DPED, thanks to their high-resolution (2K) and diversity. DIV2K\footnote{https://data.vision.ee.ethz.ch/cvl/DIV2K/} was firstly released from the NTIRE 2017 challenge on SISR, which consists of 800 training images, 100 validation images, and 100 test images. Flickr2K\footnote{http://cv.snu.ac.kr/research/EDSR/Flickr2K.tar} consists of 2650 diverse HR images, whose clean HR images were used as the unpaired references for pseudo-supervised training. DPED\footnote{http://people.ee.ethz.ch/~ihnatova/index.html} consists of photos taken synchronously in the wild by three smartphones and one professional camera. We used the DPED-iPhone from the NTIRE 2020 challenge for training and test. This dataset consists of 5614 training images, 113 validation images, and 100 test images. For \textit{ideal} SISR, we used the bicubic DIV2K where LR images were synthesized via bicubic interpolation. For \textit{realistic} SISR, we utilized the mild DIV2K where LR images were corrupted by unknown Poisson noise and random shifts. For \textit{real-world} SISR, we adopted the DPED-iPhone where LR images were corrupted by real noise.

Seven metrics were used to evaluate the performance of SISR, including five full-reference image quality assessments (IQA), i.e., the standard Peak Signal to Noise Ratio (PSNR) in HR space, the Structural Similarity (SSIM) index, the PSNR in LR space (LRPSNR), the Learned Perceptual Image Patch Similarity (LPIPS), and the Diversity (Div.) Score, and two no-reference IQAs, i.e., the Natural Image Quality Evaluator (NIQE) and the Blind/Referenceless Image Spatial QUality Evaluator (BRISQUE). To evaluate the performance of BayeSR in deterministic restoration, we obtained a restoration $ \breve{\boldsymbol \mu}_x + \breve{\boldsymbol \mu}_z $, as shown in Section \ref{sec:teststage}, from an observation $ \mathbf y $, and computed the PSNR, SSIM, LRPSNR, NIQE, and BRISQUE, where LRPSNR was computed between the degradation $ \mathbf A(\breve{\boldsymbol \mu}_x + \breve{\boldsymbol \mu}_z) $ and the observation $ \mathbf y $. To evaluate the performance of BayeSR in stochastic restoration, we obtained 10 stochastic restorations as shown in section \ref{sec:teststage}, and computed the LPIPS of each restoration to calculate the average LPIPS and Div. Score. For the task of ideal SISR, being consistent with the previous works \cite{Dong/2016}, we converted the super-resolved RGB images to YCbCr image, and computed PSNR and SSIM only on the Y channel by ignoring $ s+4 $ pixels from boundaries, where $ s $ denotes the downscaling factor. For the task of realistic SISR, being consistent with the evaluation criterion of NTIRE 2018 SR challenge on realistic SR, we computed maximal PSNR and SSIM by cropping a $ 60\times 60 $ patch from the center of an RGB image and shifting it up to 40 pixels in four directions. Since the unknown random shifts between LR images and their references make the computation of LPIPS and Div. Score inaccurate, we adopted NIQE and BRISQUE instead of LPIPS and Div. Score to evaluate models. For the task of real-world SISR, we only adopted LRPSNR, NIQE, and BRISQUE to evaluated models, since the ground truth of degraded images is inaccessible.

\begin{table*}
	\centering
	\caption{Ablation study: the effect of basic settings of BayeSR for realistic SISR ($\times 4$), including whether to use batch normalization (BN), using different network architectures, using different basic modules, increasing the depth of BayeSR, using different training strategies, and whether to train by generative learning (GL), discriminative learning (DL), and generative adversarial learning (GAL). Here, the depths of CNN$ _m $, CNN$ _z $, and CNN$ _x $ are denoted as $ (d_m, d_z, d_x) $. The bold font indicates the optimal settings in each of the sub-studies, while the underline font denotes the best model across the sub-studies.}
	\label{tab:ablation}
	\resizebox{1\linewidth}{!}{
		\begin{tabular}{|c|c|c|c|c|c|c|c|c|c|c|c|c|c|}
			\hline
			\multirow{2}{*}{Model}&  \multirow{2}{*}{Network}&  \multirow{2}{*}{BasicBlock}&
			\multirow{2}{*}{BN}& \multirow{2}{*}{$ (d_m, d_z, d_x) $}&  \multirow{2}{*}{Strategy}&  \multirow{2}{*}{GL/DL/GAL}&  \multicolumn{4}{c|}{DIV2K} &  \multirow{2}{*}{\#Para}&  Training & Test\\
			\cline{8-11}
			&  &  &  &  &  &  &  PSNR$ \uparrow $&  SSIM$ \uparrow $&  NIQE$ \downarrow $&  BRISQUE$ \downarrow $&  &  time& time\\
			\hline
			\#1  &  \multirow{3}{*}{ResNet}&  ResBlock&  Y&  \multirow{3}{*}{$ (2,2,2) $}&  \multirow{6}{*}{Sup}&  \multirow{3}{*}{Y/Y/N}&  23.83&  0.5455&  9.26&  66.78&  1.29M&  1.55h&  5.18s\\
			\#2  &  &  ResBlock&  \textbf{N}&  &  &  &  23.84&  0.5452&  9.26&  66.66&  1.29M&  1.68h&  5.15s\\
			\#3  &  &  \textbf{RCAB}&  N&  &  &  &  23.83&  0.5452&  9.31&  66.72&  1.29M&  1.68h&  5.78s\\
			\cline{1-5}\cline{7-14}
			\#4  &  ResNet&  RCAB&  \multirow{3}{*}{N}&  $ (4,4,4) $&  &  \multirow{3}{*}{Y/Y/N}&  23.92&  0.5491&  9.20&  67.13&  1.74M&  1.77h&  6.03s\\
			\underline{\#5}  &  \textbf{ResNet}&  RCAB&  &  $ \mathbf{(8,8,8)} $&  &  &  24.10&  0.5560&  8.78&  66.79&  2.63M&  2.35h&  6.21s\\
			\#6  &  U-Net&  ConvBlock&  &  $ (8,6,6) $&  &  &  23.92&  0.5510&  8.93&  63.05&  4.59M&  2.95h&  8.57s\\
			\hline\hline
			\#7  &  \textbf{ResNet}&  RCAB&  \multirow{2}{*}{N}&  $ (8,8,8) $&  \multirow{10}{*}{Unsup}&  \multirow{2}{*}{Y/Y/N}&  23.43&  0.5284&  7.23&  59.32&  2.63M&  3.23h&  6.61s\\
			\#8  &  U-Net&  ConvBlock&  &  $ (8,6,6) $&  &  &  23.01&  0.5063&  8.13&  59.46&  4.59M&  3.47h&  8.04s\\
			\cline{1-5}\cline{7-14}
			\underline{\#9}  &  \textbf{ResNet}&  RCAB&  \multirow{2}{*}{N}&  $ (8,8,8) $&  &  \multirow{2}{*}{Y/Y/GAN}&  23.67&  0.5334&  7.40&  57.83&  2.63M&  5.38h&  7.22s\\
			\#10  &  U-Net&  ConvBlock&  &  $ (8,6,6) $&  &  &  23.45&  0.5138&  7.40&  57.84&  4.59M&  5.58h&  7.25s\\
			\cline{1-5}\cline{7-14}
			\#11  &  \textbf{ResNet}&  RCAB&  \multirow{2}{*}{N}&  $ (8,8,8) $&  &  \multirow{2}{*}{Y/Y/WGAN}&  23.41&  0.5134&  5.06&  22.78&  2.63M&  5.18h&  6.17s\\
			\#12  &  U-Net&  ConvBlock&  &  $ (8,6,6) $&  &  &  23.12&  0.4956&  4.55&  17.81&  4.59M&  5.58h&  8.40s\\
			\cline{1-5}\cline{7-14}
			\#13  &  \textbf{ResNet}&  RCAB&  \multirow{2}{*}{N}&  $ (8,8,8) $&  &  \multirow{2}{*}{Y/Y/LSGAN}&  23.54&  0.5224&  5.73&  29.72&  2.63M&  5.63h&  6.44s\\
			\#14  &  U-Net&  ConvBlock&  &  $ (8,6,6) $&  &  &  23.36&  0.5083&  4.96&  14.55&  4.59M&  5.57h&  8.46s\\
			\cline{1-5}\cline{7-14}
			\#15  & \multirow{2}{*}{ResNet}&  \multirow{2}{*}{RCAB}&  \multirow{2}{*}{N}&  \multirow{2}{*}{$ (8,8,8) $}&  & Y/N/N&  21.74&  0.4707&  8.09&  67.77& 2.63M&  3.35h&  6.61s\\
			\#16  &  &  &  &  &  & N/Y/N&  22.20&  0.4849&  8.14&  56.29&  2.63M&  2.55h&  6.50s\\
			\hline
	\end{tabular}}
\end{table*}

For downsampling modules, the architecture is the same as KernelGAN. For other modules, the kernel size of convolutional layers (Convs) is $ 3\times 3 $, and that of transpose Convs is $ 5\times 5 $. The discriminator $ D_u $ consists of four Convs followed with batch normalization (BN) and Leaky ReLU, and one Conv as the output layer. The kernel sizes of five Convs are 4. The strides of the first three Convs are 2, and that of the last two Convs are 1. The numbers of kernels of five Convs are 64, 128, 256, 512, and 1, respectively. The discriminator $ D_y $ has the same structures as $ D_u $, except that only the strides of the first Conv are 2.    

For the graphical model, as shown in Fig. (\ref{fig:pipeline}) (a), the elements of $ \boldsymbol \gamma_{\upsilon} $, $ \boldsymbol \gamma_{\omega} $, and $ \boldsymbol \gamma_{\rho} $ were set to 2; the elements of $ \boldsymbol \phi_{\upsilon} $ and $ \boldsymbol \phi_{\omega} $ were set to $ 10^{-3} $; the elements of $ \boldsymbol \phi_{\rho} $ were set $ 10^{-5} $ in (\ref{eq:totalunsuploss}) and (\ref{eq:totalpsuploss}), and that were set to $ 10^{-3} $ in (\ref{eq:totalsuploss}). In the training stage, the hyperparameters $ \tau $ and $ \lambda $ in (\ref{eq:totalunsuploss}) were set to 1 and $ 10^{-4} $, respectively. Besides, we adopted the ADAM optimizer with $ \beta_1 = 0.9 $, $ \beta_2 = 0.999 $ and $ \epsilon = 1\times 10^{-8} $ to train BayeSR. The total training steps were set to be $ 1\times 10^6 $. The initial learning rate was set to be $ 1\times 10^{-4} $, and it was decreased by a factor of 0.5 every $ 2\times 10^5 $ updates. BayeSR was implemented with TensorFlow, and all models were trained and tested on a TITAN RTX GPU with 24 GB memory.

\begin{figure}[!t]
	\centering
	\includegraphics[width=0.9\linewidth]{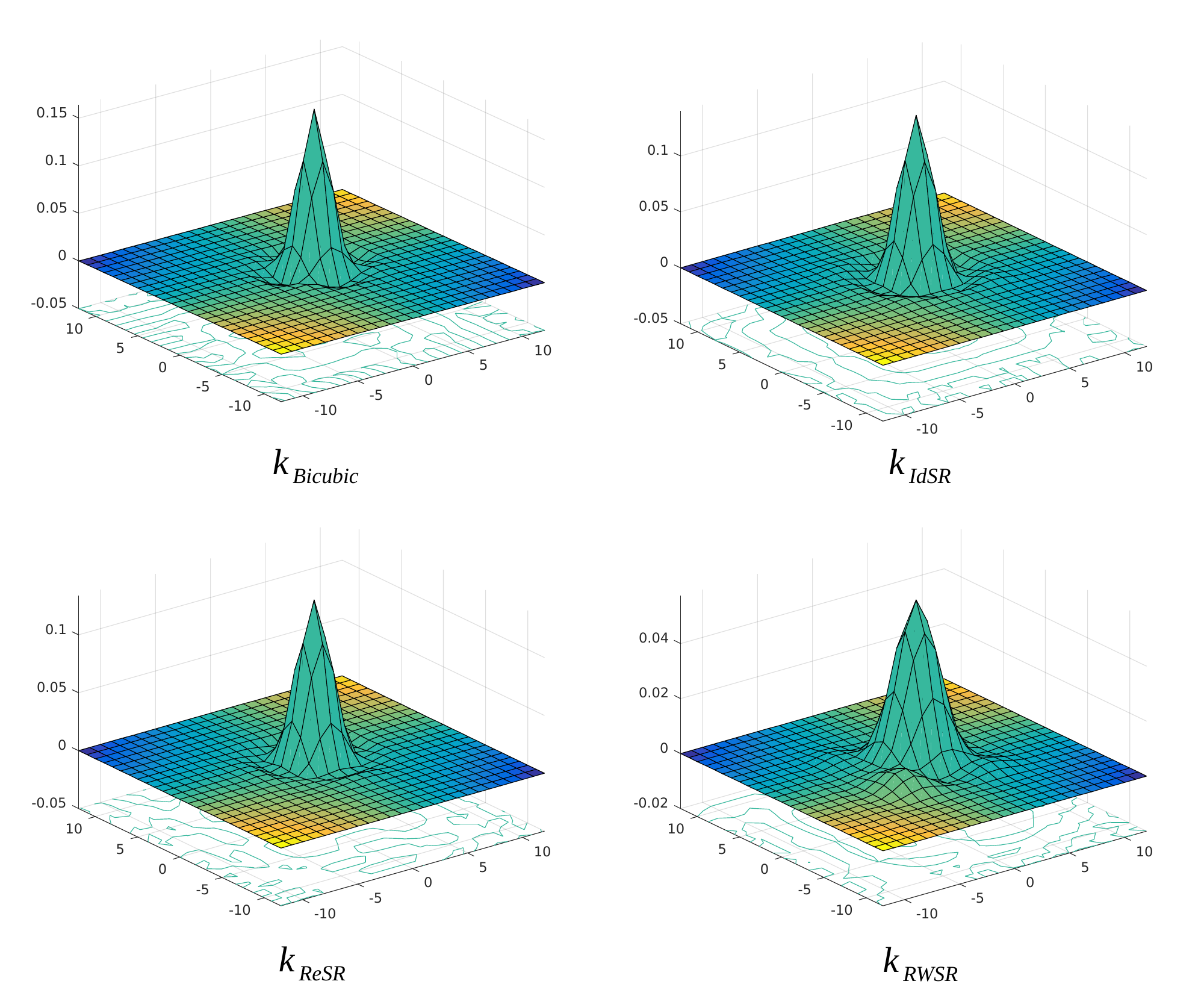}
	\caption{Visualization of degradation kernels ($ 25\times 25 $) for SISR $ \times 4 $. $ \mathbf k_{Bicubic} $ presents the true bicubic kernel. $ \mathbf k_{IdSR} $, $ \mathbf k_{ReSR} $, and $ \mathbf k_{RWSR} $ show the estimated kernels by KernelGAN for the tasks of ideal SISR, realistic SISR, and real-world SISR, respectively.  }
	\label{fig:kernels}
\end{figure}

\subsection{Preliminary study}\label{sec:ablation}
In this section, we studied appropriate settings for BayeSR and interpreted the functionality of BayeSR.
\subsubsection{Degradation kernel study}
To study degradation kernels, we pre-trained the downsampling module, i.e., $ \mathbf A $ in Fig. \ref{fig:architecture}, on three SISR ($ \times 4 $) tasks. Concretely, we first trained the module on ideal SISR using the supervised strategy as shown in Section \ref{sec:prestage}, and obtained a standard kernel notated as $ \mathbf k_{Bicubic} $. After that, we used the unsupervised strategy to train the module on the task of ideal, realistic, and real-world SISR, and obtained three kernels notated as $ \mathbf k_{IdSR} $, $ \mathbf k_{ReSR} $, and $ \mathbf k_{RWSR} $, respectively. Finally, we visualized the four kernels in Fig. \ref{fig:kernels}. This figure shows that $ \mathbf k_{IdSR} $ is very similar to $ \mathbf k_{Bicubic} $, which demonstrates the effectiveness of KernelGAN in estimating degradation kernels. Moreover, the realistic kernel is similar to the ideal kernel, but the real-world one is greatly different, which shows the challenges and necessity of estimating degradation kernels for real-world images. 

\subsubsection{Ablation study}
To study basic modules, we fixed the depth of CNN$ _m $, CNN$ _z $, and CNN$ _x $ to be $ d_m=2 $, $ d_z=2 $, and $ d_x=2 $, and used the supervised strategy in (\ref{eq:totalsuploss}) to train BayeSR on the task of realistic SISR by setting the basic modules to be different structures, as shown in Table \ref{tab:ablation}. We first trained two models to test the effect of whether to use BN or not. After that, we trained an additional model to test the effect of using RCAB. The comparisons between model \#1 and \#2 show that using BN does not improve the performance of BayeSR, and therefore we remove BN in the following studies. The comparisons between model \#2 and \#3 show that using RCAB does not improve the performance of BayeSR, but we adopt RCAB in the following studies, since the weight of skip connection in RCAB is learnable while that in ResBlock is fixed to 0.2.

\begin{figure*}[!t]
	\centering
	\includegraphics[width=1\linewidth]{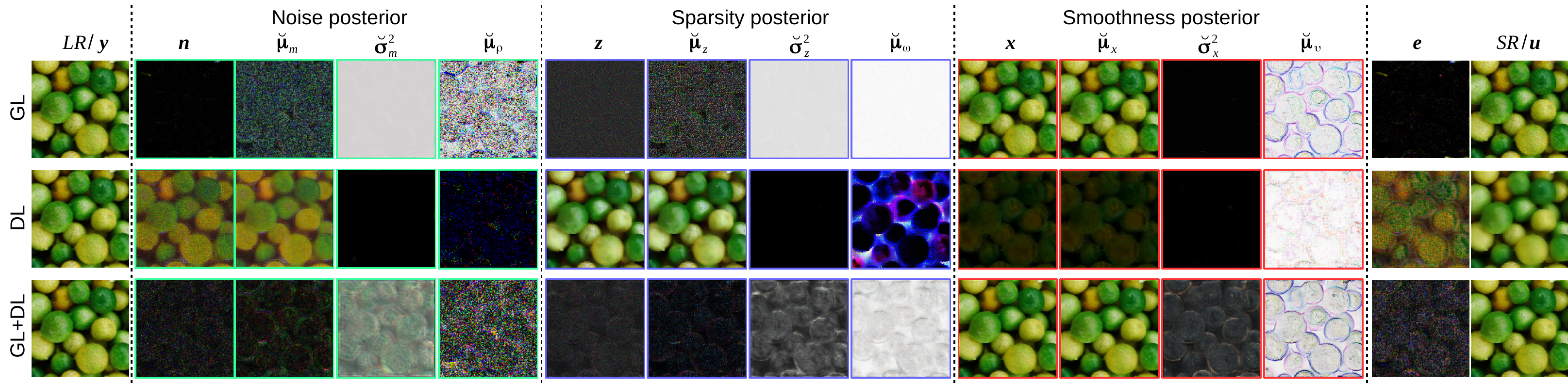}
	\caption{Visualization of variational posteriors inferred by three typical models in Table \ref{tab:ablation}, i.e., \#15 (GL), \#16 (DL), and \#7 (GL+DL). The five columns split by dotted lines represent LR observations, posteriors w.r.t. the noise $ \mathbf n $, the sparsity residual $ \mathbf z $, and the smoothness component $ \mathbf x $, and outputs. Here, $ \mathbf e = \mathbf y - \mathbf A(\mathbf x + \mathbf z) $ denotes the residual error in LR space, GL represents generative learning, and DL is discriminative learning. Please zoom in the online electronic version for more details.}
	\label{fig:posteriors}
\end{figure*} 

To study network architectures, we trained BayeSR by increasing its depth and using different backbones. We first trained two models to test the effect of increasing the depth of CNNs from 4 to 8. Then, we trained an additional model, i.e., model \#6, to test the effect of using different networks. The comparisons among model \#3, \#4, and \#5 show that deeper models deliver better performance. Table \ref{tab:ablation} shows that model \#5 built with ResNet is lighter in terms of the number of parameters and computationally cheaper according to the training and test time. Moreover, model \#5 delivers the best performance among supervised models, and thus its settings are used in the following supervised BayeSR models. 

To study generative adversarial learning (GAL), we trained four groups of unsupervised models by using different strategies, and each group contained two models with different backbones. First, we trained a group of models without using any GAL strategies. Then, we trained the second group of models, i.e., model \#9 and \#10, using the original GAL strategy, which is known as generative adversarial networks (GAN). Besides, we trained the third group of models, i.e., model \#11 and \#12, using the strategy of Wasserstein GAN. Finally, we trained the fourth group of models, i.e., model \#13 and \#14, using the strategy of the least square GAN (LSGAN). The comparisons between model \#7 and \#8 show that model \#7 using ResNet delivers better performance in SISR and is computationally more efficient, which is consistent with the supervised case. As the same as the first group, the internal comparisons of other groups confirm that the architecture of ResNet is more appropriate for our framework. Moreover, the external comparisons among four groups demonstrate that model \#9 with GAN delivers the best performance in SISR among all unsupervised models.

To study generative learning (GL) and discriminative learning (DL), we trained two models using different learning methods. First, we trained model \#15 by GL, i.e., minimizing the variation loss $ \mathcal L_{var} $ in (\ref{eq:totalunsuploss}). Then, we trained model \#16 by DL, i.e., minimizing the self-supervised loss $ \mathcal L_{self} $ in (\ref{eq:totalunsuploss}). The comparisons among model \#7, \#15, and \#16 show that training BayeSR by combing GL and DL could obtain particularly better PSNR and SSIM values. Moreover, model \#9 shows that training BayeSR by combining GL, DL, and GAL could further improve PSNR and SSIM values. Thus, we adopt the settings of model \#9 for unsupervised learning in the following sections.

\subsubsection{Interpretation of BayeSR}\label{sec:interpretation}
Fig. \ref{fig:posteriors} visualizes the posteriors inferred by model \#15, \#16, and \#7 in Table \ref{tab:ablation}, respectively denoted as GL, DL, and GL+DL for convenience. Note that $ \mathbf n $, $ \mathbf x $, and $ \mathbf z $ are sampled from their variational distributions, e.g., $ \mathbf n \sim \mathcal N(\mathbf m, \mbox{diag}(\breve{\boldsymbol \mu}_{\rho})^{-1}) $ and $ \mathbf m \sim \mathcal N(\breve{\boldsymbol \mu}_m, \mbox{diag}(\breve{\boldsymbol \sigma}_m^2)) $. Besides, we normalize all inferences by first taking absolute values and then being divided by their maximal values, except for $ \mathbf x $, $ \breve{\boldsymbol \mu}_x $, and $ \mathbf u $.

To understand the advantage of combining generative learning (GL) and discriminative learning (DL), we compared the posteriors of GL and DL in Fig. \ref{fig:posteriors}. One can see that $ \mathbf z $ and $ \mathbf x $ of GL are prone to be sparse and smooth, respectively, this is because we explicitly modeled the priors of $ \mathbf z $ and $ \mathbf x $ by (\ref{eq:zpdf}) and (\ref{eq:xpdf}), However, since the noise $ \mathbf n $ of corrupting $ \mathbf y $ was not properly estimated, maximizing the observation likelihood of $ \mathbf y $, as shown by $ \mathcal L_y $ in (\ref{eq:VBloss}), induced a particularly small error $ \mathbf e $ in LR space, and thus the restoration of GL contains a lot of artifacts induced by noise. By contrast, $ \mathbf e $ of DL is prone to approximate the noise of corrupting $ \mathbf y $, since we minimized the distance between the restoration and its reference by the self-supervised loss $ \mathcal L_{self} $ in (\ref{eq:totalunsuploss}). However, $ \mathbf z $ and $ \mathbf x $ of DL are feature maps with unknown statistics, since we did not explicitly model their priors. Moreover, the restoration of DL is prone to be over-smooth. Therefore, we trained BayeSR by combining GL and DL. That could generate the smooth $ \mathbf x $, the sparse $ \mathbf z $, and the best restoration $ \mathbf u $, as shown in Fig. \ref{fig:posteriors}.

To understand why BayeSR can produce interpretable components, we explained its functionality based on the posteriors of GL+DL in Fig. \ref{fig:posteriors}. $ \mathcal L_{\breve{\boldsymbol \mu}_x} $ in (\ref{eq:VBloss}) quantifies the smoothness of $ \breve{\boldsymbol\mu}_x $ weighted by $ \breve{\boldsymbol \mu}_{\upsilon} $. The visualized results show that small values of $ \breve{\boldsymbol \mu}_{\upsilon} $ are aligned to boundaries of $ \breve{\boldsymbol\mu}_x $, while large values are aligned to smooth areas. Therefore, minimizing $ \mathcal L_{\breve{\boldsymbol \mu}_x} $ could produce a piece-wisely smooth $ \mathbf x $ with sharp edges; $ \mathcal L_{\breve{\boldsymbol \mu}_z} $ quantifies the sparsity of $ \breve{\boldsymbol\mu}_z $ weighted by $ \breve{\boldsymbol \mu}_{\omega} $. The visualized results show that small values are aligned to pixels of $ \breve{\boldsymbol\mu}_z $ representing image details, while large values are aligned to pixels located in smooth areas. Therefore, minimizing $ \mathcal L_{\breve{\boldsymbol \mu}_z} $ could generate a sparse $ \mathbf z $ including image details; $ \mathcal L_y $ quantifies the weighted error by $ \breve{\boldsymbol\mu}_{\rho} $, and $ \mbox{diag}(\breve{\boldsymbol\mu}_{\rho})^{-1} $ denoted the variance of $ \mathbf n $. The comparisons between $ \breve{\boldsymbol \mu}_{\rho} $ and $ \mathbf e $ show that small values are aligned to large errors, in other words, strong noise is used to approximate a large error that could not be fitted by $ \mathbf A(\mathbf x + \mathbf z) $. Therefore, minimizing $ \mathcal L_y $ could generate a spatially variant noise $ \mathbf n $.

To understand why BayeSR could generate diverse restorations, we analyzed its uncertainties based on the posteriors of GL+DL in Fig. \ref{fig:posteriors}. Concretely, small values of $ \breve{\boldsymbol \sigma}_x^2 $ correspond to the smooth areas of $ \mathbf x $, while large values correspond to rich textures. Therefore, $ \breve{\boldsymbol \sigma}_x^2 $ represents the uncertainty of smoothness, namely, a large value of $ \breve{\boldsymbol \sigma}_x^2 $ indicates the pixel is more likely to be located in a non-smooth area of $ \mathbf x $. Similarly, $ \breve{\boldsymbol \sigma}_z^2 $ represents the uncertainty of sparsity, namely, a large value of $ \breve{\boldsymbol \sigma}_z^2 $ indicates the pixel of $ \mathbf z $ is more likely to be non-zero. Using the two uncertainty maps, one can generate diverse stochastic restorations. Moreover, $ \breve{\boldsymbol\mu}_{\rho}^{-1} $ represents the uncertainty of observations, namely, a large value of $ \breve{\boldsymbol\mu}_{\rho}^{-1} $ indicates the pixel of $ \mathbf y $ is more likely to be corrupted by strong noise. 

\subsection{Explicit modeling and generalization ability}
This section studies the robustness of explicit modeling via training BayeSR as an auto-encoder, and shows the generalization ability of BayeSR by supervised learning.
\begin{figure}[!t]
	\centering
	\includegraphics[width=1\linewidth]{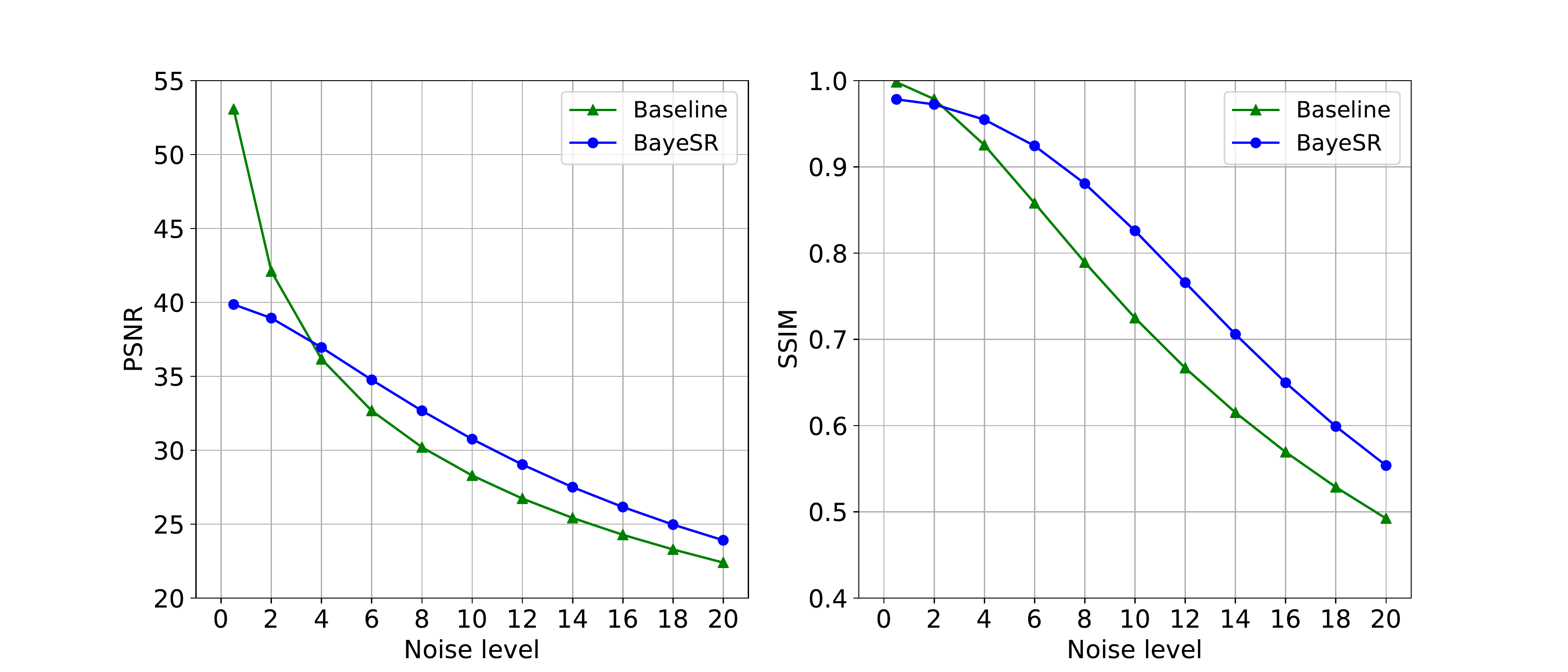}
	\caption{The performance of Baseline and BayeSR on the BSD68 with different noise levels.}
	\label{fig:psnrssim}
\end{figure}

\subsubsection{Robustness of explicit modeling}
To study the effect of explicit modeling of priors, we trained BayeSR to be an auto-encoder. Concretely, we first set up BayeSR using the same settings as model \#5 in Table \ref{tab:ablation} Section \ref{sec:ablation}, but removed the unnecessary upsampling and downsampling modules. Then, we fed clean HR patches (of size $ 128\times 128 $) from DIV2K to BayeSR, and trained it using (\ref{eq:totalsuploss}). For comparisons, we set up a baseline using the same architecture, but the network was trained without using the variational loss, $ \mathcal L_{var} $, in (\ref{eq:totalsuploss}), i.e., without explicit prior modeling. Finally, we tested the performance of baseline and BayeSR on the public BSDS68 \cite{Martin/2001}. Although the baseline and BayeSR were trained only using clean images, the test images were corrupted by adding white Gaussian noise (AWGN) with the noise level ranging in $ \left[ 0, 20 \right]  $. Fig. \ref{fig:psnrssim} shows the curves of PSNR and SSIM of the two models. One can see that BayeSR achieves lower scores than the baseline when the noise level is smaller than 2, but significantly higher PSNR and SSIM values when the noise level is bigger than 4. This demonstrates that explicit modeling of priors could improve the robustness of BayeSR against unseen noise, though the explicit modeling may handicap the performance when low-level noise or clean images are presented.

\begin{table*}[!t]
	\centering
	\caption{Evaluation of generalization ability of SRCNN \cite{Dong/2016}, VDSR \cite{Kim1/2016}, LapSRN \cite{Lai/2017}, EDSR \cite{Lim/2017}, RCAN \cite{Zhangyulun/2018}, OISR \cite{He/OISR/2019}, and S-BayeSR (ours). We test all methods on Set5 \cite{Bevilacqua/2012}, Set14 \cite{Zeyde/2012}, BSDS100 \cite{Martin/2001}, and Urban100 \cite{Huang/2015}, and report the average PSNR ($ \uparrow $), SSIM ($ \uparrow $), LRPSNR ($ \uparrow $), LPIPS ($ \downarrow $), and Div. Score ($ \uparrow $). Here, Div. Score is only used to indicate whether a model is deterministic or stochastic. The bold value denotes the best performance, and the italic value represents the second-best performance.}
	\label{tab:supIdSR}
	\resizebox{1\textwidth}{!}{
		\begin{tabular}{|c|c|c|c|c|@{}c@{}|c|c|c|@{}c@{}|c|c|c|@{}c@{}|c|c|c|@{}c@{}|c|c|}
			\hline
			\multirow{2}{*}{$ \sigma $}&  \multirow{2}{*}{Method}&  \multirow{2}{*}{\#Paras}&   \multicolumn{4}{c|}{Set5}  &  \multicolumn{4}{c|}{Set14}  &  \multicolumn{4}{c|}{BSDS100}  &  \multicolumn{4}{c|}{Urban100}&  Div.\\
			\cline{4-19}
			&  &  &  PSNR&  SSIM&  LRPSNR& LPIPS& PSNR&  SSIM&  LRPSNR& LPIPS&  PSNR&  SSIM&  LRPSNR& LPIPS&  PSNR&  SSIM&  LRPSNR& LPIPS&  Score\\
			\hline
			\multirow{9}{*}{0}&  Bicubic&  --&  28.42&  0.8105&  35.02&  0.3357&  26.10&  0.7048&  34.62&  0.4320&  25.96&  0.6676&  35.86&  0.5175&  23.15&  0.6579&  32.73&  0.4677& 0\\
			&  SRCNN&  0.07M&  30.49&  0.8629&  40.74&  0.1954&  27.61&  0.7535&  40.37&  0.3096& 26.91&  0.7104&  41.99&  0.4041&  24.53&  0.7230&  39.67&  0.3123&  0\\
			&  VDSR&  0.67M&  31.35&  0.8838&  41.18&  0.1798&  28.02&  0.7678&  40.90&  0.3002& 27.28&  0.7250&  42.85&  0.3920&  25.18&  0.7523&  39.95&  0.2730&  0\\
			&  LapSRN&  0.90M&  31.52&  0.8854&  41.19&  0.1813&  28.08&  0.7687&  41.02&  0.3014&  27.30& 0.7253 &  42.89&  0.3947&  25.20&  0.7544&  40.12&  0.2731&  0\\
			&  EDSR&  43.1M&  32.46&  0.8976&  42.89&  0.1707&  \textit{28.80}&  \textit{0.7872}&  \textit{42.66}&  0.2742&  \textit{27.72}&  0.7414&  \textit{43.91}&  0.3613&  26.64&  0.8029&  41.52&  0.2040&  0\\
			&  RCAN&  15.6M&  \textbf{32.60}&  \textbf{0.8991}&  \textit{42.93}&  \textit{0.1692}& 28.71&  0.7851&  42.56&  \textit{0.2727}&  \textbf{27.75}&  \textbf{0.7426}&  43.79&  0.3569& \textbf{26.81}&  \textbf{0.8079}&  \textit{41.53}&  \textbf{0.1953}&  0\\
			&  OISR&  44.3M&  \textit{32.51}&  \textit{0.8983}&  42.85&  0.1698& \textbf{28.85}&  \textbf{0.7878}&  \textbf{42.69}&  0.2757&  \textbf{27.75}&  \textit{0.7423}&  \textbf{43.93}&  0.3617&  \textit{26.78}&  \textit{0.8066}&  41.48&  \textit{0.2027}& 0\\
			&  S-Baseline&  2.63M&  32.07&  0.8923&  \textbf{42.94}&  0.1731&  28.38&  0.7764&  \textit{42.66}&  0.2841&  27.51&  0.7336&  43.85&  0.3737&  25.98&  0.7802&  \textbf{41.58}&  0.2322&  0\\
			&  S-BayeSR&  2.63M&  31.50&  0.8805&  39.02&  \textbf{0.1223}&  28.08&  0.7561&  38.15&  \textbf{0.2229}&  27.21&  0.7091&  38.36&  \textbf{0.3216}&  25.50&  0.7528&  37.01&  0.2336&  6.84\\
			\hline
			\multirow{9}{*}{10}&  Bicubic&  --&  25.91&  0.6715&  25.93&  0.6541&  24.21&  0.5753&  25.71&  0.7490&  24.29&  0.5449&  25.84&  0.8630&  22.18&  0.5451&  25.41&  0.8446& 0\\
			&  SRCNN&  0.07M&  24.42&  0.5711&  25.59&  0.5403&  23.11&  0.4918&  25.31&  0.6576& 23.00&  0.4572&  25.31&  0.7742&  21.70&  0.4851&  25.34&  0.7213&  0\\
			&  VDSR&  0.67M&  24.20&  0.5584&  25.53&  0.5507&  23.05&  0.4872&  25.57&  0.6613& 22.82&  0.4450&  25.27&  0.7853&  21.54&  0.4742&  25.28&  0.7184& 0\\
			&  LapSRN&  0.90M&  \textit{27.46}&  \textit{0.7164}&  \textbf{30.81}&  \textit{0.3901}&  25.33&  \textit{0.6141}&  \textbf{30.55}&  \textit{0.5354}&  \textit{25.03}&  \textit{0.5757}&  \textbf{30.74}&  0.6741&  23.38&  0.6052&  \textbf{30.61}&  \textit{0.5699}&  0\\
			&  EDSR&  43.1M&  27.20&  0.6908&  29.60&  0.4548&  25.28&  0.6089&  29.40&  0.6030& 24.94&  0.5670&  29.39&  0.7282&  23.54&  0.6027&  29.40&  0.5998&  0\\
			&  RCAN&  15.6M&  27.22&  0.6989&  29.49&  0.4413&  \textit{25.43}&  0.6170&  29.36&  0.6019&  \textit{25.03}&  0.5752&  29.32&  0.7226&  \textit{23.65}&  \textit{0.6126}&  29.33&  0.5763&  0\\
			&  OISR&  44.3M&  27.17&  0.6967&  29.56&  0.4435&  25.30&  0.6113&  29.36&  0.5930& 24.92&  0.5693&  29.35&  0.7134&  23.55&  0.6055&  29.35&  0.5801&  0\\
			&  S-Baseline&  2.63M&  27.27&  0.7027&  29.61&  0.4261&  25.39&  \textit{0.6188}&  29.49&  0.5933&  25.02&  \textit{0.5757}&  29.40&  0.7214&  23.45&  0.6041&  29.42&  0.5936& 0\\
			&  S-BayeSR&  2.63M&  \textbf{28.19}&  \textbf{0.7786}&  \textit{30.46}&  \textbf{0.2825}&  \textbf{26.11}&  \textbf{0.6717}&  \textit{30.52}&  \textbf{0.4302}&  \textbf{25.67}&  \textbf{0.6242}&  \textit{30.52}&  \textbf{0.5544}&  \textbf{23.78}&  \textbf{0.6530}&  \textit{30.12}&  \textbf{0.4614}&  4.60\\
			\hline
			\multirow{9}{*}{20}&  Bicubic&  --&  22.16&  0.4960&  20.76&  0.7769&  21.65&  \textit{0.4311}&  20.47&  0.8505&  21.71&  \textit{0.3926}&  20.49&  0.9178&  20.37&  0.4082&  20.42&  0.9370& 0\\
			&  SRCNN&  0.07M&  19.28&  0.3605&  19.98&  0.6782&  18.84&  0.2962&  19.68&  0.7605& 18.78&  0.2617&  19.64&  0.8281&  18.18&  0.3013&  19.74&  0.8228&  0\\
			&  VDSR&  0.67M&  19.00&  0.3212&  19.84&  0.6872&  18.59&  0.2817&  19.96&  0.7473& 18.45&  0.2488&  19.63&  0.8173&  17.79&  0.2852&  19.73&  0.8103&  0\\
			&  LapSRN&  0.90M&  \textit{23.05}&  \textit{0.5009}&  \textbf{25.18}&  \textit{0.5690}& 21.73&  0.4195&  \textit{24.78}&  0.6790&  \textit{21.78}&  0.3903&  \textit{24.89}&  \textit{0.7838}& \textit{20.73}&  \textit{0.4259}&  \textbf{24.97}&  \textit{0.7361}& 0\\
			&  EDSR&  43.1M&  22.40&  0.4455&  23.80&  0.6431&  21.62&  0.3992&  23.52&  0.7437& 21.51&  0.3665&  23.47&  0.8141&  20.56&  0.4041&  23.61&  0.7644&  0\\
			&  RCAN&  15.6M&  22.41&  0.4553&  23.62&  0.6363&  21.69&  0.4110&  23.32&  0.7300& 21.55&  0.3749&  23.31&  0.7993&  20.49&  0.4082&  23.39&  0.7434&  0\\
			&  OISR&  44.3M&  22.52&  0.4530&  23.76&  0.6365&  21.64&  0.4000&  23.47&  0.7338& 21.51&  0.3664&  23.43&  0.8036&  20.51&  0.4012&  23.53&  0.7509& 0\\
			&  S-Baseline&  2.63M&  22.55&  0.4759&  23.78&  0.5976&  \textit{21.82}&  0.4251&  23.48&  \textit{0.7234}&  21.67&  0.3887&  23.37&  0.8009&  20.60&  0.4209&  23.51&  0.7434& 0\\
			&  S-BayeSR&  2.63M&  \textbf{24.47}&  \textbf{0.6279}&  \textit{24.57}&  \textbf{0.4680}&  \textbf{23.79}&  \textbf{0.5481}&  \textbf{25.15}&  \textbf{0.6021}&  \textbf{23.60}&  \textbf{0.5052}&  \textbf{25.10}&  \textbf{0.7005}&  \textbf{21.98}&  \textbf{0.5316}&  \textit{24.85}&  \textbf{0.6386}&  3.44\\
			\hline
	\end{tabular}}
\end{table*}
\begin{figure*}[!t]
	\centering
	\includegraphics[width=1\linewidth]{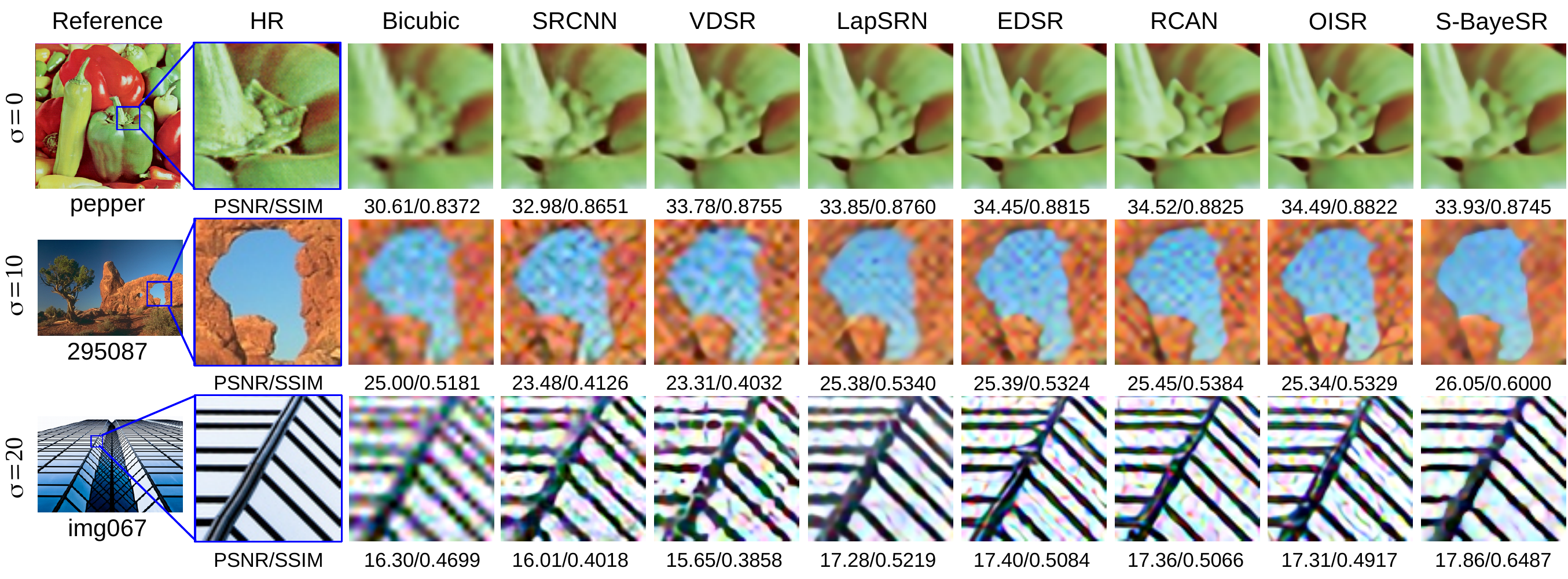}
	\caption{Visualization on the task of supervised ideal SISR $ \times 4 $: three typical examples from Set14, B100, Urban100, respectively. The first, second, and third rows denote the super-resolved results of LR images with the noise levels of $ \sigma=0 $, $ \sigma=10 $, and $ \sigma=20 $, respectively. Please refer to Supplementary Material for high-resolution images.}
	\label{fig:bisrsup}
\end{figure*}

\subsubsection{Generalization ability}\label{sec:generality}
To study the generalization ability of BayeSR, we trained BayeSR on the task of ideal SISR $ \times 4 $ via supervised learning. In the training stage, we first randomly cropped HR patches $ \mathbf u_i^* $ (of size $ 128\times 128 $) and LR patches $ \mathbf y_i $ (of size $ 32\times 32 $) from the bicubic DIV2K to generate the data $ \lbrace \mathbf y_i, \mathbf u_i^* \rbrace_{i=1}^N $ for supervised training, as shown in Section \ref{sec:psuptrain}. Then, we set up BayeSR using the same settings as model \#5 in Table \ref{tab:ablation} Section \ref{sec:ablation}, and initialized the downsampling module using the pre-trained model with respect to $ \mathbf k_{Bicubic} $. Finally, we froze the downsampling module, and fed randomly selected batches (of size 4) to train BayeSR up to $ 1\times 10^6 $ steps. This BayeSR supervisedly trained using (\ref{eq:totalsuploss}) was referred to as S-BayeSR. For comparisons, we trained a supervised baseline model referred to as S-Baseline, which had the same network architecture, but only minimized the supervised loss $ \mathcal L_{sup} $ without $ \mathcal L_{var} $ in (\ref{eq:totalsuploss}). 

In the test stage, we used four public datasets, i.e., Set5 \cite{Bevilacqua/2012}, Set14 \cite{Zeyde/2012}, BSDS100 \cite{Martin/2001}, and Urban100 \cite{Huang/2015}, to evaluate the performance of S-Baseline and S-BayeSR, and comparisons with six supervised methods, i.e., SRCNN \cite{Dong/2016}, VDSR \cite{Kim1/2016}, LapSRN \cite{Lai/2017}, EDSR \cite{Lim/2017}, RCAN \cite{Zhangyulun/2018}, and OISR \cite{He/OISR/2019}, in terms of PSNR, SSIM, LRPSNR, LPIPS, and Div. Score. Besides, we tested the methods on noisy datasets, which were corrupted by the AWGN with the noise level of $ \sigma=10,20 $, to show the generalization ability of them.

Table \ref{tab:supIdSR} summarizes the results on the task of supervised ideal SISR $ \times 4 $. The results in the cases of $ \sigma=10 $ and $ \sigma=20 $ show that S-BayeSR significantly outperforms the compared models in PSNR, SSIM, and LPIPS, which confirms that the explicit modeling of image priors could improve the generalization ability. The comparisons between S-Baseline and S-BayeSR in the case of $ \sigma=0 $ show that the imperfect modeling of image priors, i.e., by the combination of smoothness and sparsity priors, decreases the PSNR, SSIM, and LRPSNR values of S-BayeSR, but increases its LPIPS. Due to the uncertainties of S-BayeSR as shown in section \ref{sec:interpretation}, it could generate diverse stochastic restorations instead of a deterministic reconstruction, and thus achieves non-zero Div. Scores. Fig. \ref{fig:bisrsup} visualizes three typical examples. This figure shows that S-BayeSR maintains better local-similarity in $ \sigma=10 $ and $ \sigma=20 $, thanks to the explicit modeling of image priors.

\subsection{Unsupervised learning}\label{sec:unsup} 
This section studies the unsupervised performance of BayeSR via three different tasks, i.e., ideal SISR, realistic SISR, and real-world SISR. The difference among these tasks was described in Section \ref{sec:implementation}. Similar to S-BayeSR and S-Baseline, we refer to unsupervised BayeSR as U-BayeSR, Pseudo-supervised BayeSR as Ps-BayeSR, and unsupervised baseline as U-Baseline to avoid confusion.

\subsubsection{Ideal image super-resolution}\label{sec:IdSR}

\begin{table*}[!t]
	\centering
	\caption{Evaluation on the task of \textit{ideal} SISR $ \times 4 $. We test all methods on Set5 \cite{Bevilacqua/2012}, Set14 \cite{Zeyde/2012}, BSDS100 \cite{Martin/2001}, and Urban100 \cite{Huang/2015}, and report the average PSNR ($ \uparrow $), SSIM ($ \uparrow $), LRPSNR ($ \uparrow $), LPIPS ($ \downarrow $), and Div. Score ($ \uparrow $). Here, Div. Score is only used to indicate whether a model is deterministic or stochastic. The bold font indicates the best performance for the models trained without ground truth, and the italic value represents the second-best performance. Note that the supervised methods are just for reference.}
	\label{tab:unsupIdSR}
	\resizebox{1\linewidth}{!}{
		\begin{tabular}{|c|c|c|c|@{}c@{}|c|c|c|@{}c@{}|c|c|c|@{}c@{}|c|c|c|@{}c@{}|c|c|}
			\hline
			\multirow{2}{*}{Model}&  \multirow{2}{*}{\#Paras}&   \multicolumn{4}{c|}{Set5}  &  \multicolumn{4}{c|}{Set14}  &  \multicolumn{4}{c|}{BSDS100}  &  \multicolumn{4}{c|}{Urban100}& Div. \\
			\cline{3-18}
			&  &  PSNR&  SSIM&  LRPSNR&  LPIPS&  PSNR&  SSIM&  LRPSNR& LPIPS& PSNR&  SSIM&  LRPSNR& LPIPS&  PSNR&  SSIM&  LRPSNR& LPIPS&  Score\\
			\hline
			Bicubic&  --&  28.42&  0.8105&  35.02&  0.3357&  26.10&  0.7048&  34.62&  0.4320&  25.96&  0.6676&  35.86&  0.5175&  23.15&  0.6579&  32.73&  0.4677&  0\\
			ZSSR&  0.23M&  29.46&  0.8319&  42.15&  0.2062&  27.05&  0.7413&  42.33&  0.3186&  26.66&  0.7058&  43.54&  0.4101&  24.02&  0.7045&  41.13&  0.3363&  0\\
			MZSR&  0.23M&  28.12&  0.8029&  41.46&  0.2589&  25.77&  0.7089&  41.14&  0.3471& 25.83&  0.6782&  42.88&  0.4183&  23.02&  0.6659&  39.75&  0.3782&  0\\
			U-Baseline&  2.63M&  \textit{30.56}&  \textit{0.8738}&  \textbf{48.85}&  \textit{0.1785}&  \textit{27.33}&  \textit{0.7656}&  \textbf{47.30}&  \textit{0.2962}&  \textit{26.88}&  \textit{0.7267}&  \textbf{48.97}&  \textit{0.3905}&  \textit{24.58}&  \textit{0.7468}&  \textbf{46.28}&  \textit{0.2751}&  0\\
			U-BayeSR&  2.63M&  \textbf{30.86}&  \textbf{0.8809}&  \textit{45.81}&  \textbf{0.1088}& \textbf{27.51}&  \textbf{0.7679}&  \textit{45.14}&  \textbf{0.2123}&  \textbf{27.08}&  \textbf{0.7284}&  \textit{46.10}&  \textbf{0.2986}&  \textbf{24.91}&  \textbf{0.7614}&  \textit{44.33}&  \textbf{0.2171}& 7.70\\
			\hline
			\multicolumn{19}{|c|}{Supervised model for reference}\\
			\hline
			EnhanceNet&  0.85M&  28.56&  0.8093&  40.06&  0.1014&  25.04&  0.6528&  36.41&  0.1656&  24.09&  0.6006&  34.47&  0.2055&  22.30&  0.6504&  33.21&  0.1692&  0\\
			SRGAN&  2.03M&  28.19&  0.8163&  33.18&  0.0906&  25.97&  0.7001&  33.87&  0.1746& 24.63&  0.6416&  33.16&  0.2066&  23.67&  0.6984&  33.44&  0.1791&  0\\
			ESRGAN&  16.7M&  30.44&  0.8505&  40.74&  0.0750&  26.28&  0.6974&  38.65&  0.1341&  25.30&  0.6494&  40.05&  0.1615&  24.35&  0.7322&  37.70&  0.1231&  0\\
			SRFlow&  39.5M&  30.26&  0.8416&  42.17&  0.0771&  26.82&  0.7130&  42.03&  0.1314&  26.04&  0.6704&  43.13&  0.1825&  25.25&  0.7493&  40.87&  0.1271& 22.70\\
			\hline
	\end{tabular}}
\end{table*}
\begin{figure*}[!t]
	\centering
	\includegraphics[width=1\linewidth]{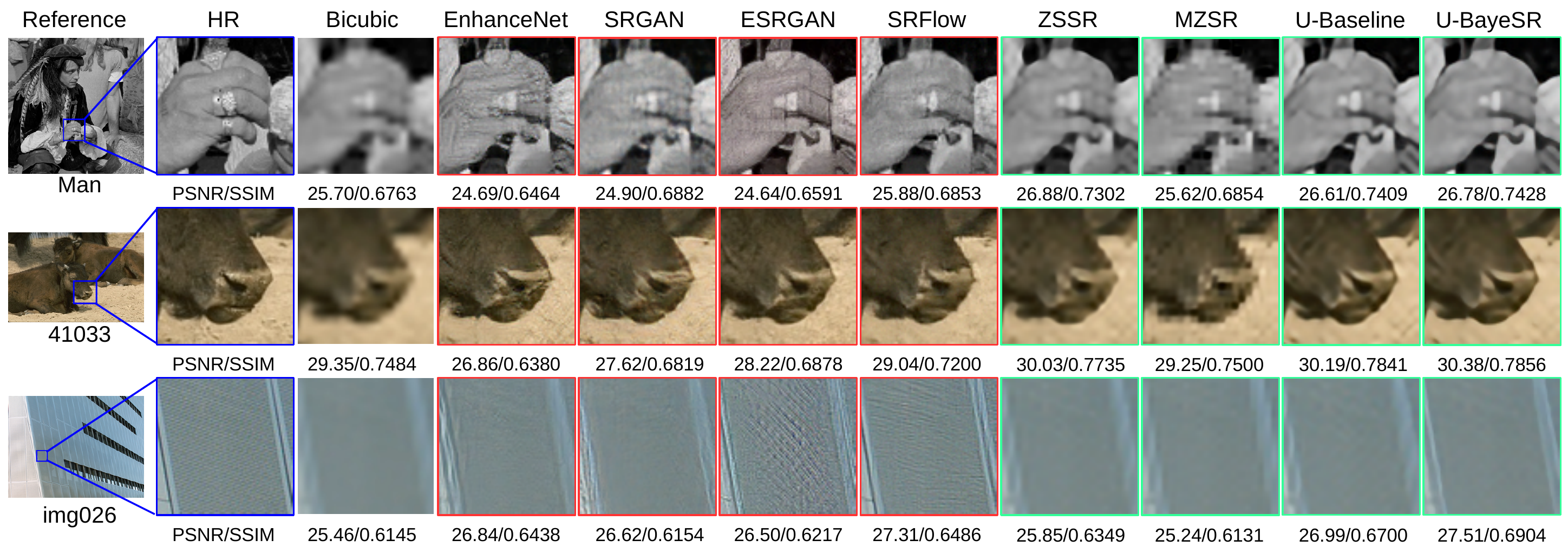}
	\caption{Visualization on the task of ideal SISR $ \times 4 $: three typical examples from Set14, B100, Urban100, respectively. The red boundary denotes the supervised method, while the green boundary represents the model trained without ground truth. Please refer to Supplementary Material for high-resolution images.}
	\label{fig:bisrunsup}
\end{figure*}

To study the unsupervised performance of BayeSR in ideal SISR $ \times 4 $, we trained BayeSR on the bicubic DIV2K. In the training stage, we first randomly cropped large LR patches $ \mathbf u_i^{lr} $ (of size $ 128\times 128 $) and small LR patches $ \mathbf y_i $ (of size $ 32\times 32 $) from the LR images of the bicubic DIV2K. Then, we obtained pseudo LR patches $ \mathbf y_i^{lr} $ from $ \mathbf u_i^{lr} $ using the strategy as shown in Section \ref{sec:unsuptrain}, and generated the data $ \lbrace \mathbf y_i, \mathbf y_i^{lr}, \mathbf u_i^{lr} \rbrace_{i=1}^N $ for unsupervised training. Besides, we set up BayeSR using the same settings as model \#9 in Table \ref{tab:ablation} Section \ref{sec:ablation}, and initialized the downsampling module of BayeSR using the pre-trained model with respect to $ k_{IdSR} $. For comparisons, we first trained U-Baseline by minimizing the self-supervised loss $ \mathcal L_{self} $ in (\ref{eq:totalunsuploss}). After that, we trained U-BayeSR using (\ref{eq:totalunsuploss}). 

In the test stage, we included two unsupervised methods, i.e., ZSSR \cite{Shocher/2018} and MZSR \cite{Soh/2020}, trained via internal learning for comparisons, and four supervised methods, i.e., EnhanceNet \cite{Lim/2017}, SRGAN \cite{Ledig/2017}, ESRGAN \cite{Wang/2018}, and SRFlow \cite{Lugmayr/2020}, oriented by perceptual quality for reference. Moreover, we used the same test datasets and criteria as the previous section to evaluate the performance of compared methods.

\begin{figure*}[!t]
	\centering
	\includegraphics[width=1\linewidth]{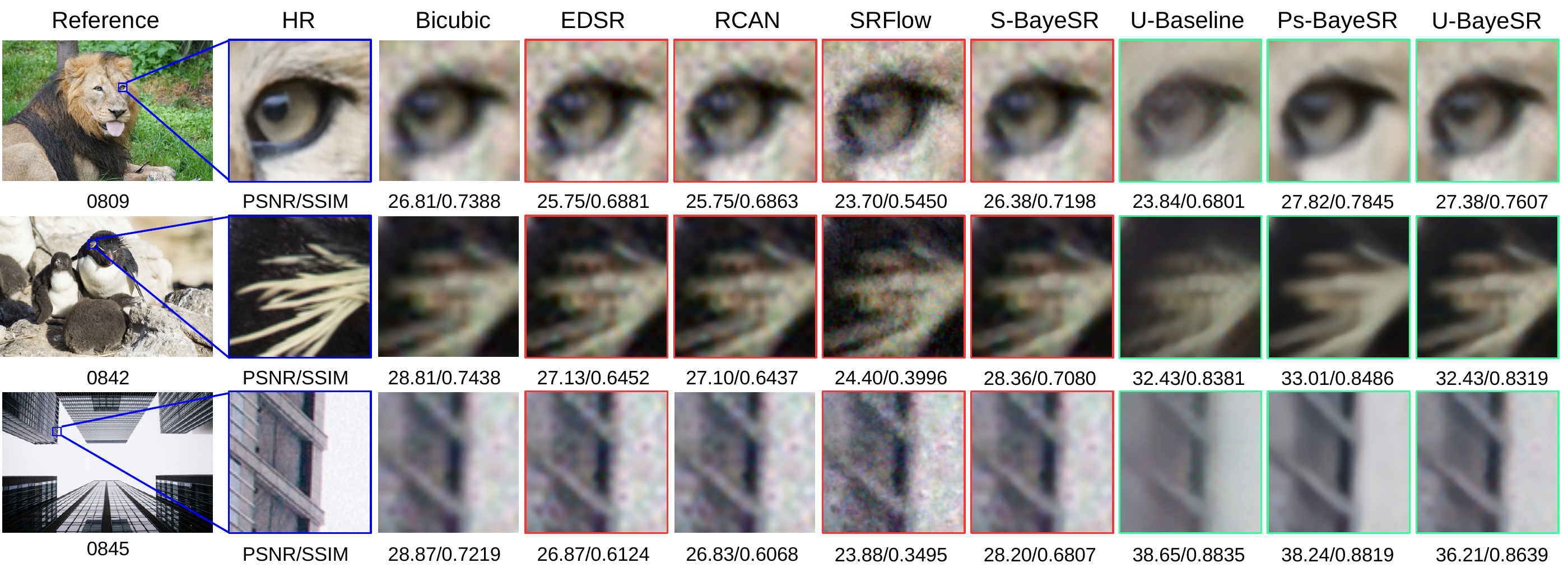}
	\caption{Visualization on the task of realistic SISR $ \times 4 $: three typical examples from DIV2K. The red boundary denotes the model transferred from ideal SISR, while the green boundary represents the model trained without ground truth. Please refer to Supplementary Material for high-resolution images.}
	\label{fig:resr}
\end{figure*}
\begin{table}[htp]
	\centering
	\caption{ Evaluation on the task of \textit{realistic} SISR $ \times 4 $. We test all methods on the validation dataset of the mild DIV2K, and report the average PSNR ($ \uparrow $), SSIM ($ \uparrow $), LRPSNR, NIQE, and BRISQUE. The bold value denotes the best PSNR and SSIM for the models transferred from ideal SISR or trained without ground truth, and the italic value represents the second-best performance.}
	\label{tab:ReSR}
	\resizebox{0.7\linewidth}{!}{
		\begin{tabular}{|c|c|c|@{}c@{}|c|@{}c@{}|c|}
			\hline
			Model& PSNR& SSIM& LRPSNR& NIQE& BRISQUE& \# Paras\\ 
			\hline
			HR&  $ \infty $&  1&  18.63&  3.07&  14.66& --\\
			Bicubic& 23.16&  0.5178&  38.98&  8.20&  62.65& --\\
			\cline{1-7}
			\hline
			\multicolumn{7}{|c|}{Model transferred from ideal SISR}\\
			\hline
			EDSR& 22.83&  0.4958&  44.46&  7.32&  58.56& 43.1M\\
			RCAN&  22.84&  0.4962&  44.51&  7.39&  57.45& 15.6M\\
			SRFlow& 21.41&  0.3688&  43.38&  3.73&  14.60& 39.5M\\
			S-BayeSR& \textbf{23.09}&  \textbf{0.5090}&  40.29&  7.40&  60.71& 2.63M\\
			\cline{1-7}
			\hline
			\multicolumn{7}{|c|}{Model trained without ground truth}\\
			\hline
			U-Baseline&  22.20&  0.4849&  27.44&  8.14&  56.29& 2.63M\\
			Ps-BayeSR&  \textbf{23.86}&  \textbf{0.5422}&  32.84&  8.28&  61.18& 2.63M\\
			U-BayeSR&  \textit{23.67}&  \textit{0.5334}&  33.37&  7.40&  57.83& 2.63M\\
			\hline
			\multicolumn{7}{|c|}{Supervised model for reference}\\
			\hline
			EDSR& 24.38&  0.5800&  24.44&  7.85&  61.35& 43.1M\\
			WDSR& 24.45&  0.5824&  24.08&  7.88&  61.89& 9.9M\\
			RCAN& 24.55&  0.5831&  25.13&  8.09&  63.77& 15.6M\\
			S-Baseline&  24.55&  0.5827&  25.97&  8.15&  63.49& 2.63M\\
			S-BayeSR&  24.10&  0.5560&  26.15&  8.78&  66.79& 2.63M\\
			\hline
	\end{tabular}}
\end{table}

Table \ref{tab:unsupIdSR} summarizes the results on the task of ideal SISR $ \times 4 $. U-BayeSR achieves the best PSNR, SSIM, and LPIPS in all studies, and gets the second-best LRPSNR. Besides, U-BayeSR outperforms U-Baseline in 12 studies (out of 16), which shows that train BayeSR by combining GL, DL, and GAL is more effective. Moreover, \textit{U-Baseline outperforms ZSSR and MZSR in all studies, which means the self-supervised learning on the LR dataset could be better than the internal learning on a single LR image}. Fig. \ref{fig:bisrunsup} visualizes three typical examples. This figure shows that U-BayeSR could restore more image details than the other unsupervised models. Also, the supervised models could generate some image artifacts, while U-BayeSR maintains better local similarity due to the explicit modeling of image priors.

\subsubsection{Realistic image super-resolution}\label{sec:ReSR}

To study the unsupervised performance of BayeSR in realistic SISR $ \times 4 $, we trained BayeSR on the mild DIV2K. For unsupervised training, we adopted the similar strategy as the ideal SISR to generate training data $ \lbrace \mathbf y_i, \mathbf y_i^{lr}, \mathbf u_i^{lr} \rbrace_{i=1}^N $ from the LR images of the mild DIV2K. For pseudo-supervised training, we cropped HR patches $ \mathbf u_i^{hr} $ (of size $ 128\times 128 $) from the HR images of Flickr2K to generate training data $ \lbrace \mathbf y_i, \mathbf y_i^{hr}, \mathbf u_i^{hr} \rbrace_{i=1}^N $ using the strategy as shown in Section \ref{sec:psuptrain}. For supervised training, we cropped the ground truth of $ \mathbf y_i $ from the HR images of the mild DIV2K to generate training data $ \lbrace \mathbf y_i, \mathbf u_i^* \rbrace_{i=1}^N $. In the training stage, we first set up BayeSR using the same settings as model \#9 (\#5) in Table \ref{tab:ablation} Section \ref{sec:ablation} for unsupervised or pseudo-supervised (supervised) training, and initialized its downsampling module using the pre-trained model with respect to $ \mathbf k_{ReSR} $. After that, we trained U-Basline and U-BayeSR as the ideal SISR, and trained Ps-BayeSR using (\ref{eq:totalpsuploss}). For reference, we trained EDSR \cite{Lim/2017}, WDSR \cite{Yu/2018}, RCAN \cite{Zhangyulun/2018}, and S-Baseline by minimizing $ \mathcal L_{sup} $ in (\ref{eq:totalsuploss}), and trained S-BayeSR using (\ref{eq:totalsuploss}). In the test stage, since the test dataset of the mild DIV2K is not public, we evaluated the performance of all methods on the validation dataset by PSNR, SSIM, LRPSNR, NIQE, and BRISQUE. Note that \textit{higher LRPSNR does not mean better performance for realistic SISR}, since LR images were corrupted by noise.

Table \ref{tab:ReSR} summarizes the quantitative results on the task of realistic SISR $ \times 4 $. Ps-BayeSR achieves the best performance in terms of PSNR and SSIM, and U-BayeSR gets the second best. Besides, U-BayeSR significantly outperforms the U-Baseline in PSNR and SSIM, which shows that training BayeSR by combining GL, DL, and GAL is better than only by DL. Among transferred models, S-BayeSR achieves the highest PSNR and SSIM values, due to its better generalization ability as shown in Section \ref{sec:generality}. Compared with other methods, the transferred models achieve higher LRPSNR, and thus more noise artifacts are included in their restoration. Fig. \ref{fig:resr} visualizes three typical examples. This figure shows that U-BayeSR and Ps-BayeSR are prone to produce images with fewer noise artifacts, while U-Baseline generates color artifacts. Among the transferred models, S-BayeSR maintains better local similarity due to the explicit modeling of image priors. Although SRFlow achieves the best NIQE and BRISQUE in Table \ref{tab:ReSR}, it generates more noisy artifacts, as one can observe from exemplar cases in Fig. \ref{fig:resr}.

\subsubsection{Real-world Image super-resolution}\label{sec:RwSR}
To study the unsupervised performance of BayeSR in real-world SISR $ \times 4 $, we trained BayeSR on the DPED-iPhone. In the training stage, we used a similar strategy as the realistic SISR to train U-Baseline, U-BayeSR, and Ps-BayeSR, except for replacing the LR dataset and degradation kernel with DPED-iPhone and $ k_{RWSR} $, respectively. In the test stage, we evaluated the performance of all models on the test dataset of DPED-iPhone by reporting LRPSNR, NIQE, and BRISQUE. Here, EDSR \cite{Lim/2017}, RCAN \cite{Zhangyulun/2018}, SRGAN \cite{Ledig/2017}, ESRGAN \cite{Wang/2018}, and SRFlow \cite{Lugmayr/2020} were transferred from ideal SISR, while RealSR \cite{Ji/2020} was trained on real-world SISR. Note that we mainly evaluated the visual quality of restorations, due to the lack of ground truth.

\begin{table*}[!t]
	\centering
	\caption{Evaluation on the task of \textit{real-world} SISR $ \times 4 $. We test all methods on the test dataset of DPED-iPhone, and report the average LRPSNE, NIQE, and BRISQUE.}
	\label{tab:RWSR}
	\resizebox{1\linewidth}{!}{
		\begin{tabular}{|c|c|c|c|c|c|c|c||c|c|c|c|}
			\hline
			Model&  Bicubic&   EDSR&  RCAN&  SRGAN&  ESRGAN&  SRFlow&  S-BayeSR&  RealSR&  U-Baseline&  Ps-BayeSR&  U-BayeSR\\
			\hline
			LRPSNR&  36.74&  37.78&  37.80&  27.89&  36.83&  37.52&  36.93&  33.08&  36.82&  38.59&  35.65\\
			NIQE& 7.99&  6.89&  6.91&  3.83&  4.02&  3.84&  7.26&  4.85&  7.69&  6.96&  6.73\\
			BRISQUE&  60.41&  55.74&  55.12&  15.68&  27.50&  25.66&  58.60&  16.42&  60.45&  47.67&  31.46\\
			\#Paras& --&  43.1M&  15.6M&  2.03M&  16.7M&  39.5M&  2.63M& 16.7M&  2.63M&  2.63M&  2.63M\\
			\hline
			--& --& \multicolumn{6}{c||}{Model transferred from ideal SISR}& \multicolumn{4}{c|}{Model trained without ground truth}\\  
			\hline
	\end{tabular}}
\end{table*}
\begin{figure*}[!t]
	\centering
	\includegraphics[width=1\linewidth]{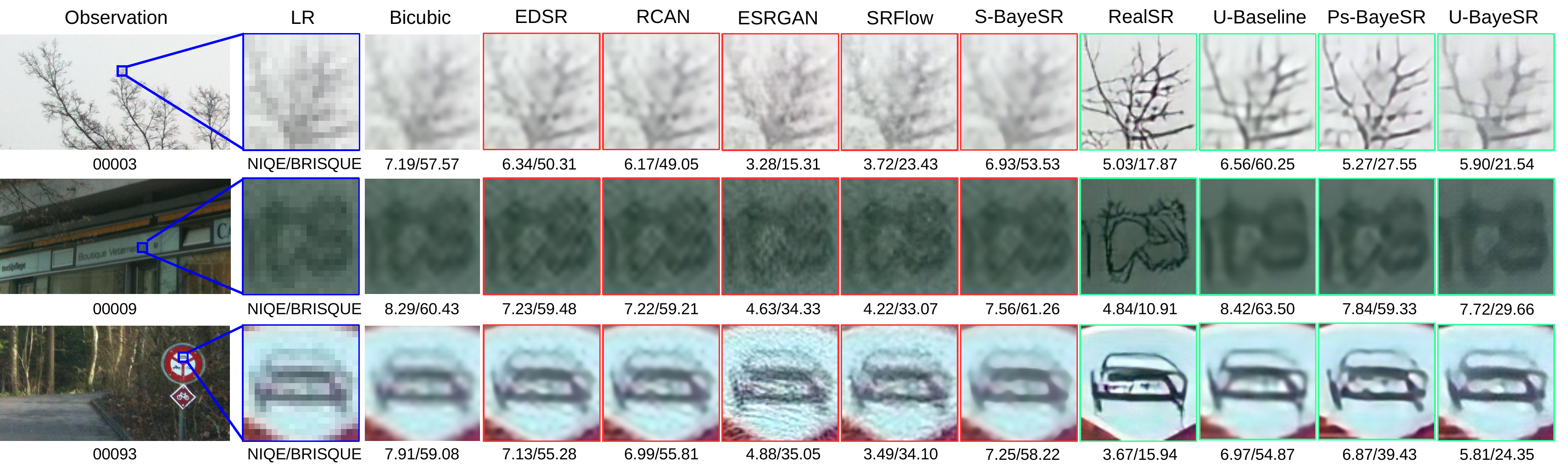}
	\caption{Visualization on the task of real-world SISR $ \times 4 $: three typical examples from DPED-iPhone. The red boundary denotes the model transferred from ideal SISR, while the green boundary represents the model trained without ground truth. Please refer to Supplementary Material for high-resolution images.}
	\label{fig:rwsr}
\end{figure*}

Table \ref{tab:RWSR} summarizes the results on the task of real-world SISR $ \times 4 $. Due to the lack of ground truth, we combine the quantitative results with visualized examples to evaluate each model. Fig. \ref{fig:rwsr} shows three typical examples from the test dataset of DPED-iPhone. Compared with the transferred models, U-BayeSR and Ps-BayeSR could generate clean images with more details. Besides, U-BayeSR outperforms U-Baseline in qualitative and quantitative results, which means training BayeSR by combining GL, DL, and GAL is more effective. Since RealSR was oriented by perceptual quality, it could generate more details than U-BayeSR oriented by PSNR, but some of them are fake. For example, the letters ``ts'' are inaccurately super-resolved by RealSR in the second row. Similarly, RealSR could generate unrealistic branches and wheels in the first and third rows, respectively. In contrast, U-BayeSR was trained by the maximum likelihood of observations, and thus could generate restorations more consistent with LR images. Overall, real-world SISR is still challenging due to diverse degradation and lack of reliable references.

\section{Discussion}\label{sec:discussion}
In this section, we discuss the generalizability of BayeSR to diverse noise levels and degradation kernels, and present our perspective regarding BayeSR in real-world applications. Similar to Setion \ref{sec:unsup}, we refer to supervised BayeSR as S-BayeSR, pseudo-supervised BayeSR as Ps-BayeSR, unsupervised BayeSR as U-BayeSR, and unsupervised Baseline as U-Baseline.

\subsection{Generalizability to diverse noise}
This study investigates the performance of BayeSR when there is difference between the pre-extracted noise and true noise in the training stage. Concretely, we simulated real-world camera sensor noise by a signal-dependent Gaussian distribution \cite{Healey/1994}, i.e., $ \mathcal N(0, \sigma_r^2 + \sigma_s y_i) $, where, $ \sigma_r $ and $ \sigma_s $ respectively denote the levels of read noise and shot noise \cite{Healey/1994}, and $ y_i $ denotes the $ i $-th pixel of an image $ \mathbf y $. 
After that, we degraded HR images of DIV2K by bicubic interpolation, and added the Gaussian noise to generate realistic LR images, with $ \sigma_r $ and $ \sigma_s $ uniformly ranging in $ [0, 25] $ and $ [0, 8] $, respectively. 
The resulting LR and HR image pairs were used to train S-BayeSR. 
Besides, we adopted the same strategy as shown in Section~\ref{sec:prestage} to extract pseudo noise from the real-world dataset DPED-iPhone to ensure the difference of distributions between the extracted noise and Gaussian noise. 
Moreover, we used the same strategy showed in Section \ref{sec:psuptrain} (and Section~\ref{sec:unsuptrain}) to generate pseudo LR images for training Ps-BayeSR (and U-Baseline and U-BayeSR) by degrading the HR images from Flickr2K (the realistic LR images generated from DIV2K) with bicubic interpolation and the pre-extracted pseudo noise. 
Due to the diversity of Gaussian noise, we increased each element of the hyperparameters, $ \boldsymbol \gamma_{\rho} $, which controls the shape of Gamma prior, for BayeSR from 2 to 8 to ensure flatter-shaped Gamma distributions. Finally, we trained U-Baseline, Ps-BayeSR, U-BayeSR, and S-BayeSR using the same settings as shown in Section \ref{sec:ReSR}.

In the test stage, we degraded the HR images from Set5, Set14, BSD100, and Urban100 by bicubic interpolation and the Gaussian noise with three different noise levels as shown in Table~\ref{tab:pracnoise}, to generate test LR images.
Note that the third  level, \textit{i.e.,} 30/9 for $ \sigma_r $/$ \sigma_s $, was out-of-scope  noise from the training stage.
We evaluated the performance of all models using the same strategy of computing PSNR and SSIM as the ideal SISR.

Table \ref{tab:pracnoise} presents the performance of U-Baseline, Ps-BayeSR, U-BayeSR, and S-BayeSR. One can see that the difference between the pre-extracted noise and true noise could greatly weaken the performance of U-Baseline. By contrast, the proposed generative learning could improve the generalizability of BayeSR to this difference, and therefore the performance of Ps-BayeSR and U-BayeSR did not degrade much. Owing to the generalizability of BayeSR to unseen noise, as shown in Table \ref{tab:supIdSR}, the performance of BayeSR dropped less than Baseline when the test noise level, \textit{i.e.,} 30/9 for $ \sigma_r $/$ \sigma_s $, was out of the scope of training noise. 
Overall, the supervised model achieved superior performance, and further improving unsupervised models of BayeSR yet remains to be explored in future work.

\begin{table}[!t]
	\centering
	\caption{Evaluation when images are corrupted by noise for SISR $\times 4$. We report the average PSNR ($ \uparrow $) and SSIM ($ \uparrow $) for different levels of read and shot noise, i.e., $ \sigma_r $ and $ \sigma_s $. Note that the third noise level, i.e., 30/9, is out-of-scope noise from the training stage.}
	\label{tab:pracnoise}
	\resizebox{1\linewidth}{!}{
		\begin{tabular}{|c|c|c|c|c|c|c|c|c|c|}
			\hline
			\multirow{2}{*}{$ \sigma_r $/$ \sigma_s $}&  \multirow{2}{*}{Method}& \multicolumn{2}{c|}{Set5}  &  \multicolumn{2}{c|}{Set14}  &  \multicolumn{2}{c|}{BSDS100}  &  \multicolumn{2}{c|}{Urban100}\\
			\cline{3-10}
			&  &  PSNR&  SSIM&  PSNR&  SSIM&  PSNR&  SSIM&  PSNR&  SSIM\\
			\hline
			\multirow{4}{*}{10/3}
			&  U-Baseline&  20.99&  0.4121&  20.06&  0.3423&  20.15&  0.3200&  19.27&  0.3561\\
			&  Ps-BayeSR&  25.53&  0.7129&  24.10&  0.6028&  24.15&  0.5666&  21.86&  0.5719\\
			&  U-BayeSR&  25.79&  0.7131&  24.00&  0.5991&  24.21&  0.5710&  21.83&  0.5675\\
			&  S-BayeSR&  27.93&  0.7978&  25.76&  0.6730&  25.41&  0.6296&  23.55&  0.6698\\
			\hline
			\multirow{4}{*}{20/6}
			&  U-Baseline&  18.08&  0.2819&  17.43&  0.2295&  17.51&  0.2071&  17.01&  0.2492\\
			&  Ps-BayeSR&  23.85&  0.6527&  22.77&  0.5520&  23.13&  0.5263&  20.74&  0.5140\\
			&  U-BayeSR&  24.27&  0.6531&  22.89&  0.5464&  23.22&  0.5228&  20.90&  0.5085\\
			&  S-BayeSR&  26.56&  0.7631&  24.84&  0.6399&  24.64&  0.5970&  22.83&  0.6369\\
			\hline
			\multirow{4}{*}{30/9}
			&  U-Baseline&  {16.39}&  {0.2141}&  {15.88}&  {0.1736}&  {15.93}&  {0.1534}&  {15.59}&  {0.1921}\\
			&  Ps-BayeSR&  {22.94}&  {0.6139}&  {22.02}&  {0.5164}&  {22.33}&  {0.4926}&  {20.20}&  {0.4777}\\
			&  U-BayeSR&  {23.22}&  {0.6042}&  {22.10}&  {0.5033}&  {22.42}&  {0.4821}&  {20.23}&  {0.4642}\\
			&  S-BayeSR&  {25.56}&  {0.7351}&  {24.12}&  {0.6160}&  {24.06}&  {0.5753}&  {22.25}&  {0.6106}\\
			\hline
	\end{tabular}}
\end{table}

\subsection{Generalizability to kernel estimation}

This section studies the performance of BayeSR when there is evident difference between the estimated degradation kernels and the true ones in the test stage. 
To this end, we trained a new BayeSR model, referred to as K-BayeSR, using the similar network architecture in Fig. \ref{fig:architecture} and training strategy of S-BayeSR in Section \ref{sec:generality}.

Concretely, we set the downsampling operator of K-BayeSR, i.e., $\mathbf A$ in (\ref{eq:model}), to be an explicit one dependent on the input, instead of a trainable module as S-BayeSR used.
This was implemented by replacing the downsampling module of the network in Fig.~\ref{fig:architecture} by the input degradation operation.
Therefore, in the training stage of K-BayeSR we adopted random Gaussian kernel and noise degradation, referred to as $\mathbf A_{Gaussian}$, and used $\mathbf A_{Gaussian}$ to degrade HR images from DIV2K to generate realistic LR images.
This Gaussian degradation $\mathbf A_{Gaussian}$ used two parameters ranging within $ [0.7, 4] $ for generating Gaussian blur kernels, and  valued $ \sigma_r $ and $ \sigma_s $ respectively ranging within $ [0, 12] $ and $ [0, 4] $ for Gaussian noise.
Note that the training images of K-BayeSR were different from that of S-BayeSR which adopted solely bicubic interpolation \textbf{K}$\mathbf k_{Bicubic}$ as the degradation kernel to generate training images.
For comparisons, we also trained the Baseline model adopting the same settings as K-BayeSR but without using the proposed generative learning loss.

In the test stage, to be consistent with USRNet \cite{Zhang_USRnet/2020}, we used twelve kernels, including four for isotropic Gaussian, four for anisotropic Gaussian, and four motion blur kernels, to generate test LR images from Set14.
Then, four groups of methods were evaluated for comparisons. 
The first group included RCAN \cite{Zhangyulun/2018} and S-BayeSR. They were directly transferred from the resulting models in Section \ref{sec:generality} and did not need an explicit input of blur kernels. 
The second group, \textit{i.e.,} Baseline, K-BayeSR, and USRNet \cite{Zhang_USRnet/2020}, were tested by feeding bicubic interpolation \textbf{K}$_{Bicubic}$  as the degradation input for super-resolving LR images. 
The third group consisted of four methods, \textit{i.e.,} DIP-FKP \cite{Liang_FKP/2021}, DIP-FKP+Baseline, DIP-FKP+K-BayeSR, and DIP-FKP+USRNet. 
DIP-FKP is a state-of-the-art blind SR method for jointly estimating kernels and super-resolving LR images, and the latter three took the estimated kernels from DIP-FKP as inputs for super-resolving LR images.
Finally, the fourth group, \textit{i.e.,} GT+Baseline, GT+K-BayeSR, and GT+USRNet, were tested by feeding the true degradation kernel of each LR image. 
As the evaluation criteria in VIRNet \cite{Yue_VIRNet/2020} are different from ours, we solely cited their test results in the paper for reference.

\begin{table*}[!t]
	\centering
	\caption{Evaluation when images are blurred by diverse kernels for SISR $\times 4$. Here, we report the average PSNR/SSIM on Set14. The bold values denote the best performance in each group. 
		Here, the right arrow ($\rightarrow$) indicates the input of degradation kernels; \textbf{K}$_{Bi}$ denotes the bicubic interpolation degradation and GT means the ground truth kernel.
		Note that the results of the last group, indicated as gray values, are directly cited from VIRNet \cite{Yue_VIRNet/2020} for reference, as the evaluation criteria are different.}
	\label{tab:diversekernel}
	\resizebox{1\linewidth}{!}{
		\begin{tabular}{|c| cccc| cccc| cccc|}
			\hline
			\multirow{4}{*}{Method}&  \multicolumn{12}{c|}{Test images are degraded by diverse blur kernels without noise corruption}\\
			\cline{2-13}
			&
			\begin{minipage}[b]{0.1\columnwidth}
				\centering
				\vspace{4pt}
				\includegraphics[width=0.8\linewidth]{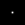}
				\vspace{1pt}
			\end{minipage}  
			&
			\begin{minipage}[b]{0.1\columnwidth}
				\centering
				\vspace{4pt}
				\includegraphics[width=0.8\linewidth]{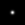}
				\vspace{1pt}
			\end{minipage}  
			&
			\begin{minipage}[b]{0.1\columnwidth}
				\centering
				\vspace{4pt}
				\includegraphics[width=0.8\linewidth]{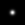}
				\vspace{1pt}
			\end{minipage}  
			&  
			\begin{minipage}[b]{0.1\columnwidth}
				\centering
				\vspace{4pt}
				\includegraphics[width=0.8\linewidth]{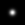}
				\vspace{1pt}
			\end{minipage}
			&  
			\begin{minipage}[b]{0.1\columnwidth}
				\centering
				\vspace{4pt}
				\includegraphics[width=0.8\linewidth]{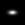}
				\vspace{1pt}
			\end{minipage}
			&  
			\begin{minipage}[b]{0.1\columnwidth}
				\centering
				\vspace{4pt}
				\includegraphics[width=0.8\linewidth]{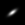}
				\vspace{1pt}
			\end{minipage}
			&  
			\begin{minipage}[b]{0.1\columnwidth}
				\centering
				\vspace{4pt}
				\includegraphics[width=0.8\linewidth]{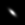}
				\vspace{1pt}
			\end{minipage}
			&  
			\begin{minipage}[b]{0.1\columnwidth}
				\centering
				\vspace{4pt}
				\includegraphics[width=0.8\linewidth]{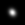}
				\vspace{1pt}
			\end{minipage}
			&  
			\begin{minipage}[b]{0.1\columnwidth}
				\centering
				\vspace{4pt}
				\includegraphics[width=0.8\linewidth]{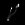}
				\vspace{1pt}
			\end{minipage}
			& 
			\begin{minipage}[b]{0.1\columnwidth}
				\centering
				\vspace{4pt}
				\includegraphics[width=0.8\linewidth]{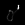}
				\vspace{1pt}
			\end{minipage}
			&  
			\begin{minipage}[b]{0.1\columnwidth}
				\centering
				\vspace{4pt}
				\includegraphics[width=0.8\linewidth]{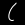}
				\vspace{1pt}
			\end{minipage}
			&  
			\begin{minipage}[b]{0.1\columnwidth}
				\centering
				\vspace{4pt}
				\includegraphics[width=0.8\linewidth]{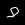}
				\vspace{1pt}
			\end{minipage}
			\\
			\hline
			RCAN&  20.04/0.5391&  21.65/0.5958&  22.85/\textbf{0.6242}&  \textbf{23.34/0.6288}&  23.07/\textbf{0.6189}&  22.40/0.5787&  23.48/\textbf{0.6463}&  \textbf{23.36/0.6202}&  20.70/0.4920&  20.29/0.4994&  21.77/0.5439&  21.37/0.5251\\
			S-BayeSR&  \textbf{21.41/0.5984}&  \textbf{22.46/0.6143}&  \textbf{23.08}/0.6240&  23.32/0.6230&  \textbf{23.16}/0.6165&  \textbf{22.69/0.5871}&  \textbf{23.51}/0.6344&  23.29/0.6134&  \textbf{21.00/0.5067}&  \textbf{20.54/0.5058}&  \textbf{21.99/0.5498}&  \textbf{21.60/0.5317}\\
			\hline
			\textbf{K}$_{Bi}$ $\rightarrow$ Baseline&  22.47/0.6199&  23.15/0.6299&  23.49/0.6326&  23.64/0.6299&  23.51/0.6241&  23.14/0.6051&  23.85/0.6401&  23.65/0.6225&  21.12/0.5134&  20.57/0.5020&  \textbf{22.08/0.5497}&  21.88/0.5451\\
			\textbf{K}$_{Bi}$ $\rightarrow$ K-BayeSR&  \textbf{26.36/0.7323}&  \textbf{26.91/0.7296}&  \textbf{26.78/0.7172}&  \textbf{26.41/0.6997}&  \textbf{26.12/0.6918}&  \textbf{26.04/0.6931}&  \textbf{26.12/0.6917}&  \textbf{25.88/0.6786}&  \textbf{21.61/0.5304}&  \textbf{21.40/0.5224}&  21.73/0.5279&  \textbf{23.67/0.5994}\\
			\textbf{K}$_{Bi}$ $\rightarrow$ USRNet&  20.56/0.5666&  21.99/0.6065&  22.98/0.6280&  23.36/0.6287&  23.14/0.6208&  22.56/0.5849&  23.61/0.6474&  23.37/0.6192&  20.79/0.4957&  20.36/0.5015&  21.86/0.5474&  21.50/0.5279\\
			\hline
			DIP-FKP&  21.39/0.5652&  22.25/0.5889&  22.63/0.5987&  22.83/0.6020&  22.73/0.5964&  22.54/0.5845&  23.15/0.6132&  23.02/0.6034&  19.97/0.4340&  20.16/0.4417&  19.91/0.4284&  22.79/0.5682\\
			DIP-FKP$\rightarrow$ Baseline&  22.17/0.6093&  22.78/0.6200&  23.13/0.6233&  23.33/0.6245&  23.31/0.6220&  22.87/0.6008&  23.58/0.6365&  23.50/0.6226&  20.11/0.4611&  20.34/0.4685&  20.10/0.4499&  23.37/0.5950\\
			DIP-FKP$\rightarrow$ K-BayeSR&  \textbf{26.35/0.7320}&  \textbf{26.90/0.7294}&  \textbf{26.77/0.7170}&  \textbf{26.39/0.6997}&  \textbf{26.11/0.6917}&  \textbf{26.04/0.6934}&  \textbf{26.11/0.6917}&  \textbf{25.88/0.6788}&  \textbf{21.59/0.5297}&  \textbf{21.40/0.5221}&  \textbf{21.68/0.5255}&  \textbf{23.73/0.6013}\\
			DIP-FKP$\rightarrow$ USRNet&  21.90/0.6070&  22.26/0.6178&  22.73/0.6212&  22.85/0.6207&  22.84/0.6188&  22.38/0.5975&  23.17/0.6324&  23.12/0.6205&  19.92/0.4687&  20.33/0.4807&  19.98/0.4615&  22.88/0.5886\\
			\hline
			GT$\rightarrow$ Baseline&  26.63/0.7408&  27.25/0.7484&  27.29/0.7431&  27.07/0.7308&  26.81/0.7228&  26.75/0.7251&  26.84/0.7237&  26.64/0.7111&  20.52/0.5384&  20.01/0.5108&  20.71/0.5153&  22.36/0.5872\\
			GT$\rightarrow$ K-BayeSR&  26.40/0.7299&  26.90/0.7276&  26.77/0.7160&  26.40/0.6993&  26.12/0.6914&  26.06/0.6931&  26.11/0.6914&  25.89/0.6791&  21.58/0.5298&  21.37/0.5210&  21.68/0.5254&  23.71/0.6010\\
			GT$\rightarrow$ USRNet&  \textbf{27.47/0.7678}&  \textbf{28.35/0.7807}&  \textbf{28.61/0.7832}&  \textbf{28.70/0.7833}&  \textbf{28.61/0.7824}&  \textbf{28.53/0.7796}&  \textbf{28.52/0.7771}&  \textbf{28.68/0.7809}&  \textbf{28.32/0.7711}&  \textbf{27.72/0.7656}&  \textbf{28.06/0.7684}&  \textbf{28.02/0.7649}\\
			\hline
			\color{gray}{RCAN}&  \color{gray}{20.08/0.5403}&  \color{gray}{21.73/0.5982}&  \color{gray}{22.97/0.6274}&  \color{gray}{23.47/0.6324}&  \color{gray}{23.20/0.6226}&  \color{gray}{22.52/0.5820}&  \color{gray}{23.61/0.6499}&  \color{gray}{23.51/0.6241}&  --&  --&  --&  --\\
			\color{gray}{GT$\rightarrow$ VIRNet}&  \color{gray}{27.18/0.7546}&  \color{gray}{27.84/0.7650}&  \color{gray}{28.01/0.7668}&  \color{gray}{28.03/0.7652}&  \color{gray}{27.87/0.7610}&  \color{gray}{27.69/0.7571}&  \color{gray}{27.71/0.7548}&  \color{gray}{27.91/0.7594}&  --&  --&  --&  --\\
			\hline
	\end{tabular}}
\end{table*}

Table \ref{tab:diversekernel} presents the results for SISR $ \times $4. 
One can see that the BayeSR-based methods demonstrated better generalizability than others when the input kernels were different from the ground truth (GT). 
Note that when the GT kernels were given, USRNet set superior performance in all categories of the fourth group; by contrast when the input changed to the estimated ones or bicubic interpolation, its performance dropped down dramatically, to much poorer results compared to K-BayeSR. 
This confirmed neither bicubic nor DIP-FKP could represent or estimate the kernels of test LR images accurately enough for USRNet. 
By contrast, K-BayeSR performed consistently in the second, third and fourth groups with three sources of kernel inputs.

K-BayeSR demonstrated good robustness to the estimated kernels,
while Baseline and USRNet could be more sensitive. 
The robustness could be attributed to the advantageous statistical modeling. 
Concretely, given the observation $ \mathbf y $ and the blur kernel $ \mathbf k $, the degraded term, $ \mathbf y - \mathbf m - \mathbf A \mathbf z $ is deterministic for Baseline, since the distributions of $ \mathbf z $ and $ \mathbf m $ are degraded into one-point distributions without the constraints in Table \ref{tab:varloss}. By contrast, $ \mathbf y - \mathbf m - \mathbf A \mathbf z $ is stochastic for K-BayeSR, since $ \mathbf z $ and $ \mathbf m $ follow their own priors. 
In other word, Baseline is aimed to learn a point-to-point mapping from $ \mathbf y - \mathbf m - \mathbf A \mathbf z $ to $ \mathbf x $, where $  \mathbf x $ is also deterministic, but K-BayeSR is conducted to learn a distribution-to-distribution mapping from $ \breve{q}(\mathbf y - \mathbf m - \mathbf A \mathbf z) $ to $ \breve{q}(\mathbf x) $. Since the prior corresponding to $ \breve{q}(\mathbf x) $ is kernel-independent, as shown in (\ref{eq:xpdf}), K-BayeSR is less sensitive to given kernels.
Nevertheless, it is worth mentioning that K-BayeSR delivered much poor results when the LR images were degraded by the motion kernels, a group of different degradation to the Gaussian kernels. Therefore, how to improve the generalizability when the distributions of degradation kernels are different remains to be further explored.
Furthermore, K-BayeSR did not match the best results when the GT kernels were given, due to the limitation of explicit modeling and generative learning. Nevertheless, in real-world image super resolution the GT kernels could not be available, and improving the modeling capacity and accuracy should be considered in future work.

\begin{figure}[!t]
	\centering
	\includegraphics[width=0.9\linewidth]{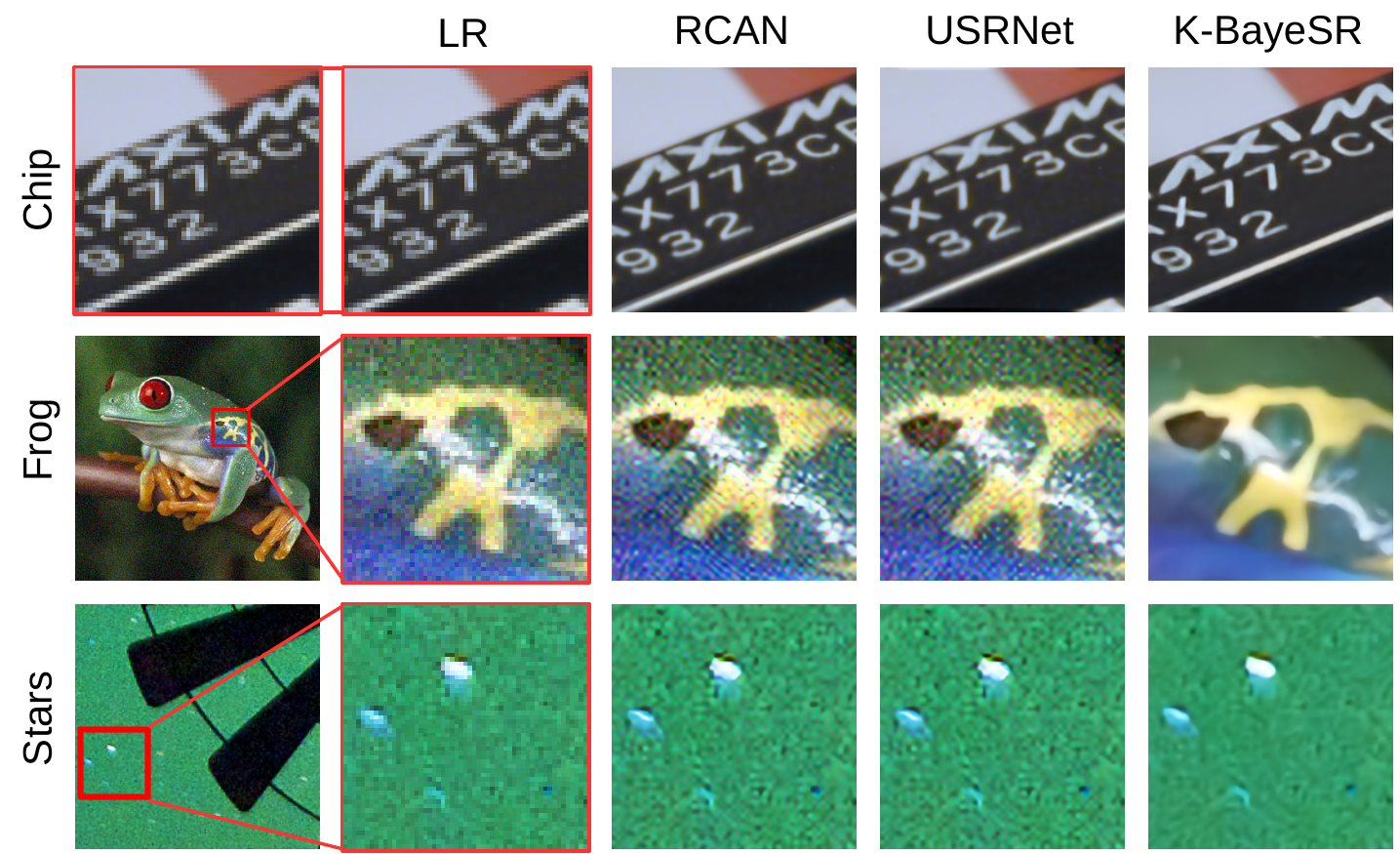}
	\caption{Visualization of three typical real-world examples for SISR $\times 4$. Here, the bicubic interpolation kernel is fed into USRNet and K-BayeSR. Please refer to Supplement Material for high-resolution images.}
	\label{fig:realworld}
\end{figure}

\subsection{Super-resolution on real-world images}
Here, we studied the real-world image SR, where the degradation procedures of images are unknown. 
We used three examples, \textit{i.e.,} chip, frog, and stars, and compared the results from RCAN \cite{Zhangyulun/2018}, K-BayeSR, and USRNet \cite{Zhang_USRnet/2020}. 
Note that K-BayeSR and USRNet require a blur kernel as the input, which is simply set as the bicubic interpolation. 
Figure \ref{fig:realworld} visualizes the results. 
Due to the difference between the bicubic interpolation and real-world degradation, RCAN and USRNet generated over-smooth results for chip.
By contrast, K-BayeSR overcame the difference of degradation and performed well. 
Besides, the real-world noise  in frog and stars was mapped into artifacts by RCAN and USRNet due to the difference between simulated noise and real-world noise. By contrast, K-BayeSR could super resolve images with less artifacts, demonstrating better generalizability in real-world scenarios.

\section{Conclusion}
In this work, we proposed a Bayesian image restoration framework, and implemented it for SISR by neural networks. Concretely, we first modeled image statistics using the smoothness and sparsity priors, and presented the variational inference framework of estimating the smoothness component and sparsity residual from an observation. Then, we built neural networks to implement the framework for SISR, and proposed the unsupervised strategies to train the networks. Finally, we showed the superior generalization ability of our method, and demonstrated its effectiveness in unsupervised SISR.

In our future work, we can jointly infer blur kernels and restorations by simultaneously modeling kernel and image priors. Besides, modeling image priors and quantifying uncertainties of IR models is opening. We adopted the smoothness and sparsity priors to model image features, but this method cannot represent particularly complex image structures. How to accurately model image priors is still opening and worth further exploring. Moreover, quantifying uncertainties of deep learning models has arisen as one of the new requirements in many applications \cite{Gal_Uncertianty/2016}. As low-level computer vision, IR could be further considered as an estimation of stochastic mappings, to explore any possible solutions of this ill-posed inverse problem. After that, one can evaluate uncertainties of deep IR models, which is helpful for AI safety \cite{Gal_Uncertianty/2016} in computer vision systems.

\section*{Acknowledgments}
This work was funded by the National Natural Science Foundation of China (grant no. 61971142, 62111530195 and 62011540404) and the development fund for Shanghai talents (no. 2020015). The authors are grateful to all anonymous reviewers for their insightful comments, which improve many aspects of this work, especially the discussion of BayeSR in Section \ref{sec:discussion}. We also would like to thank Fuping Wu, Hangqi Zhou, and Ke Zhang for useful comments and proofread of the manuscript.

\section*{Appendix A}\label{appendA}
Suppose $ z $ is a variable follows normal distribution, namely, 
\begin{equation}\label{Gauss}
	p(z|\mu, \omega)= \mathcal N(z|\mu, \omega^{-1}) = \frac{1}{\sqrt{2\pi/\omega}}\exp^{-\frac{\omega}{2}(z-\mu)^2},
\end{equation}
and $ \omega $ is a variable follow Gamma distribution, namely
\begin{equation}\label{Gamma}
	p(\omega|\phi,\gamma)= \mathcal G(\omega|\phi, \gamma) = \frac{\phi^\gamma}{\Gamma(\gamma)} \omega^{\gamma - 1} e^{-\phi \omega},
\end{equation}
where, $ \Gamma(\cdot) $ denotes the Gamma function. Then, $ p(z, \omega) = p(z|\mu, \omega)p(\omega|\phi,\gamma) $ is known as Normal-Gamma distribution. Moreover, the marginal distribution of $ z $ is Student's t distribution, namely,
\begin{equation}\label{Student}
	p(z|\mu, \phi, \gamma) = \int_{\mathbb R} p(z, \omega)d\omega = \mathcal S(z|\mu, \gamma^{-1}\phi, 2\gamma),
\end{equation}
where,
\begin{equation}
	\mathcal S(z|\mu, \lambda, \alpha) = \frac{\Gamma((\alpha + 1)/2)}{\Gamma(\alpha/2)\Gamma(1/2)}\left( \frac{\lambda}{\alpha}\right)^{1/2} \left[ 1 + \frac{\lambda}{\alpha}(z - \mu)^2 \right]^{-(\alpha + 1)/2}.  
\end{equation}

\section*{Appendix B}\label{appendB}
Let $ \boldsymbol \psi = \left\lbrace \mathbf m, \boldsymbol \rho, \mathbf x, \boldsymbol \upsilon, \mathbf z, \boldsymbol \omega \right\rbrace  $, then the joint distribution $ p(\boldsymbol \psi, \mathbf y) $ could be expressed as,
\begin{equation}\label{jointpdf}
	p(\boldsymbol\psi,\mathbf y)= p(\mathbf y|\boldsymbol \psi)p(\boldsymbol \psi)=p(\mathbf y|\mathbf A, \mathbf x,\mathbf z,\mathbf m,\boldsymbol \rho)p(\mathbf m)p(\boldsymbol \rho)p(\mathbf x|\boldsymbol \upsilon)p(\boldsymbol \upsilon)p(\mathbf z| \boldsymbol \omega)p(\boldsymbol \omega), 
\end{equation}
where, 
\begin{equation}\label{likelihood}
	p(\mathbf y|\mathbf A, \mathbf x,\mathbf z,\mathbf m,\boldsymbol \rho) =\mathcal N(\mathbf y|\mathbf A(\mathbf x + \mathbf z) + \mathbf m,\mbox{diag}(\boldsymbol \rho
	)^{-1}),
\end{equation}
\begin{equation}\label{noisypdf}
	p(\mathbf m)p(\boldsymbol \rho) = \mathcal N(\mathbf m|\boldsymbol \mu_0,\sigma_0^{-1} I)\cdot\prod_{i=1}^{d_y}\mathcal G(\rho_i|\phi_{\rho i},\gamma_{\rho i}).
\end{equation}
\begin{equation}\label{smoothpdf}
	p(\mathbf x|\boldsymbol \upsilon)p(\boldsymbol \upsilon) = \mathcal N(\mathbf x|\mathbf 0, [ \mathbf D_h^\top \mbox{diag}(\boldsymbol \upsilon)\mathbf D_h + \mathbf D_v^\top \mbox{diag}(\boldsymbol \upsilon)\mathbf D_v] ^{-1})\cdot\prod_{i=1}^{d_u}\mathcal G(\upsilon_i|\phi_{\upsilon i},\gamma_{\upsilon i}),
\end{equation}
\begin{equation}\label{sparsepdf}
	p(\mathbf z| \boldsymbol \omega)p(\boldsymbol \omega) = \mathcal N(\mathbf z|\mathbf 0, \mbox{diag}(\boldsymbol \omega)^{-1})\cdot\prod_{i=1}^{d_u}\mathcal G(\omega_i|\phi_{\omega i},\gamma_{\omega i}),
\end{equation}
It is intractable to directly compute the posterior distribution of one variable by marginalizing $ p(\boldsymbol \psi| \mathbf y) $ over other variables, since some variables are conditionally dependent. Motivated by the mean-field theory, we adopt the following variational distribution to approximate $ p(\boldsymbol \psi| \mathbf y) $,
\begin{equation}\label{variationaldistribution}
	q(\boldsymbol \psi) = q(\mathbf m)q(\boldsymbol \rho) \prod_{i=1}^{d_u}q(x_i)q(\boldsymbol \upsilon) \prod_{i=1}^{d_u}q(z_i)q(\boldsymbol \omega).
\end{equation} 
The variational posterior distribution of minimizing KL divergence could be obtained via the VB theorem. Concretely, the minimum $ \mbox{KL}(\breve{q}(\boldsymbol \psi)\parallel p(\boldsymbol \psi | y)) $ is reached for
\begin{equation}\label{eq:optimalpdf:a}
	\breve{q}(\boldsymbol \psi_i) \propto \exp\left(\mathbb E_{\breve{q}(\boldsymbol \psi \setminus \boldsymbol \psi_i)}\left[ \log p(\boldsymbol \psi, y) \right] \right) 
\end{equation}
where, $ \boldsymbol \psi \setminus \boldsymbol \psi_i $ denotes the complement of $ \boldsymbol \psi_i $ in $ \boldsymbol \psi $. According to the VB theorem, the optimal variational posterior distributions could be expressed as follows,
\begin{align}
	\breve{q}(\mathbf m) &= \mathcal N(\mathbf m|\breve{\boldsymbol\mu}_m, \mbox{diag}(\breve{\boldsymbol\sigma}_m^2))\label{eq:mvarpdf}\\
	\breve{q}(\boldsymbol \rho) &= \prod_{i=1}^{d_y}\mathcal G(\rho_i | \breve{\beta}_{\rho i}, \breve{\alpha}_{\rho i})\label{eq:rhovarpdf}\\
	\breve{q}(\mathbf x) &= \mathcal N(\mathbf x|\breve{\boldsymbol \mu}_x, \mbox{diag}(\breve{\boldsymbol\sigma}_x^2))\label{eq:xvarpdf}\\
	\breve{q}(\boldsymbol \upsilon) &= \prod_{i=1}^{d_u}\mathcal G(\upsilon_i | \breve{\beta}_{\upsilon i}, \breve{\alpha}_{\upsilon i})\label{eq:upsilonvarpdf}\\
	\breve{q}(\mathbf z) &= \mathcal N(\mathbf z|\breve{\boldsymbol \mu}_z, \mbox{diag}(\breve{\boldsymbol\sigma}_z^2))\label{eq:zvarpdf}\\
	\breve{q}(\boldsymbol \omega) &= \prod_{i=1}^{d_u}\mathcal G(\omega_i | \breve{\beta}_{\omega i}, \breve{\alpha}_{\omega i})\label{eq:omegavarpdf},
\end{align} 
and the optimal variational parameters $ \breve{\boldsymbol \mu}_{\cdot} $, $ \breve{\boldsymbol \sigma}_{\cdot}$, $ \breve{\boldsymbol \alpha}_{\cdot} $, and $ \breve{\boldsymbol \beta}_{\cdot} $ satisfy the following non-linear equations, 
\begin{equation}\label{equations}
	\begin{cases}
		\breve{\boldsymbol\mu}_m = \breve{\boldsymbol\sigma}_m^2 \odot \left[ \breve{\boldsymbol \mu}_{\rho}\odot (\mathbf y - \mathbf A(\breve{\boldsymbol \mu}_x-\breve{\boldsymbol \mu}_z)) + \sigma_0\boldsymbol \mu_0 \right]\\ 
		\breve{\sigma}_{mi}^2 = (\breve{\mu}_{\rho i} + \sigma_0)^{-1}\\
		\breve{\boldsymbol \mu}_x = \breve{\boldsymbol \sigma}_x^2 \odot \left[ \mathbf A^\top \mbox{diag}(\breve{\boldsymbol \mu}_{\rho})(\mathbf y - \mathbf A\breve{\boldsymbol\mu}_z - \breve{\boldsymbol \mu}_m)\right]\\ 
		\breve{\sigma}_{x i}^2 = \left[ \left\langle \breve{\boldsymbol \mu}_{\rho}, \mathbf a_i^2 \right\rangle  + \left\langle \breve{\boldsymbol \mu}_{\upsilon}, \mathbf d_{hi}^2 + \mathbf d_{vi}^2\right\rangle  \right]^{-1}\\
		\breve{\boldsymbol \mu}_z=\breve{\boldsymbol\sigma}_z^2\odot\left[ \mathbf A^\top \mbox{diag}(\breve{\boldsymbol \mu}_{\rho})(\mathbf y - \mathbf A\breve{\boldsymbol\mu}_x-\breve{\boldsymbol\mu}_m)\right]\\ 
		\breve{\sigma}_{zi}^2 = (\left\langle \breve{\boldsymbol\mu}_{\rho}, \mathbf a_i^2\right\rangle  + \breve{\mu}_{\omega i})^{-1}\\
		\breve{\boldsymbol\alpha}_{\upsilon} =\boldsymbol \gamma_{\upsilon} + \frac{1}{2}\\
		\breve{\beta}_{\upsilon i}= \frac{1}{2}\left[  (\mathbf D_h\breve{\boldsymbol \mu}_x)_i^2 + (\mathbf D_v\breve{\boldsymbol \mu}_x)_i^2 + \left\langle \breve{\boldsymbol \sigma}_x^2, \mathbf d_{hi}^2 + \mathbf d_{vi}^2\right\rangle  \right] + \phi_{\upsilon i}\\
		\breve{\boldsymbol \mu}_{\upsilon} = \breve{\boldsymbol \alpha}_{\upsilon} / \breve{\boldsymbol \beta}_{\upsilon}\\
		\breve{\boldsymbol\alpha}_{\omega} = \boldsymbol\gamma_{\omega}+\frac{1}{2}\\
		\breve{\beta}_{\omega i} =  \frac{1}{2}(\breve{\mu}_{z i}^2+\breve{\sigma}_{z i}^2) + \phi_{\omega i}\\
		\breve{\boldsymbol \mu}_{\omega} = \breve{\boldsymbol \alpha}_{\omega} / \breve{\boldsymbol \beta}_{\omega}\\
		\breve{\boldsymbol\alpha}_{\rho} = \boldsymbol\gamma_{\rho} + \frac{1}{2}\\
		\breve{\beta}_{\rho i} = \frac{1}{2}\left[ (y_i - \mathbf a_i^\top (\breve{\boldsymbol \mu}_x-\breve{\boldsymbol \mu}_z)-\breve{\mu}_{mi})^2 + \left\langle \mathbf a_i^2, \breve{\boldsymbol \sigma}_x^2 + \breve{\boldsymbol \sigma}_z^2\right\rangle  + \breve{\sigma}_{mi}^2  \right] + \phi_{\rho i}\\
		\breve{\boldsymbol \mu}_{\rho} = \breve{\boldsymbol \alpha}_{\rho} / \breve{\boldsymbol \beta}_{\rho}\\
	\end{cases}
\end{equation}
where, $ \odot $ denotes the element-wise multiplication, and $ \left\langle\cdot, \cdot \right\rangle  $ denotes the tensor product. Besides, \[ \mathbf D_h^\top = \left[ \mathbf d_{h1}, \mathbf d_{h2}, \dots, \mathbf d_{hd_u} \right], \mathbf D_v^\top = \left[ \mathbf d_{v1}, \mathbf d_{v2}, \dots, \mathbf d_{vd_u}\right] , \mathbf A^\top = \left[ \mathbf a_1, \mathbf a_2, \dots, \mathbf a_{d_y} \right]. \] Since explicitly solving the equations in (\ref{equations}) is difficult, one can compute the variational parameters iteratively. However, iterative methods present heavy computational burden due to the high-dimension of variational parameters.

\section*{Appendix C}\label{append02}
In practice, we do not directly compute the KL divergence from $ \breve{q}(\boldsymbol \psi) $ to $ p(\boldsymbol \psi|\mathbf y) $ due to heavy computational burden as aforementioned, but convert it to an easily derived formula, 
\begin{align}
	\mbox{\footnotesize KL}(\breve{q}(\boldsymbol \psi)||p(\boldsymbol \psi|\mathbf y)) &= \mathbb E\left[ \log \breve{q}(\boldsymbol \psi)\right]  - \mathbb E\left[ \log p(\boldsymbol \psi|\mathbf y)\right] \label{eq:KLtoELBO} \\
	&= \mathbb E\left[ \log \breve{q}(\boldsymbol \psi)\right]  - \mathbb E\left[ \log p(\boldsymbol \psi,\mathbf y)\right]  + \log p(\mathbf y),\nonumber
\end{align}
where, all expectations are taken with respect to $ \breve{q}(\boldsymbol \psi) $, and the evidence $ p(\mathbf y) $ only depends on the priors. This formula shows that minimizing KL divergence is equivalent to
\begin{equation}
	\min_{\breve{q}(\boldsymbol \psi)} \mathbb E\left[ \log \breve{q}(\boldsymbol \psi)\right] - \mathbb E\left[ \log p(\boldsymbol \psi,\mathbf y)\right] = \mbox{KL}(\breve{q}(\boldsymbol \psi)||p(\boldsymbol \psi)) - \mathbb E\left[ \log p(\mathbf y|\boldsymbol \psi)\right]
\end{equation} 
The second term on the right hand side can be expressed as
\begin{equation}\label{eq:paralikelihood}
	\begin{aligned}
		- \mathbb E\left[ \log p(\mathbf y|\boldsymbol \psi)\right]
		= &\frac{d_y}{2}\log (2\pi) - \frac{1}{2}\textstyle\sum_{i=1}^{d_y}\left( \Psi(\breve{\alpha}_{\rho i}) + \log(\breve{\mu}_{\rho i}) - \log(\breve{\alpha}_{\rho i}) \right)\\
		+ &\frac{1}{2} [ (\mathbf y-\mathbf A(\breve{\boldsymbol \mu}_x +\breve{\boldsymbol \mu}_z) -\breve{\boldsymbol \mu}_m)^\top \mbox{diag}(\breve{\boldsymbol \mu}_{\rho}) (\mathbf y-\mathbf A(\breve{\boldsymbol \mu}_x +\breve{\boldsymbol \mu}_z) -\breve{\boldsymbol \mu}_m) ]\\
		+ &\frac{1}{2} [ \langle \mathbf A^\top \mbox{diag}(\breve{\boldsymbol \mu}_{\rho})\mathbf A, \mbox{diag}(\breve{\boldsymbol \sigma}_x^2 )  + \mbox{diag}(\breve{\boldsymbol \sigma}_z^2) \rangle  + \langle \mbox{diag}(\breve{\boldsymbol \mu}_{\rho}), \mbox{diag}(\breve{\boldsymbol \sigma}_m^2 \rangle  ],
	\end{aligned}
\end{equation}
where, $ \Psi(\cdot) $ denotes the Digamma function. In reality, we learn the downsampling operator $ \mathbf A $ via GANs, and thus $ \mathbf A^\top $ is unavailable. That mean we cannot directly compute the expectation. To tackle the difficulty, we use the reparameterization trick to avoid the computation of $ \mathbf A^\top $. Concretely, let $ \boldsymbol \epsilon $ denote white Gaussian noise sampled from $ \mathcal N(\mathbf 0, \mathbf I) $, then we have $ \mathbf x = \breve{\boldsymbol \sigma}_x \odot \boldsymbol \epsilon + \breve{\boldsymbol \mu}_x $, $ \mathbf z = \breve{\boldsymbol \sigma}_z \odot \boldsymbol \epsilon + \breve{\boldsymbol \mu}_z $, and $ \mathbf m = \breve{\boldsymbol \sigma}_m \odot \boldsymbol \epsilon + \breve{\boldsymbol \mu}_m $. Therefore, the formula (\ref{eq:paralikelihood}) could be converted to
\begin{equation}
	\begin{aligned}
		-\mathbb E\left[ \log p(\mathbf y|\boldsymbol \psi)\right] &= -\mathbb E_{\breve{q}(\boldsymbol \rho)}\left[ \mathbb E_{\breve{q}(\boldsymbol \psi \setminus \boldsymbol\rho)} \left[ \log p(\mathbf y|\boldsymbol \psi)\right] \right] \\
		&\approx -\mathbb E_{\breve{q}(\boldsymbol \rho)}\left[ \frac{1}{N}\textstyle\sum_{i=1}^N\log p(\mathbf y|\mathbf A, \mathbf x_i,\mathbf z_i,\mathbf m_i,\boldsymbol \rho) \right],\\
	\end{aligned} 
\end{equation}
where, $ \left\lbrace \mathbf x_i,\mathbf z_i,\mathbf m_i \right\rbrace_{i=1}^N  $ are reparameterized samples as aforementioned. In practice, $ N $ is often set to 1, namely
\begin{equation}\label{eq:relaxedlikelihood}
	-\mathbb E\left[ \log p(\mathbf y|\boldsymbol \psi)\right]=-\mathbb E_{\breve{q}(\boldsymbol \rho)}\left[ \mathbb E_{\breve{q}(\boldsymbol \psi \setminus \rho)} \left[ \log p(\mathbf y|\boldsymbol \psi)\right] \right] \approx -\mathbb E_{\breve{q}(\boldsymbol \rho)}\left[ \log p(\mathbf y|\mathbf A, \mathbf x,\mathbf z,\mathbf m,\boldsymbol \rho) \right] . 
\end{equation}
Overall, we will optimize the following problem to infer the variational distribution  $ \breve{q}(\boldsymbol \psi) $,
\begin{equation}\label{eq:ELBO}
	\min_{\breve{q}(\boldsymbol \psi)} \mbox{KL}(\breve{q}(\boldsymbol \psi)||p(\boldsymbol \psi)) - \mathbb E_{\breve{q}(\boldsymbol \rho)}\left[ \log p(\mathbf y|\mathbf A, \mathbf x,\mathbf z,\mathbf m,\boldsymbol \rho) \right]
\end{equation} 

\noindent\textbf{Step 1: Infer $ \breve{\boldsymbol \mu}_{\upsilon} $, $ \breve{\boldsymbol \mu}_{\omega} $, and $ \breve{\boldsymbol \mu}_{\rho} $}

The first term of (\ref{eq:ELBO}) could be expressed as
\begin{equation}\label{eq:KLs}
	\begin{aligned}
		\mbox{KL}(\breve{q}(\boldsymbol \psi)||p(\boldsymbol \psi)) &= \mbox{KL}(\breve{q}(\mathbf x)\breve{q}(\boldsymbol \upsilon)||p(\mathbf x|\boldsymbol \upsilon)p(\boldsymbol \upsilon)) + \mbox{KL}(\breve{q}(\mathbf z)\breve{q}(\boldsymbol \omega)||p(\mathbf z|\boldsymbol \omega)p(\boldsymbol \omega))\\
		&+ \mbox{KL}(\breve{q}(\mathbf m)||p(\mathbf m)) + \mbox{KL}(\breve{q}(\boldsymbol \rho)||p(\boldsymbol \rho)). 
	\end{aligned} 
\end{equation}
One can see that the variable $ \boldsymbol \upsilon $ is only related to the first term on the right hand side of (\ref{eq:KLs}). Therefore, its parameters can be computed by minimizing $ \mbox{KL}(\breve{q}(\mathbf x)\breve{q}(\boldsymbol \upsilon)||p(\mathbf x|\boldsymbol \upsilon)p(\boldsymbol \upsilon)) $. 
According to VB theorem as shown in (\ref{eq:optimalpdf:a}),
\begin{equation}
	\breve{q}(\boldsymbol \upsilon) \propto \exp\left(\left\langle \log \left[ p(\mathbf x|\boldsymbol \upsilon)p(\boldsymbol \upsilon)\right] \right\rangle_{\breve{q}(\mathbf x)} \right) \propto \exp\left(\left\langle \log p(\boldsymbol \psi, y) \right\rangle_{\breve{q}(\boldsymbol \psi\setminus \upsilon)}  \right).
\end{equation}
That means the computation of $ \breve{\boldsymbol \alpha}_{\upsilon} $ and $ \breve{\boldsymbol \beta}_{\upsilon} $ is the same as (\ref{equations}). Therefore, we have
\begin{equation}
	\breve{\mu}_{\upsilon i} = \frac{\breve{\alpha}_{\upsilon i}}{\breve{\beta}_{\upsilon i}} =  \frac{2 \gamma_{\upsilon i} + 1}{(\mathbf D_h\breve{\boldsymbol \mu}_x)_i^2 + (\mathbf D_v\breve{\boldsymbol \mu}_x)_i^2 + \left\langle \breve{\boldsymbol \sigma}_x^2, (\mathbf d_{hi}^2 + \mathbf d_{vi}^2\right\rangle  + 2 \phi_{\upsilon i}},
\end{equation}   
For simplify computation, we approximate $ \left\langle \breve{\boldsymbol \sigma}_x^2, \mathbf d_{hi}^2 + \mathbf d_{vi}^2\right\rangle  $ with $ 4\breve{\sigma}_{xi}^2 $. Then, the parameter $ \breve{\boldsymbol \mu}_{\upsilon} $ is given by
\begin{equation}\label{eq:muupsilon}
	\breve{\boldsymbol \mu}_{\upsilon} = \frac{\breve{\boldsymbol \alpha}_{\upsilon}}{\breve{\boldsymbol \beta}_{\upsilon}} \approx \frac{2 \boldsymbol \gamma_{\upsilon} + 1}{(\mathbf D_h\breve{\boldsymbol \mu}_x)^2 + (\mathbf D_v\breve{\boldsymbol \mu}_x)^2 + 4\breve{\boldsymbol\sigma}_x^2  + 2\boldsymbol\phi_{\upsilon}}.
\end{equation}
Similarly, according to (\ref{equations}), the parameter $ \breve{\boldsymbol \mu}_{\omega} $ is given by
\begin{equation}\label{eq:muomega}
	\breve{\boldsymbol \mu}_{\omega} = \frac{\breve{\boldsymbol \alpha}_{\omega}}{\breve{\boldsymbol \beta}_{\omega}} = \frac{2 \boldsymbol \gamma_{\omega} + 1}{\breve{\boldsymbol \mu}_z^2 + \breve{\boldsymbol\sigma}_z^2  + 2\boldsymbol\phi_{\omega}}.
\end{equation}
The parameters of $ \boldsymbol \rho $ can be computed by minimizing 
\begin{equation}
	\begin{aligned}
		&\mbox{KL}(\breve{q}(\boldsymbol \rho)||p(\boldsymbol \rho)) -\mathbb E_{\breve{q}(\boldsymbol \rho)}\left[ \log p(\mathbf y|\mathbf A, \mathbf x,\mathbf z,\mathbf m,\boldsymbol \rho) \right]\\
		= &\mbox{KL}(\breve{q}(\boldsymbol \rho)||p(\boldsymbol \rho)p(\mathbf y|\mathbf A, \mathbf x,\mathbf z,\mathbf m,\boldsymbol \rho)).
	\end{aligned}
\end{equation}
This induces the formula of computing the parameter $ \breve{\boldsymbol \mu}_{\rho} $ as follows,
\begin{equation}\label{eq:murho}
	\breve{\boldsymbol \mu}_{\rho} = \frac{\breve{\boldsymbol \alpha}_{\rho}}{\breve{\boldsymbol \beta}_{\rho}} = \frac{2 \boldsymbol \gamma_{\rho} + 1}{(\mathbf y - \mathbf A (\mathbf x + \mathbf z) - \mathbf m)^2 + 2\boldsymbol\phi_{\rho}}.
\end{equation}

\noindent\textbf{Step 2: Infer $ \left\lbrace \breve{\boldsymbol \mu}_x, \breve{\boldsymbol \sigma}_x \right\rbrace $, $ \left\lbrace \breve{\boldsymbol \mu}_z, \breve{\boldsymbol \sigma}_z \right\rbrace $ and $ \left\lbrace \breve{\boldsymbol \mu}_m, \breve{\boldsymbol \sigma}_m \right\rbrace $ }

Given $ \breve{\boldsymbol \mu}_{\rho} $, minimizing $ - \mathbb E_{\breve{q}(\boldsymbol \rho)}\left[ \log p(\mathbf y|\mathbf A, \mathbf x,\mathbf z,\mathbf m,\boldsymbol \rho) \right] $ in (\ref{eq:ELBO}) induces a loss as follows,
\begin{equation}\label{lossy}
	\mathcal L_y = \frac{1}{2}\left\| \mathbf y - \mathbf A (\mathbf x + \mathbf z) - \mathbf m \right\|_{\mathbf M_{\rho}}^2.
\end{equation}
where, $ \mathbf M_{\rho} = \mbox{diag}(\breve{\boldsymbol \mu}_{\rho}) $. This is developed to maximize the likelihood given $ \mathbf y $ and ensures the consistency between restorations and observations. 

Given $ \breve{\boldsymbol \mu}_{\upsilon} $, minimizing $ \mbox{KL}(\breve{q}(\mathbf x)\breve{q}(\boldsymbol \upsilon)||p(\mathbf x|\boldsymbol \upsilon)p(\boldsymbol \upsilon)) $ in (\ref{eq:KLs}) induces two losses as follows,
\begin{equation}\label{lossx}
	\begin{aligned}
		\mathcal L_{\breve{\mu}_x} &= \frac{1}{2}\| \mathbf D_h\breve{\boldsymbol\mu}_x \|^2_{\mathbf M_{\upsilon}} + \frac{1}{2}\| \mathbf D_v\breve{\boldsymbol\mu}_x \|^2_{\mathbf M_{\upsilon}}\\
		\mathcal L_{\breve{\sigma}_x} &= \frac{1}{2} [ \langle 4\breve{\boldsymbol \mu}_{\upsilon},\breve{\boldsymbol\sigma}_x^2 \rangle  - \langle \mathbf 1, \log(\breve{\boldsymbol \sigma}_{x}^2) \rangle  ]
	\end{aligned} 
\end{equation}
where, $ \mathbf M_{\upsilon} = \mbox{diag}(\breve{\boldsymbol \mu}_{\upsilon})$, $ \mathbf 1 $ is a vector with all elements to be one, and $ 4\breve{\mu}_{\upsilon i} $ is an approximation of $ \left\langle \breve{\boldsymbol \mu}_{\upsilon}, \mathbf d_{hi}^2 + \mathbf d_{vi}^2\right\rangle $ in (\ref{equations}). $ \mathcal L_{\breve{\mu}_x} $ is aimed to regularize $ \breve{\boldsymbol \mu}_x $ to be piece-wisely smooth, and $ \mathcal L_{\breve{\sigma}_x} $ could prevent $ \breve{q}(\mathbf x) $ from degrading to a one-point distribution. 

Given $ \breve{\boldsymbol \mu}_{\omega} $, minimizing $ \mbox{KL}(\breve{q}(\mathbf z)\breve{q}(\boldsymbol \omega)||p(\mathbf z|\boldsymbol \omega)p(\boldsymbol \omega)) $ in (\ref{eq:KLs}) induces two losses as follows, 
\begin{equation}\label{lossz}
	\begin{aligned}
		\mathcal L_{\breve{\mu}_z} &= \frac{1}{2} \left\| \breve{\boldsymbol\mu}_z \right\|^2_{\mathbf M_{\omega}}\\
		\mathcal L_{\breve{\sigma}_z} &= \frac{1}{2} [ \langle \breve{\boldsymbol \mu}_{\omega},\breve{\boldsymbol\sigma}_z^2\rangle  - \langle \mathbf 1, \log(\breve{\boldsymbol \sigma}_{z}^2) \rangle],
	\end{aligned} 
\end{equation} 
where, $ \mathbf M_{\omega} = \mbox{diag}(\breve{\boldsymbol \mu}_{\omega}) $. $ \mathcal L_{\breve{\mu}_z} $ is to impose on $ \breve{\boldsymbol \mu}_z $ to be sparse, and $ \mathcal L_{\breve{\sigma}_z} $ could prevent $ \breve{q}(\mathbf z) $ from degrading to a one-point distribution.

Given $ \boldsymbol \mu_0 = \mathbf 0 $ and $ \sigma_0 $, minimizing $ \mbox{KL}(\breve{q}(\mathbf m)||p(\mathbf m)) $ in (\ref{eq:KLs}) induces two losses as follows,
\begin{equation}\label{lossm}
	\mathcal L_{\breve{\mu}_m} = \frac{\sigma_0}{2} \left\|\breve{\boldsymbol\mu}_m \right\|_2^2 \mbox{ and } \mathcal L_{\breve{\sigma}_m} = \frac{1}{2} [ \langle \sigma_0 \mathbf 1, \breve{\boldsymbol\sigma}_m^2\rangle  - \langle \mathbf 1, \log(\breve{\boldsymbol \sigma}_{m}^2) \rangle ],
\end{equation}
where, $ \mathcal L_{\breve{\mu}_m} $ is aimed to constraint the mean of noise, and $ \mathcal L_{\breve{\sigma}_m} $ could prevent $ \breve{q}(\mathbf m) $ from degrading to a one-point distribution. 

Overall, \textit{given} $ \breve{\boldsymbol \mu}_{\upsilon} $, $ \breve{\boldsymbol \mu}_{\omega} $, $ \breve{\boldsymbol \mu}_{\rho} $, $ \boldsymbol \mu_0 = \mathbf 0 $, and $ \sigma_0 $, we infer $ \left\lbrace \breve{\boldsymbol \mu}_x, \breve{\boldsymbol \sigma}_x \right\rbrace $, $ \left\lbrace \breve{\boldsymbol \mu}_z, \breve{\boldsymbol \sigma}_z \right\rbrace $ and $ \left\lbrace \breve{\boldsymbol \mu}_m, \breve{\boldsymbol \sigma}_m \right\rbrace $ from an observation $ \mathbf y $ by minimizing the following variational loss,
\begin{equation}\label{eq:VBloss}
	\mathcal L_{var}(\mathbf y) = \mathcal L_y + \mathcal L_{\breve{\mu}_x} + \mathcal L_{\breve{\mu}_z} + \mathcal L_{\breve{\mu}_m} + \mathcal L_{\breve{\sigma}_x} + \mathcal L_{\breve{\sigma}_z} + \mathcal L_{\breve{\sigma}_m},
\end{equation} 

The element-wise expectations with respect to $ \breve{q}(\boldsymbol \psi) $ are given as follows,

\begin{eqnarray}
	&&
	\begin{split}\small\label{elbo:01}
		\mathbb E\left[  \log p(\mathbf y|\boldsymbol \psi)\right]  = &-\frac{d_y}{2}\log(2\pi) + \frac{1}{2}\textstyle\sum_{i=1}^{d_y} \left[ \Psi(\breve{\alpha}_{\rho i}) - \log(\breve{\beta}_{\rho i}) \right]\\
		&- \frac{1}{2} \left\| \mathbf y-\mathbf A(\breve{\boldsymbol \mu}_x +\breve{\boldsymbol \mu}_z) -\breve{\boldsymbol \mu}_m\right\|_{ diag(\breve{\boldsymbol \mu}_{\rho})}^2\\
		&- \frac{1}{2} [ \langle \mathbf A^\top \mbox{diag}(\breve{\boldsymbol \mu}_{\rho})\mathbf A, \breve{\boldsymbol \Sigma}_x + \breve{\boldsymbol \Sigma}_z \rangle  + \langle \mbox{diag}(\breve{\boldsymbol \mu}_{\rho}), \breve{\boldsymbol \Sigma}_m \rangle  ],
	\end{split}\\
	&&
	\begin{split}\small\label{elbo:02}
		\mathbb E\left[ \log p(\mathbf x|\boldsymbol \upsilon)\right]  = &-\frac{d_u}{2}\log(2\pi) + \frac{1}{2}\textstyle\sum_{i=1}^{d_u} \left[  \Psi(\breve{\alpha}_{\upsilon i}) - \log(\breve{\beta}_{\upsilon i}) \right]  \\
		&- \frac{1}{2}\left[ \left\| \mathbf D_h\breve{\boldsymbol \mu}_x\right\| _{diag(\breve{\boldsymbol \mu}_{\upsilon})}^2 + \left\|  \mathbf D_v\breve{\boldsymbol \mu}_x\right\|_{diag(\breve{\boldsymbol \mu}_{\upsilon})}^2 \right]\\
		&- \frac{1}{2} \langle  \mathbf D_h^\top\mbox{ diag}(\breve{\boldsymbol \mu}_{\upsilon})\mathbf D_h + \mathbf D_v^\top\mbox{ diag}(\breve{\boldsymbol \mu}_{\upsilon})\mathbf D_v, \breve{\boldsymbol \Sigma}_x \rangle, 
	\end{split}\\
	&&
	\begin{split}\small\label{elbo:03}
		\mathbb E\left[ \log p(\boldsymbol \upsilon|\boldsymbol \phi_{\upsilon},\boldsymbol \gamma_{\upsilon})\right]  &=\textstyle\sum_{i=1}^{d_u} \left[  \gamma_{\upsilon i}\log(\phi_{\upsilon i}) - \log(\Gamma(\gamma_{\upsilon i})) \right] \\
		&+ \textstyle\sum_{i=1}^{d_u} \left[ (\gamma_{\upsilon i} -1) (\Psi(\breve{\alpha}_{\upsilon i}) - \log(\breve{\beta}_{\upsilon i}) ) - \phi_{\upsilon i}\breve{\alpha}_{\upsilon i}/\breve{\beta}_{\upsilon i} \right],
	\end{split}\\
	&&
	\begin{split}\small\label{elbo:04}
		\mathbb E\left[ \log p(\mathbf z|\boldsymbol \omega)\right]  = &-\frac{d_u}{2} \log(2\pi) + \frac{1}{2}\textstyle\sum_{i=1}^{d_u} \left[ \Psi(\breve{\alpha}_{\omega i}) - \log(\breve{\beta}_{\omega i}) \right]\\
		&- \frac{1}{2}\left[ \left\|\breve{\boldsymbol \mu}_z \right\|_{diag(\breve{\boldsymbol \mu}_{\omega})}^2 + \langle \mbox{diag}(\breve{\boldsymbol \mu}_{\omega}), \breve{\boldsymbol \Sigma}_z \rangle  \right], 
	\end{split}\\
	&&
	\begin{split}\small\label{elbo:05}
		\mathbb E\left[ \log p(\boldsymbol \omega|\boldsymbol \phi_{\omega},\boldsymbol \gamma_{\omega})\right]  &= \textstyle\sum_{i=1}^{d_u}\left[ \gamma_{\omega i}\log(\phi_{\omega i}) -\log(\Gamma(\gamma_{\omega i})) \right] \\
		&+ \textstyle\sum_{i=1}^{d_u} \left[ (\gamma_{\omega i} - 1) (\Psi(\breve{\alpha}_{\omega i}) - \log(\breve{\beta}_{\omega i}) ) - \phi_{\omega i}\breve{\alpha}_{\omega i}/\breve{\beta}_{\omega i}  \right],
	\end{split}\\
	&&
	\begin{split}\small\label{elbo:06}
		\mathbb E\left[ \log p(\mathbf m|\boldsymbol \mu_0,\sigma_0)\right]  = -\frac{d_y}{2}\log(2\pi) + \frac{d_y}{2}\log(\sigma_0) -\frac{\sigma_0}{2}(\left\|\breve{\boldsymbol \mu}_m -\boldsymbol \mu_0 \right\|_2^2 + \mathbf \langle I, \breve{\boldsymbol \Sigma}_m \rangle  ),
	\end{split}\\
	&&
	\begin{split}\small\label{elbo:07}
		\mathbb E\left[ \log p(\boldsymbol \rho|\boldsymbol \phi_{\rho},\boldsymbol \gamma_{\rho})\right]  &= \textstyle\sum_{i=1}^{d_y}\left[ \gamma_{\rho i}\log(\phi_{\rho i}) - \log(\Gamma(\gamma_{\rho i})) \right] \\
		&+ \textstyle\sum_{i=1}^{d_y} \left[ (\gamma_{\rho i} -1) (\Psi(\breve{\alpha}_{\rho i}) - \log(\breve{\beta}_{\rho i}) ) -\phi_{\rho i}\breve{\alpha}_{\rho i}/\breve{\beta}_{\rho i} \right],
	\end{split}\\
	&&
	\begin{split}\small\label{elbo:08}
		\mathbb E\left[ \log \breve{q}(\mathbf x)\right]  = -\frac{d_u}{2} \log(2\pi e) - \frac{1}{2}\log(|\breve{\boldsymbol \Sigma}_x|),
	\end{split}\\
	&&
	\begin{split}\small\label{elbo:09}
		\mathbb E\left[ \log \breve{q}(\boldsymbol \upsilon)\right]  =\textstyle\sum_{i=1}^{d_u}\left[  -\breve{\alpha}_{\upsilon i} + \log(\breve{\beta}_{\upsilon i}) - \log(\Gamma(\breve{\alpha}_{\upsilon i}))+(\breve{\alpha}_{\upsilon i}-1)\Psi(\breve{\alpha}_{\upsilon i})\right],
	\end{split}\\
	&&
	\begin{split}\small\label{elbo:10}
		\mathbb E\left[ \log \breve{q}(\mathbf z)\right]  = -\frac{d_u}{2} \log(2\pi e) - \frac{1}{2}\log(|\breve{\boldsymbol \Sigma}_z|),
	\end{split}\\
	&&
	\begin{split}\small\label{elbo:11}
		\mathbb E\left[ \log \breve{q}(\boldsymbol \omega)\right] =\textstyle\sum_{i=1}^{d_u} \left[ -\breve{\alpha}_{\omega i} + \log(\breve{\beta}_{\omega i}) - \log(\Gamma(\breve{\alpha}_{\omega i}))+(\breve{\alpha}_{\omega i}-1)\Psi(\breve{\alpha}_{\omega i}) \right],
	\end{split}\\
	&&
	\begin{split}\small\label{elbo:12}
		\mathbb E\left[ \log \breve{q}(\mathbf m)\right] =-\frac{d_y}{2} \log(2\pi e) - \frac{1}{2}\log(|\breve{\boldsymbol \Sigma}_m|),
	\end{split}\\
	&&
	\begin{split}\small\label{elbo:13}
		\mathbb E\left[ \log \breve{q}(\boldsymbol \rho)\right] =\textstyle\sum_{i=1}^{d_y}\left[ -\breve{\alpha}_{\rho i} + \log(\breve{\beta}_{\rho i}) - \log(\Gamma(\breve{\alpha}_{\rho i}))+(\breve{\alpha}_{\rho i}-1)\Psi(\breve{\alpha}_{\rho i})\right],
	\end{split}
\end{eqnarray}
where, $ \Gamma(\cdot) $ and $ \Psi(\cdot) $ denote Gamma and Digamma functions, respectively.	

\bibliographystyle{abbrv}
\bibliography{references.bib}

\begin{thebibliography}{10}

\bibitem{Ayasso/2012}
H.~Ayasso, T.~Rodet, and A.~Abergel.
\newblock A variational bayesian approach for unsupervised super-resolution
  using mixture models of point and smooth sources applied to astrophysical
  map-making.
\newblock {\em Inverse Problems}, 28(12):125005, 2012.

\bibitem{Babacan/2009}
S.~D. Babacan, R.~Molina, and A.~K. Katsaggelos.
\newblock Variational bayesian blind deconvolution using a total variation
  prior.
\newblock {\em IEEE Trans. Image Process.}, 18(1):12--26, 2009.

\bibitem{Batson/2019}
J.~Batson and L.~Royer.
\newblock {Noise2Self}: blind denoising by self-supervision.
\newblock In {\em Proc. Int. Conf. Mach. Learn.}, pages 524--533, 2019.

\bibitem{Bell/2019}
S.~Bell-Kligler, A.~Shocher, and M.~Irani.
\newblock Blind super-resolution kernel estimation using an internal-gan.
\newblock In {\em Proc. Adv. Neural Inf. Process. Syst.}, 2019.

\bibitem{Bevilacqua/2012}
M.~Bevilacqua, A.~Roumy, C.~Guillemot, and M.~L. AlberMorel.
\newblock Low-complexity single-image super-resolution based on nonnegative
  neighbor embedding.
\newblock In {\em Proc. British Mach. Vis. Conf.}, 2012.

\bibitem{Bigdeli/2017}
S.~A. Bigdeli, M.~Jin, P.~Favaro, and M.~Zwicker.
\newblock Deep mean-shift priors for image restoration.
\newblock In {\em Proc. Adv. Neural Inf. Process. Syst.}, pages 763--772, 2017.

\bibitem{Bulat/2018}
A.~Bulat, J.~Yang, and G.~Tzimiropoulos.
\newblock To learn image super-resolution, use a gan to learn how to do image
  degradation first.
\newblock In {\em Proc. Eur. Conf. Compute. Vis.}, pages 187--202, 2018.

\bibitem{Burger/2012}
H.~Burger, C.~Schuler, and S.~Harmeling.
\newblock Image denoising: can plain neural networks compete with {BM3D}?
\newblock In {\em Proc. IEEE Conf. Compute. Vis. Pattern Recognit.}, pages
  2392--2399, 2012.

\bibitem{Candes_tight/2011}
E.~J. Cand\`{e}s and Y.~Plan.
\newblock Tight oracle inequalities for low-rank matrix recovery from a minimal
  number of noisy random measurements.
\newblock {\em IEEE Trans. Inf. Theory}, 57(4):2342--2359, 2011.

\bibitem{Candes/2011}
E.~Cand$\grave{e}$s, X.~Li, Y.~Ma, and J.~Wright.
\newblock Robust principal component analysis?
\newblock {\em J. ACM}, 58(3):11, 2011.

\bibitem{Celebi/2014}
M.~E. Celebi and B.~Smolka.
\newblock {\em Advances in low-level color image}.
\newblock Springer, Dordrecht, 2014.

\bibitem{Chambolle/1997}
A.~Chambolle and P.~Lions.
\newblock Image recovery via total variation minimization and related problems.
\newblock {\em Numer. Math.}, 76(2):167--188, 1997.

\bibitem{Chan/2003}
R.~Chan, T.~Chan, L.~Shen, and Z.~Shen.
\newblock Wavelet algorithms for high-resolution image reconstruction.
\newblock {\em SIAM J. Sci. Comput.}, 24(4):1408--1432, 2003.

\bibitem{Chan/2006}
T.~Chan, S.~Esedoglu, F.~Park, and A.~Yip.
\newblock Total variation image restoration: Overview and recent developments.
\newblock In {\em Handbook of Mathematical Models in Computer Vision}, pages
  17--31, 2006.

\bibitem{Chantas/2008}
G.~Chantas, N.~Galatsanos, A.~Likas, and M.~Saunders.
\newblock Variation bayesian image restoration based on a product of
  t-distributions image prior.
\newblock {\em IEEE Trans. Image Process.}, 17(10):1795--1805, 2008.

\bibitem{Chen/2020}
H.~Chen, Y.~Wang, T.~Guo, C.~Xu, Y.~Deng, Z.~Liu, S.~Ma, C.~Xu, C.~Xu, and
  W.~Gao.
\newblock Pre-trained image processing transformer.
\newblock {\em arXiv e-print, arXiv:2012.00364}, 2020.

\bibitem{Chen/2018}
J.~Chen, J.~Chen, H.~Chao, and M.~Yang.
\newblock Image blind denoising with generative adversarial network based noise
  modeling.
\newblock In {\em Proc. IEEE Conf. Compute. Vis. Pattern Recognit.}, pages
  3155--3164, 2018.

\bibitem{Diaconis/1979}
R.~Diaconis and D.~Ylvisaker.
\newblock Conjugate priors for exponential families.
\newblock {\em The Annals of Statistics}, 7(2):269--281, 1979.

\bibitem{Dong/2016}
C.~Dong, C.~Loy, K.~He, and X.~Tang.
\newblock Image super-resolution using deep convolutional networks.
\newblock {\em IEEE Trans. Pattern Anal. Mach. Intell.}, 38(2):259--307, 2016.

\bibitem{Dong/2019}
W.~Dong, P.~Wang, W.~Yin, G.~Shi, F.~Wu, and X.~Lu.
\newblock Denoising prior driven deep neural network for image restoration.
\newblock {\em IEEE Trans. Pattern Anal. Mach. Intell.}, 41(10):2305--2318,
  2019.

\bibitem{Figueiredo/2003}
M.~Figueiredo and R.~Nowak.
\newblock An em algorithm for wavelet-based image restoration.
\newblock {\em IEEE Trans. Image Process}, 12(8):906--916, 2003.

\bibitem{Gal_Uncertianty/2016}
Y.~Gal.
\newblock {\em Uncertainty in Deep Learning}.
\newblock PhD thesis, University of Cambridge, 2016.

\bibitem{Guerrero/2008}
J.~A. Guerrero-Col\'{o}n, L.~Mancera, and J.~Portilla.
\newblock Image restoration using space-variant gaussian scale mixtures in
  overcomplete pyramids.
\newblock {\em IEEE Trans. Image Process.}, 17(1):27--41, 2008.

\bibitem{He/2016}
K.~He, X.~Zhang, J.~Sun, and S.~Ren.
\newblock Deep residual learning for image recognition.
\newblock In {\em Proc. IEEE Conf. Compute. Vis. Pattern Recognit.}, pages
  770--778, 2016.

\bibitem{He/OISR/2019}
X.~He, Z.~Mo, P.~Wang, Y.~Liu, M.~Yang, and J.~Cheng.
\newblock {ODE}-inspired network design for single image super-resolution.
\newblock In {\em Proc. IEEE Int. Conf. Comput. Vis.}, 2019.

\bibitem{Healey/1994}
G.~Healey and R.~Kondepudy.
\newblock Radiometric ccd camera calibration and noise estimation.
\newblock {\em IEEE Trans. Pattern Anal. Mach. Intell.}, 16(3):267--276, 1994.

\bibitem{Helou/2020}
M.~E. Helou and S.~S\"{u}sstrunk.
\newblock Blind universal bayesian image denoising with gaussian noise level
  learning.
\newblock {\em IEEE Trans. Image Process.}, 29:4885--4897, 2020.

\bibitem{Huang/2015}
J.~B. Huang, A.~Singh, and N.~Ahuja.
\newblock Single image super-resolution from transformed self-exemplars.
\newblock In {\em Proc. IEEE Conf. Compute. Vis. Pattern Recognit.}, 2015.

\bibitem{Hunt/1977}
B.~R. Hunt.
\newblock Bayesian methods in nonlinear digital image restoration.
\newblock {\em IEEE Trans. Comput.}, c-26(3):219--229, 1977.

\bibitem{Izadi/2020}
S.~Izadi and G.~Hamarneh.
\newblock Patch-based non-local bayesian networks for blind confocal microscopy
  denoising.
\newblock {\em arXiv e-print, arXiv:2003.11177}, 2020.

\bibitem{Jalobeanu/2004}
A.~Jalobeanu, L.~Blanc-F\'{e}raud, and J.~Zerubia.
\newblock An adaptive gaussian model for satellite image deblurring.
\newblock {\em IEEE Trans. Image Process.}, 13(4):613--621, 2004.

\bibitem{Jan/2006}
J.~Jan.
\newblock {\em Medical image processing, reconstruction and restoration}.
\newblock Crc press, 2006.

\bibitem{Jensen/1995}
J.~R. Jensen.
\newblock {\em Introductory digital image processing: a remote sensing
  perspective}.
\newblock Prentice Hall PTR, Upper Saddle River, NJ, USA, 1995.

\bibitem{Ji/2020}
X.~Ji, Y.~Cao, Y.~Tai, C.~Wang, J.~Li, and F.~Huang.
\newblock Real-world super-resolution via kernel estimation and noise injetion.
\newblock In {\em Proc. IEEE Conf. Compute. Vis. Pattern Recognit. Workshops},
  2020.

\bibitem{Johnson/2016}
J.~Jonhson, A.~Alahi, and L.~Fei-Fei.
\newblock Perceptual losses for real-time style transfer and super-resolution.
\newblock In {\em Proc. Eur. Conf. Compute. Vis. Workshops}, pages 694--711,
  2016.

\bibitem{Kim1/2016}
J.~Kim, J.~Lee, and K.~Lee.
\newblock Accurate image super-resolution using very deep convolutional
  networks.
\newblock In {\em Proc. IEEE Conf. Compute. Vis. Pattern Recognit.}, pages
  1646--1654, 2016.

\bibitem{Kim/2016}
J.~Kim, J.~Lee, and K.~Lee.
\newblock Deeply-recursive convolutional network for image super-resolution.
\newblock In {\em Proc. IEEE Conf. Compute. Vis. Pattern Recognit.}, pages
  1637--1645, 2016.

\bibitem{Kingma/2019}
D.~P. Kingma and M.~Welling.
\newblock An introduction to variational antoencoders.
\newblock {\em Foundations and Trends in Machine Learning}, 12(4):307--392,
  2019.

\bibitem{Koltchinskii/2011}
V.~Koltchinskii, K.~Lounici, and A.~B. Tsybakov.
\newblock Nuclear-norm penalization and optimal rates for noisy low-rank matrix
  completion.
\newblock {\em The Annals of Statistics}, 39(5):2302--2329, 2011.

\bibitem{Krull/2019}
A.~Krull, T.-O. Buchholz, and F.~Jug.
\newblock {Noise2Void}-learning denoising from single noisy images.
\newblock In {\em Proc. IEEE Conf. Compute. Vis. Pattern Recognit.}, pages
  2129--2137, 2019.

\bibitem{Lai/2017}
W.~Lai, J.~Huang, N.~Ahuja, and M.~Yang.
\newblock Deep laplacian pyramid networks for fast and accurate
  super-resolution.
\newblock In {\em Proc. IEEE Conf. Compute. Vis. Pattern Recognit.}, pages
  624--632, 2017.

\bibitem{Laine/2019}
S.~Laine, T.~Karras, J.~Lehtinen, and T.~Aila.
\newblock High-quality self-supervised deep image denoising.
\newblock In {\em Proc. Adv. Neural Inf. Process. Syst.}, volume~32, pages
  6970--6980, 2019.

\bibitem{Ledig/2017}
C.~Ledig, L.~Theis, F.~Huszar, J.~Caballero, A.~Cunningham, A.~Acosta,
  A.~Aitken, A.~Tejani, J.~Totz, Z.~Wang, and W.~Shi.
\newblock Photo-realistic single image super-resolution using a generative
  adversarial network.
\newblock In {\em Proc. IEEE Conf. Compute. Vis. Pattern Recognit.}, pages
  4681--4690, 2017.

\bibitem{Lehtinen/2018}
J.~Lehtinen, J.~Munkberg, J.~Hasselgren, S.~Laine, T.~karas, M.~Aittala, and
  T.~Aila.
\newblock {Noise2Noise:} learning image restoration without clean data.
\newblock In {\em Proc. Int. Conf. Mach. Learn.}, 2018.

\bibitem{Liang_FKP/2021}
J.~Liang, K.~Zhang, S.~Gu, L.~V. Gool, and R.~Timofte.
\newblock Flow-based kernel prior with application to blind super-resolution.
\newblock In {\em Proc. IEEE Conf. Compute. Vis. Pattern Recognit.}, pages
  10596--10605, 2021.

\bibitem{Lim/2017}
B.~Lim, S.~Son, H.~Kim, S.~Nah, and K.~Lee.
\newblock Enhanced deep residual networks for single image super-resolution.
\newblock In {\em Proc. IEEE Conf. Compute. Vis. Pattern Recognit. Workshops},
  pages 136--144, 2017.

\bibitem{Lugmayr/2020}
A.~Lugmayr, M.~Danelljan, L.~V. Gool, and R.~Timofte.
\newblock Srflow: Learning the super-resolution space with normalizing flow.
\newblock In {\em Proc. Eur. Conf. Compute. Vis.}, pages 715--732, 2020.

\bibitem{Lugmayr/2019}
A.~Lugmayr, M.~Danelljan, and R.~Timofte.
\newblock Unsupervised learning for real-world super-resolution.
\newblock In {\em Proc. IEEE Int. Conf. Comput. Vis. Workshop}, 2019.

\bibitem{Maeda/2020}
S.~Maeda.
\newblock Unpaired image super-resolution using pseudo-supervision.
\newblock In {\em Proc. IEEE Conf. Compute. Vis. Pattern Recognit.}, 2020.

\bibitem{Martin/2001}
D.~Martin, C.~Fowlkes, D.~Tal, and J.~Malik.
\newblock A database of human segmented natural images and its application to
  evaluating segmentation algorithms and measuring ecological statistics.
\newblock In {\em Proc. IEEE Int. Conf. Comput. Vis.}, 2001.

\bibitem{Molina/1994}
R.~Molina.
\newblock On the hierarchical bayesian approach to image restoration:
  applications to astronomical images.
\newblock {\em IEEE Trans. Pattern Anal. Mach. Intell.}, 16(11):1122--1128,
  1994.

\bibitem{Molina/2003}
R.~Molina, J.~Mateos, A.~K. Katsaggelos, and M.~Vega.
\newblock Bayesian multichannel image restoration using compound gauss-markov
  random fields.
\newblock {\em IEEE Trans. Image Process.}, 12(12):1642--1654, 2003.

\bibitem{Nah/2017}
S.~Nah, T.~H. Kim, and K.~M. Lee.
\newblock Deep multi-scale convolutional neural network for dynamic scene
  deblurring.
\newblock In {\em Proc. IEEE Conf. Compute. Vis. Pattern Recognit.}, pages
  257--265, 2017.

\bibitem{Osher/2005}
S.~Osher, M.~Burger, D.~Goldfarb, J.~Xu, and W.~Yin.
\newblock An iterative regularization method for total variation-based image
  restoration.
\newblock {\em Multiscale Model. Simulation}, 4(2):460--489, 2005.

\bibitem{Pan/2006}
R.~Pan and S.~J. Reeves.
\newblock Efficient huber-markov edge-preserving image restoration.
\newblock {\em IEEE Trans. Image Process.}, 15(12):3728--3735, 2006.

\bibitem{Portilla/2015}
J.~Portilla, A.~Trist\'{a}n-Vega, and I.~W. Selesnick.
\newblock Efficient and robust image restoration using multiple-feature
  l2-relaxed sparse analysis priors.
\newblock {\em IEEE Trans. Image Process.}, 24(12):5046--5059, 2015.

\bibitem{Qian/1993}
W.~Qian and D.~M. Titterington.
\newblock Bayesian image restoration: an application to edge-preserving surface
  recovery.
\newblock {\em IEEE Trans. Pattern Anal. Mach. Intell.}, 15(7):748--752, 1993.

\bibitem{Ronneberger/2015}
O.~Ronneberger, P.~Fischer, and T.~Brox.
\newblock {U-Net}: Convolutional networks for biomedical image segmentation.
\newblock In {\em Med. Image Comput. Assist. Interv.}, pages 234--241, 2015.

\bibitem{Rudin/1992}
L.~Rudin, S.~Osher, and E.~Fatemi.
\newblock Nonlinear total variation based noise removal algorithms.
\newblock {\em Physica D}, 60(1-4):259--168, 1992.

\bibitem{Sajjadi/2017}
M.~Sajjadi, B.~Scholkopf, and M.~Hirsch.
\newblock {EnhanceNet}: single image super-resolution through automated texture
  synthesis.
\newblock In {\em Proc. IEEE Int. Conf. Comput. Vis.}, pages 4491--4500, 2017.

\bibitem{Shocher/2018}
A.~Shocher, N.~Cohen, and M.~Irani.
\newblock Zero-shot super-resolution using deep internal learning.
\newblock In {\em Proc. IEEE Conf. Compute. Vis. Pattern Recognit.}, 2018.

\bibitem{Soh/2020}
J.~W. Soh, S.~Cho, and N.~I. Cho.
\newblock Meta-transfer learning for zero-shot super-resolution.
\newblock In {\em Proc. IEEE Conf. Compute. Vis. Pattern Recognit.}, 2020.

\bibitem{Tikhonov/1977}
A.~Tikhonov and V.~Arsenin.
\newblock {\em Solutions of ill-posed problems}.
\newblock Springer-Verlag, 1986.

\bibitem{Timofte/2014}
R.~Timofte, V.~{De Smet}, and L.~{Van Gool}.
\newblock {A+:} adjusted anchored neighborhood regression for fast
  super-resolution.
\newblock In {\em Proc. Asian Conf. Comput. Vis.}, pages 111--126, 2014.

\bibitem{Tong/2017}
T.~Tong, G.~Li, X.~Liu, and Q.~Gao.
\newblock Image super-resolution using dense skip connections.
\newblock In {\em Proc. IEEE Int. Conf. Comput. Vis.}, pages 4799--4807, 2017.

\bibitem{Ulyanov/2018}
D.~Ulyanov, A.~Vedaldi, and V.~Lempitsky.
\newblock Deep image prior.
\newblock In {\em Proc. IEEE Conf. Compute. Vis. Pattern Recognit.}, 2018.

\bibitem{Wang/2018}
X.~Wang, S.~Wu, J.~Gu, Y.~Liu, C.~Dong, Y.~Qiao, and C.~Loy.
\newblock {ESRGAN}: Enhanced super-resolution generative adversarial networks.
\newblock In {\em Proc. Eur. Conf. Compute. Vis. Workshops}, 2018.

\bibitem{Yang/2008}
J.~Yang, J.~Wright, T.~Huang, and Y.~Ma.
\newblock Image super-resolution as sparse representation of raw image patches.
\newblock In {\em Proc. IEEE Conf. Compute. Vis. Pattern Recognit.}, pages
  1--8, 2008.

\bibitem{Yang/2010}
J.~Yang, J.~Wright, T.~S. Huang, and Y.~Ma.
\newblock Image super-resolution via sparse representation.
\newblock {\em IEEE Trans. Image Process.}, 19(11):2861--2873, 2010.

\bibitem{Yu/2018}
J.~Yu, Y.~Fan, J.~Yang, N.~Xu, Z.~Wang, X.~Wang, and T.~Huang.
\newblock Wide activation for efficient and accurate image super-resolution.
\newblock {\em arXiv e-print, arXiv:1808.08718}, 2018.

\bibitem{Yue/2019}
Z.~Yue, H.~Yong, Q.~Zhao, D.~Meng, and L.~Zhang.
\newblock Variational denoising network: Toward blind noise modeling and
  removal.
\newblock In {\em Proc. Adv. Neural Inf. Process. Syst.}, pages 1690--1701,
  2019.

\bibitem{Yue_VIRNet/2020}
Z.~Yue, H.~Yong, Q.~Zhao, L.~Zhang, and D.~Meng.
\newblock Variational image restoration network.
\newblock {\em arXiv e-print, arXiv:2008.10796v1}, 2020.

\bibitem{Zeyde/2012}
R.~Zeyde, M.~Elad, and M.~Protter.
\newblock On single image scale-up using sparse-representations.
\newblock In {\em Proc. 7th Int. Conf. Curves Surfaces}, pages 711--730, 2012.

\bibitem{Zhang_USRnet/2020}
K.~Zhang, L.~V. Gool, and R.~Timofte.
\newblock Deep unfolding network for image super-resolution.
\newblock In {\em Proc. IEEE Conf. Compute. Vis. Pattern Recognit.}, pages
  3217--3226, 2020.

\bibitem{Zhang_beyond/2017}
K.~Zhang, W.~Zuo, Y.~Chen, D.~Meng, and L.~Zhang.
\newblock Beyond a {Gaussian} denoiser: residual learning of deep {CNN} for
  image denoising.
\newblock {\em IEEE Trans. Image Process.}, 26(7):3142--3155, 2017.

\bibitem{Zhang_ffdnet/2018}
K.~Zhang, W.~Zuo, and L.~Zhang.
\newblock {FFDNet}: toward a fast and flexible solution for {CNN}-based image
  denoising.
\newblock {\em IEEE Trans. Image Process.}, 27(9):4608--4622, 2018.

\bibitem{Zhangyulun/2018}
Y.~Zhang, K.~Li, K.~Li, L.~Wang, B.~Zhong, and Y.~Fu.
\newblock Image super-resolution using very deep residual channel attention
  networks.
\newblock In {\em Proc. Eur. Conf. Compute. Vis.}, 2018.

\bibitem{Zhang/2018}
Y.~Zhang, Y.~Tian, Y.~Kong, B.~Zhong, and Y.~Fu.
\newblock Residual dense network for image super-resolution.
\newblock In {\em Proc. IEEE Conf. Compute. Vis. Pattern Recognit.}, pages
  2472--2481, 2018.

\end{thebibliography}
\end{document}